\documentclass[preprint,prd,showpacs,nobibnotes,preprintnumbers,superscriptaddress,nofootinbib,floatfix]{revtex4}
\usepackage{amsmath,amssymb,amsbsy,color}
\usepackage{epsfig}
\usepackage{psfrag}
\usepackage{epstopdf}
\usepackage{grffile} 
\usepackage{multirow}
\usepackage{hyperref}
\usepackage[ruled,vlined,noline]{algorithm2e}
\usepackage{algorithmic}

\begin{document}

\preprint{MITP/14-005}
\preprint{RBRC-1059}
\preprint{BNL-103723-2014-JA}

\title{Covariant approximation averaging}

\author{Eigo~Shintani}
\email{shintani@kph.uni-mainz.de}
\affiliation{
PRISMA Cluster of Excellence,
Institut f{\"u}r Kernphysik and Helmholtz Institute Mainz,
Johannes Gutenberg-Universit{\"a}t Mainz, D-55099 Mainz, Germany
}
\affiliation{
  RIKEN-BNL Research Center, Brookhaven National Laboratory, Upton, NY 11973, USA
}

\author{Rudy~Arthur}
\affiliation{
CP$^3$-Origins and the Danish Institute for Advanced Study DIAS,
University of Southern Denmark, Campusvej 55, DK-5230 Odense M, Denmark
}

\author{Thomas~Blum}
\affiliation{
  Physics Department, University of Connecticut, Storrs, CT 06269-3046, USA
}
\affiliation{
  RIKEN-BNL Research Center, Brookhaven National Laboratory, Upton, NY 11973, USA
}

\author{Taku~Izubuchi}
\affiliation{
 High Energy Theory Group, Brookhaven National Laboratory, Upton, NY 11973, USA
}
\affiliation{
  RIKEN-BNL Research Center, Brookhaven National Laboratory, Upton, NY 11973, USA
}

\author{Chulwoo~Jung}
\affiliation{
 High Energy Theory Group, Brookhaven National Laboratory, Upton, NY 11973, USA
}

\author{Christoph~Lehner}
\affiliation{
 High Energy Theory Group, Brookhaven National Laboratory, Upton, NY 11973, USA
}

\begin{abstract}
We present a new class of statistical error reduction techniques for 
Monte-Carlo simulations. 
Using covariant symmetries, we show that correlation functions 
can be constructed from inexpensive approximations
without introducing any systematic bias in the final result.
We introduce a new class of covariant approximation averaging techniques, 
known as all-mode averaging (AMA), 
in which the approximation takes account of contributions of all eigenmodes 
through the inverse of the Dirac operator computed from the
conjugate gradient method with a relaxed stopping condition. 
In this paper we compare the performance
and computational cost of our new method with traditional methods 
using correlation functions and masses of the pion, nucleon, 
and vector meson in $N_f=2+1$ lattice QCD using domain-wall fermions. 
This comparison indicates that AMA significantly reduces statistical errors 
in Monte-Carlo calculations over conventional methods for the same cost.

\end{abstract}
\maketitle

\section{Introduction}
In order to increase the confidence we have in the
results of a Monte-Carlo simulation, 
a huge number of independent ensembles is always required.
In lattice QCD many important observables suffer from 
notoriously large statistical errors due to
fluctuations induced by the gauge fields used to compute expectation values, 
{\it e.g.}, the neutron electric dipole moment (EDM) 
\cite{Shintani:2005xg,Berruto:2005hg,Shintani:2006xr,Shintani:2008nt}, 
the hadronic contributions to the muon anomalous magnetic moment (g-2) 
\cite{Blum:2013qu}, the
$\eta$-$\eta'$ mass and mixing angle 
\cite{Christ:2010dd}, among others. 
The precise determination of these observables, which provide
important ingredients for the Standard Model (SM) and 
models beyond the SM, is a challenging task for lattice QCD.
In this paper we present a detailed study of a new technique to 
efficiently evaluate correlation functions in a Monte-Carlo simulation.
An earlier publication by some of us already described the method 
and provided a few examples~\cite{Blum:2012uh}.

In lattice QCD, the numerical path integral is evaluated
by Monte-Carlo simulation to 
compute the expectation value of an observable $\mathcal O[U]$ given as 
the weighted average over configurations of gauge (gluon) fields, link variables $U$ 
generated under probability distribution $P[U]$ on a lattice, in an ensemble, 
\begin{equation}
  \langle \mathcal O\rangle = \sum_U \mathcal O[U]P[U]
  = \sum_{i=1}^{N_{\rm conf}} \frac{1}{N_{\rm conf}}\mathcal O[U_i] + O(1/\sqrt{N_{\rm conf}})
  \textrm{ as }N_{\rm conf}\rightarrow \infty.
  \label{eq:expectval}
\end{equation}
To increase the accuracy of the 
ensemble average given the statistics of $N_{\rm conf}$ configurations,
the development of numerical algorithms to efficiently compute 
observables is an important task.
Traditionally translational symmetry of the correlation function is 
exploited to increase statistics,
\begin{equation}
  \langle \mathcal O(x,y)\rangle = \langle \mathcal O(x^g,y^g)\rangle,
  \label{eq:symg}
\end{equation}
where the distance between operators
on the shifted lattice sites is held constant,
$||x-y||=||x^g-y^g||$. 
Ignoring statistical correlations between operators on shifted sites, 
the different $N_G$ sets of $\mathcal O(x^g,y^g)$ with 
sink location $x^g$ and source location $y^g$
can be regarded as independent measurements.
However this naively requires $N_G$ times the computational cost of a single measurement.

The original idea to avoid the cost of $N_{G}$ measurements while still performing $N_{G}$ shifts
is low-mode averaging (LMA)
\cite{Giusti:2002sm,Giusti:2004yp,DeGrand:2004qw,DeGrand:2005vb},
in which the inverse of the Dirac operator for each of $g\in G$ is computed
from its low-lying eigenvectors.
The benefit of LMA is that, once the low-modes have been computed, 
the construction of the LMA estimator
is not only low-cost but also 
useful for low-mode deflation \cite{Luscher:2007es}, 
i.e. as a preconditioner in the conjugate gradient (CG) method.
There have been many lattice studies using LMA, primarily focused on 
low-mode dominated observables, 
for example low-energy constants 
in the $\varepsilon$-regime \cite{Fukaya:2007fb}, or
the chiral behavior of pseudoscalar mesons 
in the $p$-regime \cite{Noaki:2008iy}. 
They have shown that there is some benefit from LMA
for observables related to the pion.
On the other hand, attempts to use LMA for baryons
or heavy mesons \cite{Giusti:2005sx,Li:2010pw,Bali:2010se}, 
were not as successful,
presumably because these states are not dominated 
by a relatively small number of low-modes
(we also found recent attempts to use extended method 
called as low-mode substitution for baryon spectroscopy in \cite{Gong:2013vja}).

Recently we extended the LMA idea to efficiently handle the vast 
majority of hadronic states that are not dominated by low-modes \cite{Blum:2012uh}.
The idea is to include  
all modes of the Dirac operator but with
dramatically reduced computational cost compared to the usual conjugate gradient method.
By using covariant symmetries, approximate (and therefore inexpensive) 
correlation functions are used to compute expectation values without introducing any 
systematic error (bias).
{\it All-mode-averaging} (AMA) in which 
a relaxed stopping condition of the CG is employed as in \cite{Bali:2009hu} 
takes the contributions of all modes into account.
The method is broadly applicable to other fields using Monte-Carlo simulation, 
{\it e.g.} many-body systems, atomic systems and cold gas systems
(see \cite{0034-4885-74-2-026502,0034-4885-75-9-094501,Foulkes:2001zz,
Carlson:2012mh,Leidemann:2012hr}). 
This paper gives a detailed description of the covariant approximation 
averaging (CAA) with primary examples, LMA and AMA \cite{Blum:2012uh}.
We also present several numerical results with high precision and 
cost-performance comparison with standard methods. 

This paper is organized as follows:
in the next section we describe the CAA procedure and compare LMA and AMA. 
In Section \ref{sec:result} we show numerical results for AMA using
domain-wall fermions and compare to LMA and the standard multi-source method. 
In Section \ref{sec:other} we present several 
examples extending the approximation and the results of some numerical tests.
In the last section we summarize and discuss further extensions of AMA.
In Appendix \ref{sec:bias}, possible small bias of AMA due to finite precision floating 
point arithmetic are discussed, and we present how to remove them completely in Appendix \ref{sec:notCAA}.

\section{Covariant approximation averaging}
\label{seq:caa}
\subsection{General argument}
Under a symmetry transformation $g\in G$, 
the expectation value of the transformed functional $\mathcal O[U]$
(for example a hadron propagator) 
is equivalent to that computed on the transformed configuration $U^g$
\begin{equation}
  \langle \mathcal O^g[U]\rangle = \langle \mathcal O[U^g]\rangle,
  \label{eq:latsym}
\end{equation}
where $U^g(x) = U(x^g)$, while translational symmetry of the observable is  
expressed as $\mathcal O^g[U](x,y)=\mathcal O[U](x^g,y^g)$. 
If $\mathcal O[U]$ is covariant under the symmetry, 
on each gauge configuration
\begin{equation}
  \mathcal O^g[U] = \mathcal O[U^g],
  \label{eq:cov1}
\end{equation}
then there is the trivial identity
\begin{equation}
   \sum_{g\in G}\mathcal O^{g}[U] = \sum_{g\in G}\mathcal O[U^{g}],
   \label{eq:cov2}
\end{equation}
for a set of transformations $g\in G$ whose number of elements is $N_G$.
From Eq.~(\ref{eq:latsym}), (\ref{eq:cov1}) and (\ref{eq:cov2}), 
an average over a set of symmetry transformations is defined as
\begin{equation}
  \mathcal O_G[U] \equiv \frac{1}{N_G}\sum_{g\in G}\mathcal O^{g}[U]
  = \frac{1}{N_G}\sum_{g\in G}\mathcal O[U^{g}],
  \label{eq:o_g}
\end{equation}
and one sees that $\langle O_G[U]\rangle$ is identical to $\langle O[U]\rangle$, 
since any transformed configuration $U^{g}$ appears with the same probability as 
$U$ in the Monte-Carlo simulation with an action invariant w.r.t. $g$.
Note the statistical error of $\mathcal O_G$ decreases by a factor 
$1/\sqrt{N_G}$ times smaller, 
while its computational cost increases by a factor $N_G$ times more.

In order to reduce the computational cost implied by Eq.~(\ref{eq:o_g}),
we introduce an approximation for $\mathcal O$, 
which is called as $\mathcal O^{\rm (appx)}$. 
Averaging over $g\in G$ as in Eq.~(\ref{eq:o_g}) for 
$\mathcal O^{\rm (appx)}$ yields 
\begin{equation}
  \mathcal O^{\rm (appx)}_G = \frac{1}{N_G}\sum_{g\in G}\mathcal O^{{\rm (appx)}\,g}.
  \label{eq:appx}
\end{equation}
Using $\mathcal O^{\rm (appx)}$ and the original $\mathcal O$, 
an improved estimator for $\mathcal O$ is defined by
\begin{eqnarray}
\mathcal O^{\rm (imp)}
&=& \mathcal O - \mathcal O^{{\rm (appx)}} + \mathcal O_G^{\rm (appx)} 
  \nonumber\\
&\equiv& \mathcal O^{\rm (rest)} + \mathcal O_G^{\rm (appx)},
  \label{eq:imp} \\
\mathcal O^{\rm (rest)} &=& \mathcal O - \mathcal O^{{\rm (appx)}},
\end{eqnarray}
(In the definition of $\mathcal O^{\rm(rest)}$, we used the unit element
of $G$, however, any other elements would serve the purpose just as well.).
Since $\mathcal O^{\rm (appx)}$ in $\mathcal O^{\rm (imp)}$
is canceled by $\mathcal O_G^{\rm (appx)}$
after performing the path integral 
and using the covariance of $\mathcal O^{\rm (appx)}$ as in Eq.~(\ref{eq:cov1}), 
one easily sees that 
the expectation value of the improved estimator agrees
with the original,
\begin{equation}
  \langle \mathcal O^{\rm (imp)}\rangle = \langle \mathcal O\rangle.
\end{equation}

As shown in Appendix \ref{sec:sigma}, 
using the standard deviations of $\mathcal O$, $\sigma$, the 
approximation $\mathcal O^{\rm (appx)}$, $\sigma^{\rm (appx)}$, 
and the transformed approximation $\mathcal O^{{\rm (appx)}\,g}$, $\sigma^{\rm (appx)\,g}$, 
where $\sigma^X = \sqrt{\langle (\Delta\mathcal O^X)^2\rangle}$,
and $\Delta\mathcal O^X = \mathcal O^X - \langle O^X\rangle$, 
and the correlations defined by
\begin{eqnarray}
  r_g &=& \frac{\langle\Delta\mathcal O\Delta\mathcal O^{{(\rm appx)}\,g}\rangle}
      {\sigma\sigma_g^{\rm (appx)}},
  \label{eq:corr_r}\\
  r_{gg'}^{\rm corr} &=& \frac{\langle\Delta\mathcal O^{{\rm (appx)}\,g}
          \Delta\mathcal O^{{\rm (appx)}\,g'}\rangle}
             {\sigma^{{\rm (appx)}\,g}\sigma^{{\rm (appx)}\,g'}},
  \label{eq:corr_rg}
\end{eqnarray}
the standard deviation of the improved estimator is
\begin{eqnarray}
  &&\sigma^{\rm (imp)} \simeq \sigma\Big[2\Delta r + \frac{1}{N_G}
  - \frac{2}{N_G}\Delta r
  + R^{\rm corr}\Big]^{1/2},\label{eq:sigma_imp}\\
  &&R^{\rm corr} = \frac{1}{N_G^2}\sum_{g\ne g'}r^{\rm corr}_{gg'},
  \label{eq:r_corr}
\end{eqnarray}
with $\Delta r=1-r$, $r\equiv r_{g=I}$.
Note that, in Eq.~(\ref{eq:sigma_imp}), 
we approximate $\sigma\simeq\sigma^{\rm (appx)}$, and 
the correlation between $\mathcal O$ and $\mathcal O^{{\rm (appx)}\,g}$
is similar to that for $\mathcal O^{{\rm (appx)}}$, 
{\it i.e.} $r_{g\ne I}^{\rm corr}\simeq r_{g\ne I}$
(which assumes that there is strong correlation between 
$\mathcal O$ and $\mathcal O^{\rm (appx)}$.).
In \cite{Blum:2012uh,Blum:2012my},
we also ignored the third and fourth terms in (\ref{eq:sigma_imp}).
In the equation above, $\Delta r$ and $r_g^{\rm corr}$ indicate 
the quality of the approximation and 
the magnitude of the correlation among the $\{\mathcal O^{\rm (appx)\,g}\}_{g\in G}$,
respectively.
To achieve a reduction of the statistical error of magnitude
$\sim 1/\sqrt{N_G}$, an $\mathcal O^{\rm (appx)}$ with small $\Delta r$
and small positive $r_{gg'}^{\rm corr}$ 
is necessary. 
Furthermore, the cost of computing $\mathcal O^{{\rm (appx)}}$ 
should be much cheaper than $\mathcal O$.

Taking the consideration above into account, we impose the following conditions on 
$\mathcal O^{\rm (appx)}$ and the choice of transformation, $g\in G$, 
for $\mathcal O^{{\rm (appx)}\,g}$, 

{\bf CAA-1}: $\mathcal O^{\rm (appx)}$ is covariant 
under $G$ as in Eq.~(\ref{eq:cov1})
\footnote{As explained in Appendix \ref{sec:bias},
this condition is not necessary to fulfill  Eq.~(\ref{eq:latsym}) 
if we introduce a randomly chosen shift of source location 
in Appendix \ref{sec:notCAA}. 
}.
\\

{\bf CAA-2}: $\mathcal O^{\rm (appx)}$ is strongly correlated with 
$\mathcal O$, {\it i.e.} $\Delta r\ll 1$.\\

{\bf CAA-3}: The computational cost of $\mathcal O^{{\rm (appx)}}$ is much smaller than 
$\mathcal O$.\\

{\bf CAA-4}: The transformation $g\in G$ is 
chosen to give small (compared to $1/N_G$) positive correlations among 
$\{\mathcal O^{{\rm (appx)}\,g}\}_{g\in G}$, {\it i.e.} $R^{\rm corr} \ll 1/N_G$.
\\

Note that the last condition is not necessary if 
the cost of constructing $\mathcal O^{\rm (appx)}$ is negligible
(so that, in \cite{Blum:2012uh}, we have not included the last condition).
The tuning of the most appropriate 
$\mathcal O^{\rm (appx)}$ for the target observable is important 
to maximize the reduction of the statistical error.
In the following, we show two examples of CAA
in lattice QCD. 

\subsection{Example: Low-mode-averaging (LMA)}
In lattice QCD, $\mathcal O$ is a hadron correlator, 
given as the product of inverses of the Dirac operator ($S[U]$). 
In LMA, the approximation defined as 
$\mathcal O^{\rm (appx)}=\mathcal O^{\rm (LMA)}$ 
is constructed by 
\begin{eqnarray}
&&\mathcal O^{\rm (LMA)} = \mathcal O[S^{\rm (low)}],\quad
  \mathcal O_G^{\rm (LMA)} = \frac{1}{N_G}\sum_{g\in G}\mathcal O[S^{\rm (low)\,g}],
\label{eq:LMA}\\
&&S^{\rm (low)}(x,y) = \sum_{k=1}^{N_\lambda} \lambda_k^{-1}\psi_k(x)\psi^\dag_k(y),
\label{eq:low-mode}
\end{eqnarray}
with low-lying eigenmodes $\psi_k$ and eigenvalues $\lambda_k$ 
of the Hermitian Dirac matrix $H(x,y)$, 
$\sum_y H(x,y)\psi_k(y)=\lambda_k\psi_k(x)$. 
For low-mode dominant observables,
like the pion propagator and related form factors, 
the eigenmodes with small $|\lambda_k|$ 
saturate the observable, and thus
$r$ in Eq.~(\ref{eq:corr_r}) may be close to unity ({\bf CAA-2}). 
$\mathcal O^{\rm (LMA)}$ is covariant
since $H[U^g](x,y)=H[U](x^g,y^g)$;
we have $\mathcal O^g[S^{\rm (low)}[U]] = \mathcal O[S^{\rm (low)}[U^g]]$
({\bf CAA-1}).
The construction of $\mathcal O_G^{\rm (appx)}$ 
requires an inner product of the low-mode 
and source(sink) vectors and a complex times vector multiply.
Since the construction of $\mathcal O_G^{\rm (appx)}$ is cheap, 
the statistical error of low-mode dominant observables
is significantly reduced ({\bf CAA-3})
\cite{DeGrand:2004qw,DeGrand:2005vb} (because the computational cost of $O^{g,\rm (LMA)}$ 
is small, condition ({\bf CAA-4}) is not 
so important.)

\subsection{Example: All-mode-averaging (AMA)}
AMA is similarly defined as
\begin{eqnarray}
&&\mathcal O^{\rm (AMA)} = \mathcal O[S^{\rm (all)}],\quad
  \mathcal O_G^{\rm (AMA)} = \frac{1}{N_G}\sum_{g\in G}\mathcal O[S^{\rm (all)\,g}],
  \label{eq:AMA}\\
&&S^{\rm (all)}b = \sum_{k=1}^{N_\lambda} \lambda_k^{-1}(\psi^\dag_k b)\psi_k
  + f_\varepsilon(H)b,  \label{eq:all-mode}\\
&&f_\varepsilon(H)b = \sum_{i=1}^{N_{\rm CG}}(H)^ic_i,
  \label{eq:all-mode2}
\end{eqnarray}
where $f_\varepsilon b$ is a polynomial of $H$ with vector ``coefficients'' $c_i$. 
In practice this combination is obtained from the CG, depending on the
source vector $b$ and initial guess $x_0$. 
The subscript $\varepsilon$ indicates the norm of the residual vector
after $N_{\rm CG}$ iterations, or steps, of the CG. 

In AMA, the (exact) low-mode contribution to the propagator within the range 
$[\lambda_{1},\lambda_{N_\lambda}]$
is taken into account by projecting the source vector $b$ onto the orthogonal subspace,
\begin{equation}
  b_{\rm proj} \equiv \Big(1-\sum_{k=1}^{N_\lambda}\psi_k\psi_k^\dag\Big)b,
  \label{eq:deflation}
\end{equation}
where the low-mode is normalized as $\sum_x\psi_k^\dag(x)\psi_k(x) = 1$.
By adopting the above projected source vector into the CG process 
(see Algorithm \ref{alg:CG}), 
we obtain the solution $x_{\rm CG}$, 
\begin{equation}
  x_{\rm CG} + \sum_{k=1}^{N_\lambda} \lambda_k^{-1}(\psi^\dag_k b)\psi_k = S^{\rm (all)}b.
  \label{eq:deflation2}
\end{equation}
Notice that the CG is deflated at the same time. Further, the higher mode contribution 
($\lambda_{N_\lambda} < \lambda \le \lambda_\text{max}$)
is treated approximately, $f_\varepsilon(\lambda)\approx 1/\lambda$,
by using the relaxed stopping criterion in the CG. 
Therefore the computational cost of $f_\varepsilon(H)$ is 
significantly smaller than 
the usual CG used in $\mathcal O$ ({\bf CAA-3}).
Compared to LMA, in which eigenmodes with $\lambda>\lambda_{N_\lambda}$
are ignored, AMA introduces $f_\varepsilon$ to take into account 
the contribution of all higher modes, and thus the quality 
of the approximation to $\mathcal O$ 
is greatly improved ({\bf CAA-2}).  
In Eq.~(\ref{eq:AMA}) the covariance 
({\bf CAA-1}) is also fulfilled since $f_\varepsilon(H)$ is covariant 
under the transformation $g$; 
$f_\varepsilon^g(H(x,y)) = f_\varepsilon(H(x^g,y^g))$.

Here we consider two choices of the stopping condition in the CG,
\begin{itemize}
\item the norm of the residual vector is smaller than some prescribed value,
\item a fixed number of CG iterations.
\end{itemize}
The first condition naturally controls the accuracy of 
the CG and thus the approximation $\mathcal O^{\rm (appx)}$, and in this paper 
we have employed it as the stopping condition.
However, it may happen 
that this criterion introduces a violation of covariant symmetry as systematic bias 
due to numerical round-off error, for example, 
because of the order of operations in one's code
\footnote{We thank both M. L\"uscher and S. Hashimoto who, independently, pointed this out.}.
As described in detail in Appendix B, this bias is orders of magnitude smaller than the statistical error in practice. 
In the same appendix, we also present an argument to reduce the bias by fixing the number of CG iterations 
instead of fixing the CG stopping condition for the residual vector norm. Note that $f_\varepsilon$ can also be computed 
directly from a polynomial with fixed coefficients rather than dynamically computed in the CG.  

We emphasize, as in \cite{Blum:2012uh} and demonstrated in Appendix \ref{sec:AccRoundoff}, 
when using AMA it is mandatory to compute 
the size of the violation of covariance on a small number of configurations to ensure that the bias is negligible.  
Alternatively, one can {\it completely} remove the bias by using randomly selected source locations 
as described in Appendix \ref{sec:notCAA}.


Figure \ref{fig:approx} illustrates the spectral decomposition of 
$\mathcal O^{\rm (LMA)}$ defined in Eq.~(\ref{eq:low-mode})
and  $\mathcal O^{\rm (AMA)}$ defined in Eq.~(\ref{eq:all-mode}). 
In AMA, because we use the exact low-lying eigenvectors, the behavior 
in the low-mode region is consistent with LMA. 
The number of intersections with the exact solution corresponds to 
the polynomial degree in the approximation which is equal to the number of CG iterations. 
The discrepancy with the exact solution 
can be controlled by the number of low-modes used in deflation 
and the degree of the polynomial (see Eq.~(\ref{eq:all-mode})). 

\begin{figure}
\begin{center}
\includegraphics[width=120mm]{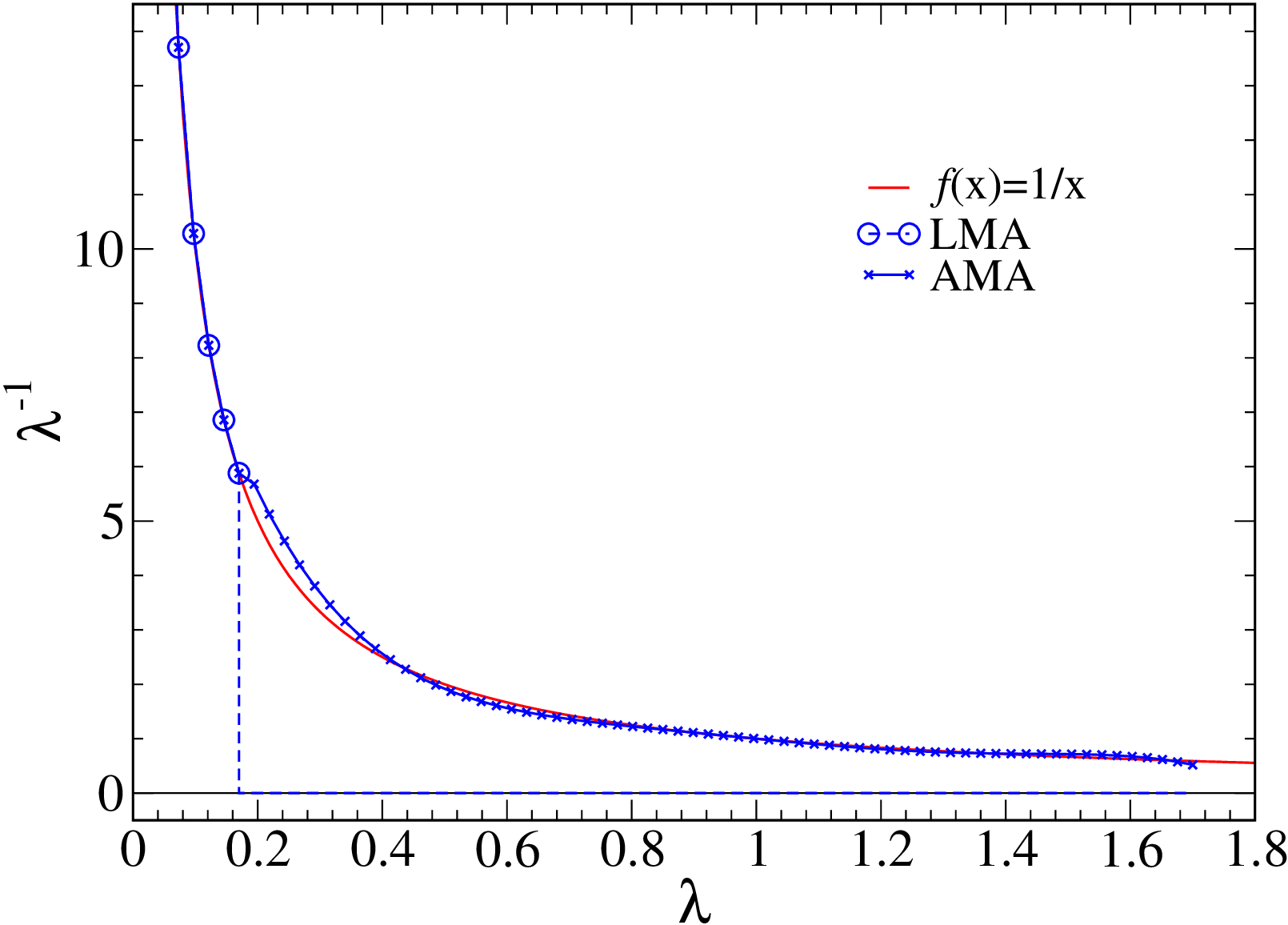}
\caption{
A sketch of approximations for the spectral decomposition 
of the quark propagator in LMA(circle-dashed line) and AMA(cross-solid line). 
The x-axis denotes the eigenvalue of the Hermitian Dirac matrix.
The circle symbol corresponds to $\mathcal O^{\rm (LMA)}$ and 
blue solid line corresponds to $\mathcal O^{\rm (AMA)}$. 
The red solid line shows the exact solution. 
}\label{fig:approx}
\end{center}
\end{figure}

\begin{table}
\begin{center}
\caption{ LMA and AMA algorithms}
\label{alg_LM_AM}
{
\begin{tabular}{ll|ll}
\hline
 & {\bf LMA algorithm} & & {\bf AMA algorithm} \\
\hline
1: & Compute low-modes $\psi_k$ of $H$ &
1: & if $\lambda_{N_\lambda}\ne 0, N_\lambda>0$\\
& & & Compute low-modes $\psi_k$ of $H$ \\\hline
2: & \multicolumn{3}{l}{Set source vector $b$ and $G$-invariant inital guess $x_0$}\\\hline
3: & \multicolumn{3}{l}{Compute accurate $x_{\rm CG}$ and $\mathcal O[S]$ precisely }\\
   & \multicolumn{3}{l}{(use deflation method in Eq.(\ref{eq:deflation}) and (\ref{eq:deflation2})
     if $\psi_k$ exits)}\\\hline
4: & Compute $S^{\rm (low)}b$ in (\ref{eq:low-mode}) & 
4: & Compute $S^{\rm (all)}b$  in (\ref{eq:all-mode}) \\
   & and $\mathcal O^{\rm (LMA)} = \mathcal O[S^{\rm (low)}]$ &
   & and $\mathcal O^{\rm (AMA)} = \mathcal O[S^{\rm (low)}]$  using\\
& & & deflated CG (if $\lambda_{N_\lambda}\ne 0$)\\\hline
5: & $\mathcal O^{\rm (rest)} = \mathcal O[S] - \mathcal O[S^{\rm (low)}]$&
5: & $\mathcal O^{\rm (rest)} = \mathcal O[S] - \mathcal O[S^{\rm (all)}]$; \\\hline
6: & \multicolumn{3}{l}{Set shifted source $b^{g}$ and  
     $G$-invariant inital guess $x_0^{g}$}\\\hline
7: & Average $\mathcal O_G^{\rm (LMA)} = \mathcal O[S^{\rm (low)}]$ & 
7: & Average $\mathcal O_G^{\rm (AMA)} = \mathcal O[S^{\rm (all)}]$ \\
   &   over $g \in G$  to get $\mathcal O^{\rm (LMA)}_G$ &
   &   over  $g \in G$ to get $\mathcal O^{\rm (AMA)}_G$  \\\hline
8:   & \multicolumn{3}{l}{ $\mathcal O^{\rm (imp)} = \mathcal O^{\rm (rest)} + \mathcal O^{\rm (appx)}_G$}\\\hline
\end{tabular}
}
\end{center}
\end{table}

The correlation among $\mathcal O^g$ will not be significant if 
we choose appropriate transformations, $g\in G$, for instance, by widely separating 
source points among $\{\mathcal O^g\}_{g\in G}$, so that the $r_{gg'}^{\rm corr}$ term 
in Eq.~(\ref{eq:sigma_imp}) is negligible
({\bf CAA-4}).
Unlike LMA, AMA entails non-negligible additional cost
to construct $S^{\rm (all)}$ (fourth step of the AMA algorithm in Table \ref{alg_LM_AM}), 
and hence the judicious tuning of $N_G$ and choice of $g\in G$ is 
important to reduce the computational cost. 

\section{Numerical results}\label{sec:result}

In this section we show the numerical comparison between the standard method and AMA/LMA 
for the hadron spectrum and the form factors of the nucleon 
using realistic lattice QCD parameters.

\subsection{Set up}
We use the $N_f=2+1$ domain-wall fermion (DWF) configurations 
generated by the RBC/UKQCD collaboration on a 24$^3\times$64 lattice, with gauge coupling 
$\beta=2.13$ for the Iwasaki gauge action \cite{Aoki:2010dy}. 
The CG algorithm with four dimensional even-odd preconditioning 
(see Appendix \ref{sec:4Deo}) was used 
to compute quark propagators at quark mass $m=0.005$ and $0.01$, 
corresponding to $0.33$ and $0.42$ GeV pion masses, respectively, 
and the 5th dimension size for DWF is $L_s=16$.

To calculate the eigenvectors of the 
Hermitian even-odd preconditioned DWF operator, 
we implement the implicitly restarted Lanczos algorithm 
with Chebychev polynomial acceleration 
\cite{Saad:1984,Sorensen:1992:IAP:131879.131903,calvetti1994implicitly,Neff:2001zr}. 
In Appendix \ref{sec:Lanczos} we describe the detailed implementation.
The degree of the Chebychev polynomial in the Lanczos method is 100, 
and the parameters $(\alpha,\beta)=(0.04,1.68)$ for $m=0.005$ and 
$(\alpha,\beta)=(0.025,1.68)$ for $m=0.01$ are chosen to rapidly converge the ``wanted'' 
part of the spectrum, here the lowest few hundred modes (see Eqs.(\ref{eq:qH}) and (\ref{eq:qHmu})). 
In the implicitly restarted Lanczos method,
we label $N_\lambda$ the number of wanted eigenvectors and 
$p=40$ the number of unwanted vectors (see Appendix \ref{sec:Lanczos}).
We compute the exact low-modes of Hermitian 4D even-odd preconditioned 
DWF Dirac operator, $H_{4Deo}$, to better than  
$10^{-10}$ numerical accuracy, 
$||(H_{4Deo}-\lambda_k)\psi_k||/||\psi_k||<10^{-10}$. 
In table \ref{tab:param} we summarize the parameters in the 
Lanczos method, the number of gauge configurations $N_{\rm conf}$ in each ensemble,
and the number of low-modes $N_\lambda$ computed on each configuration.

In AMA/LMA, the set of transformations $g\in G$ in Eq.~(\ref{eq:appx}) 
are taken as translational symmetry.
The estimator ${\cal O}_G^{\rm (appx)}$ is obtained 
with $N_G=32$ different source locations,  
separated by 12 sites for spatial directions and 16 sites 
for the temporal direction, starting from the origin, {\it i.e.} at positions
(0,0,0,0), (12,0,0,0), (12,12,0,0), $\dots$, (12,12,12,48) 
in lattice units. This setup is used for measurements on
configurations separated by 40 HMC trajectories. 
In addition, measurements are made on a second set of configurations, 
also separated by 40 trajectories, but lying in between configurations 
of the first set. On the second set, all source locations are shifted by the lattice vector 
(6,6,6,0) with respect to the original functional ${\cal O}$.
In the CG, the norm of the residual vector is defined as 
$||H_{4Deo}x-b||/||b||$ with source vector $b$
and solution vector $x_{\rm CG}$ (see also Table \ref{tab:param}).
For the stopping conditions for the exact CG and the relaxed CG we have
$\varepsilon = 10^{-8}$ and $\varepsilon = 0.003$, respectively
\footnote{Note that when using an even-odd basis, one needs to choose
the four dimensional shift vector of the source point to avoid 
breaking {\bf CAA-3}. Shifts that end on an even(odd) point for even(odd) sites are sufficient).}.

We use gauge-invariant Gaussian smeared sources with the same parameters
as in Ref.\cite{Yamazaki:2009zq} to compare the performance of LMA and AMA.
In \cite{Yamazaki:2009zq}, the authors measured three- and two-point functions for 
four source locations in the temporal direction to extract
the nucleon isovector form factors and axial charge, 
and thus  $4\times N_{\rm conf}$ samples were accumulated.
For $m=0.005$, quark sources set on two time-slices separated 
by 32 sites were used (double source method) to 
efficiently double the statistics.
\cite{Yamazaki:2009zq} also employed non-relativistic nucleon sources
(2 quark spins rather than 4) to reduce the computational cost further, 
while in our case 
we use relativistic sources. 
Therefore, in the analysis below, we account for these two factors to ensure 
a fair comparison of statistical errors.

\begin{table}
\begin{center}
\caption{Parameters of LMA/AMA in each ensembles.
$(\alpha,\beta)$ is the input range of Chebychev polynomial
in the Lanczos method with $N_{\lambda}$ wanted and 40 unwanted eigenmodes.
We present the absolute value of the minimum eigenvalue as $|\lambda_1|$ 
and $N_\lambda$-th eigenvalue $|\lambda_{N_\lambda}|$ up to the first significant figure
in each ensemble.
``\#Restart'' column shows the range of number of restarted Lanczos iterations.}
\label{tab:param}
\begin{tabular}{ccccccccccc}
\hline\hline
$m$ & $N_{\rm conf}$ & $N_\lambda$ & $(\alpha,\beta)$ & $|\lambda_{1}|$ 
& $|\lambda_{N_\lambda}|$ & \#Restart \\
\hline
0.005 & 398 & 400 & (0.04,1.68)  & 0.004 & 0.04 & 5--6 \\
0.01  & 348 & 180 & (0.025,1.68) & 0.006 & 0.02 & 5--6 \\
\hline\hline
\end{tabular}
\end{center}
\end{table}

\subsection{Computational cost estimate}

\begin{table}
\begin{center}
\caption{
The table of the number of multiplications of kernel $H_{4Deo}^2$.
``\#Mult$_{\rm Lanczos}$'' is its number in 5 restarting Lanczos process.
We also show the range of \#Mult with and without deflation method 
for exact calculation (\#Mult$_{\rm defl.CG(org)}$, \#Mult$_{\rm CG(org)}$)
and approximation in AMA (\#Mult$_{\rm defl.CG(AMA)}$)
using low-mode of $H_{4Deo}^2$.}
\label{tab:param2}
\begin{tabular}{ccccccccccc}
\hline\hline
$m$ & \#Mult$_{\rm Lanczos}$ & \#Mult$_{\rm CG(org)}$ & \#Mult$_{\rm defl.CG(org)}$ 
    & \#Mult$_{\rm defl.CG(AMA)}$\\
\hline
0.005 & 64K & 3K & 350--360 & 70--90\\
0.01  & 42K & 2K & 600--630 & 90--130\\
\hline\hline
\end{tabular}
\end{center}
\end{table}

In order to compare the computational cost between the standard method 
and LMA/AMA, we use the number of 
applications of $H_{4Deo}^2$ 
(\#Mult in Table \ref{tab:param2}) to estimate total costs in each case. 
In the standard method, the cost without deflation 
is \#Mult$_{\rm CG(org)}$ times the 
number of color and spin sources used per configuration, 
\begin{equation}
  {\rm Cost(org)} = \textrm{\#Mult}_{\rm CG(org)}\times 12 \times N_{\rm conf}.
\end{equation}
On the other hand, when deflating the Dirac operator, the cost is
\begin{equation}
  {\rm Cost}_{\rm w/defl.}{\rm (org)} = \big(\textrm{\#Mult}_{\rm Lanczos} 
  + \textrm{\#Mult}_{\rm defl.CG(org)}\times 12 \big) \times N_{\rm conf},
\end{equation}
where we add the cost of the Lanczos process to obtain the low-modes.
We note that, based on wall-clock timing, the time for multiplication of the Dirac operator 
dominates the Lanczos step, and  Gram-Schmidt reorthogonalization is negligible 
due to the O(100) degree of the Dirac matrix polynomial. 
Therefore, we use the number of multiplications of the polynomial of 
the Dirac operator as a good representative of the computational cost.

In LMA, ignoring the small cost of constructing the approximation $\mathcal O^{\rm (LMA)}$ 
and $\mathcal O^{\rm (LMA)}_G$ from the low-modes,
the total cost is the same as Cost$_{\rm w/defl.}$(org), 
\begin{equation}
  {\rm Cost(LMA)} = {\rm Cost}_{\rm w/defl.}{\rm (org)}.
\end{equation}
In AMA, there are three parts to the total cost, the eigenvector computation, 
the exact CG solve, and $N_G$ relaxed CG solves, 
so the total cost reads
\begin{equation}
  {\rm Cost(AMA)} = \big(\textrm{\#Mult}_{\rm Lanczos} 
  + \big(\textrm{\#Mult}_{\rm defl.CG(org)} + \textrm{\#Mult}_{\rm defl.CG(AMA)}\times N_G\big)
  \times 12 \big) \times N_{\rm conf}.
\end{equation}

In the following section, to compare costs of LMA/AMA to the standard method, 
we define the cost ratio multiplied with the squares of statistical error ratio 
to obtain a normalized cost, {\it i.e.}, one that reflects 
the cost to achieve the same error,
\begin{eqnarray}
  r_{\rm Cost}^{\rm w/o\,defl} &=& \frac{\rm Cost(LMA/AMA)}{\rm Cost(org)}
  r_{\rm Error}^2,\label{eq:r_cost}\\
  r_{\rm Cost}^{\rm w/\,defl} &=& \frac{\rm Cost(LMA/AMA)}{{\rm Cost}_{\rm w/defl.}{\rm (org)}}
  r_{\rm Error}^2,\label{eq:r_cost_wdefl}\\
  r_{\rm Error} &=& \frac{\rm Error(LMA/AMA)}{\rm Error(org)}.
\end{eqnarray}

\subsection{Hadron spectrum}\label{sec:hadmass}

\begin{figure}
\begin{center}
\includegraphics[width=150mm]{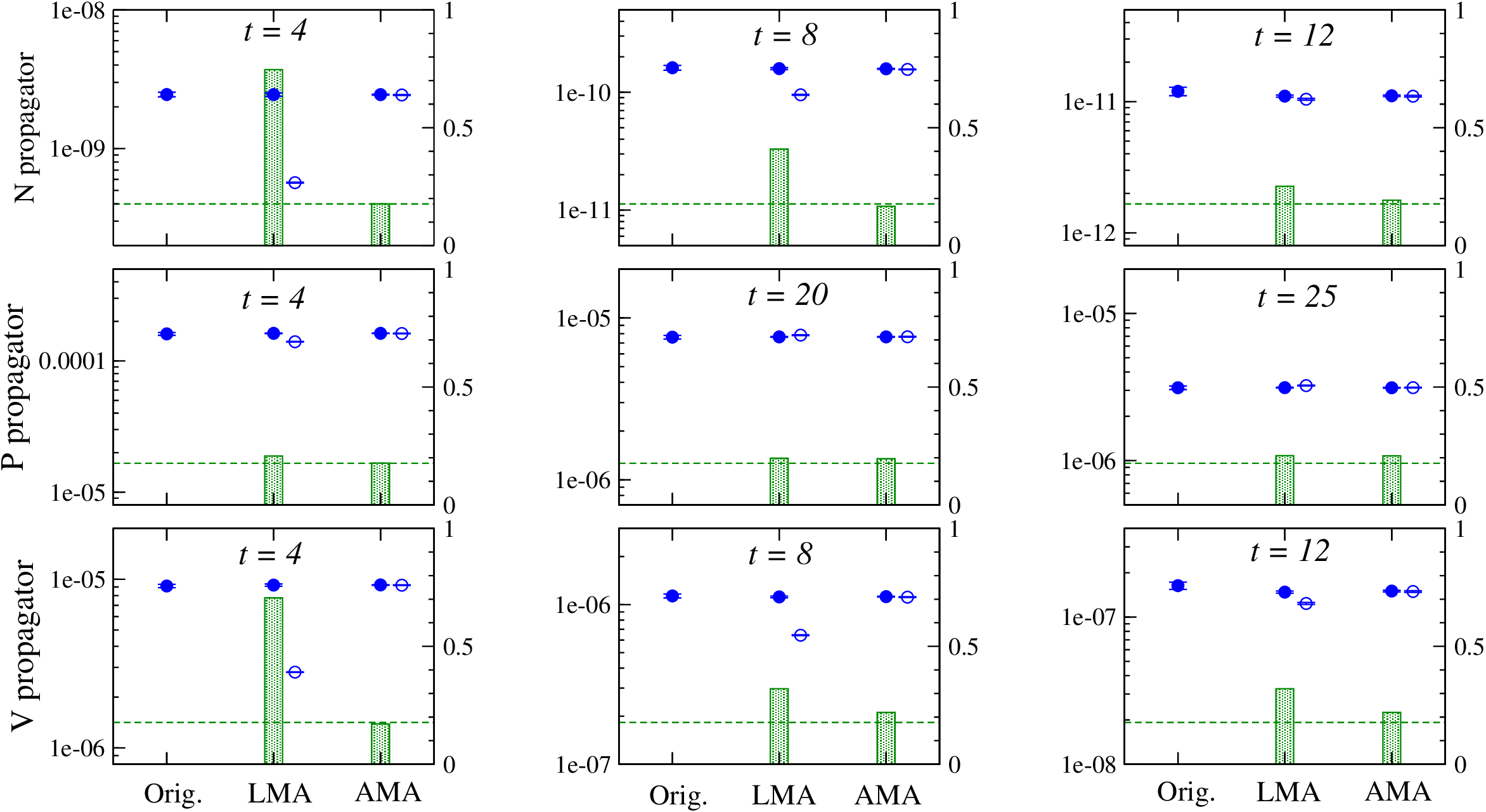}
\caption{
The propagator of nucleon (top), pion (middle) and vector meson (bottom)
at time separation $t=4,8,12$ for nucleon and vector meson, and $t=4,20,25$ for pion. 
We show the values of these propagators used in original, LMA and AMA.
The filled symbols are result of improved estimator $\mathcal O^{\rm (imp)}$
and open symbols are result of averaged approximation $\mathcal O^{\rm (appx)}_G$.
The bar in AMA/LMA shows the ratio of relative error with original one.
This value corresponds to right-perpendicular axis.
The horizontal bar shows the ideal ratio of relative error 
$1/\sqrt{32}\simeq 0.18$ in case of no correlation between spatial source 
locations.
}\label{fig:Npiv_prop}
\end{center}
\end{figure}

First we show results for hadron propagators
obtained by using the standard method and LMA/AMA 
with parameters given in the previous section. 
Figure \ref{fig:Npiv_prop} shows that the error reduction achieved with AMA 
is close to the ideal rate, $1/\sqrt{N_G}\simeq 0.18$ for nucleon, pion, and vector 
propagators, for source-sink separations $t=4$, 8, and 12 (nucleon and vector),
and $t=4$, 20, and 25 (pion), while LMA does not work well at short distance ($t=4$)
except for the pion. 
Since low-modes dominate the pion propagator, 
LMA and AMA show similar error reduction.
For AMA we see that
$\mathcal O^{\rm (imp)}$ is close in value to $\mathcal O^{\rm (appx)}_G$, 
while in LMA the difference is much larger, especially for 
short distances (except for pion propagator). 
It turns out that AMA provides a good approximation 
to the original and clearly shows that AMA can reduce statistical errors
for both long and short distances by approximating the quark propagator 
with $f_\varepsilon(H)$ obtained with the relaxed CG for the high part of the Dirac spectrum. 

\begin{figure}
\begin{center}
\includegraphics[width=150mm]{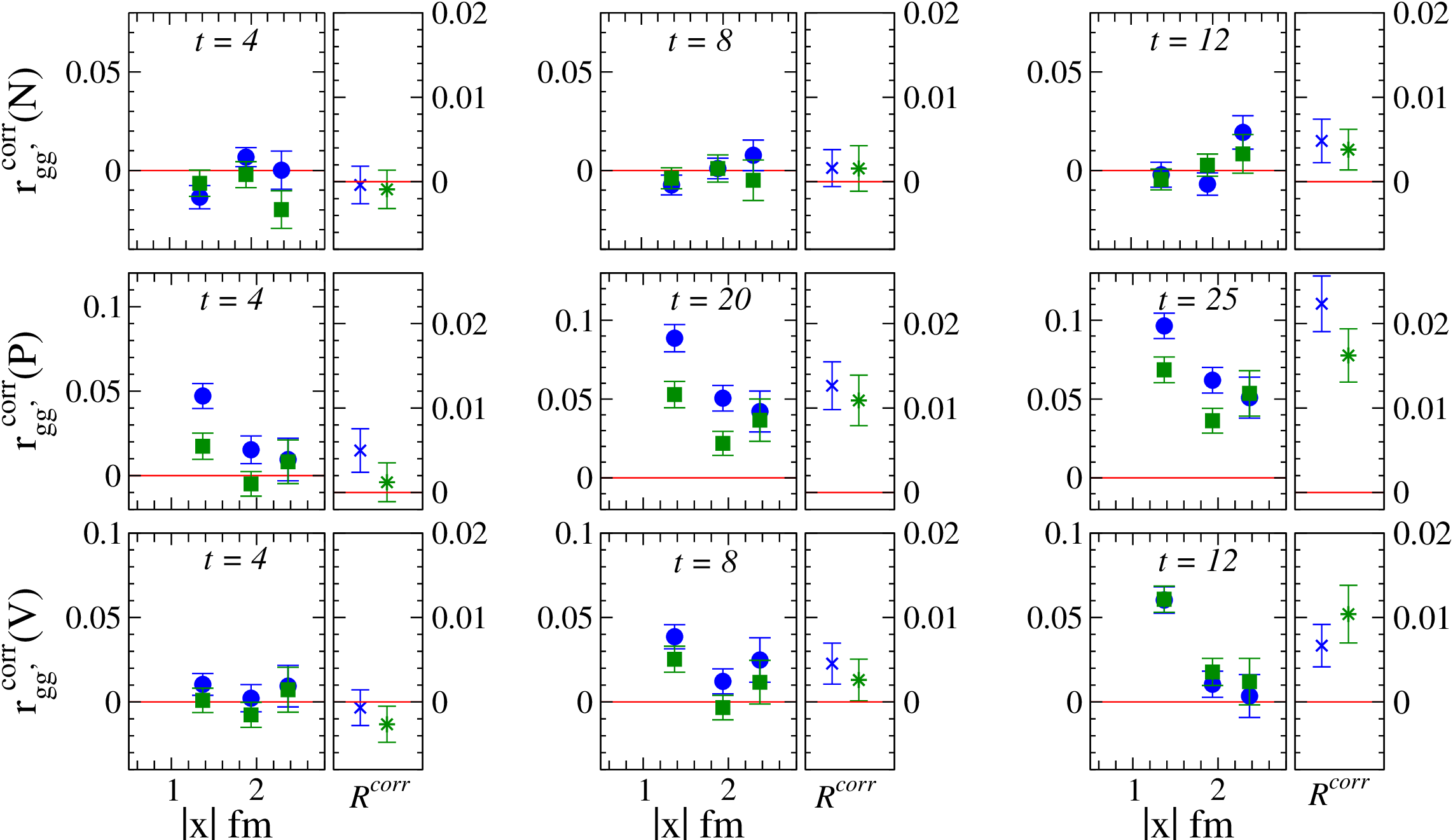}
\caption{
The correlation $r_{gg'}^{\rm corr}$ and $R^{\rm corr}$ as a function of physical spatial distance 
for source locations between $\mathcal O^{{\rm (appx)}\,g}$ and $\mathcal O^{{\rm (appx)}\,g'}$.
We take average over $r_{gg'}^{\rm corr}$ in same spatial distance 
with identical temporal source location. 
The top panel is for propagator of nucleon, middle pion 
and, bottom is of vector at time-slice $t=4,8,12$ for nucleon and vector meson
$t=4,20,25$ for pion. 
The blue (green) symbols are in $m=0.005$(0.01). 
}\label{fig:Npiv_rx_prop}
\end{center}
\end{figure}

In Figure \ref{fig:Npiv_rx_prop}, we plot $r_{gg'}^{\rm corr}$ against
the distance between source locations on a given time slice
and $R^{\rm corr}$ for zero momentum nucleon, pion and vector meson propagators. 
These quantities are important for choosing $N_G$ and 
the transformations $g\in G$ to efficiently implement 
CAA as explained in Sec.~\ref{seq:caa}. 
One sees that at the smallest separation from the origin 
(in which the source location is $(12,0,0,0)$, $(0,12,0,0)$ and $(0,0,12,0)$)
there is significant correlation compared to the case of large separation.
This behavior becomes apparent when the hadron propagates far away from source location
(large $t$). 
Comparing the different masses, especially for the pion 
propagator, $r_{gg'}^{\rm corr}$ is larger for lighter mass. 
For the nucleon and vector meson propagators $R^{\rm corr}$, 
which is the sum of $r_{gg'}^{\rm corr}$ divided by $N_{G}^2$, 
is relatively small compared to $1/N_G\simeq 0.031$ in Eq.~(\ref{eq:sigma_imp}), 
and therefore in our setting of $g\in G$ 
the reduction of statistical error is close to the ideal ratio, $1/\sqrt{N_G}\simeq 0.18$.
We notice that for the pion propagator $R^{\rm corr}$ is relatively large
since $r_{gg'}^{\rm corr}$ increases when the pion propagates a large distance. 
More details will be discussed below.

\begin{figure}
\begin{center}
\includegraphics[width=140mm]{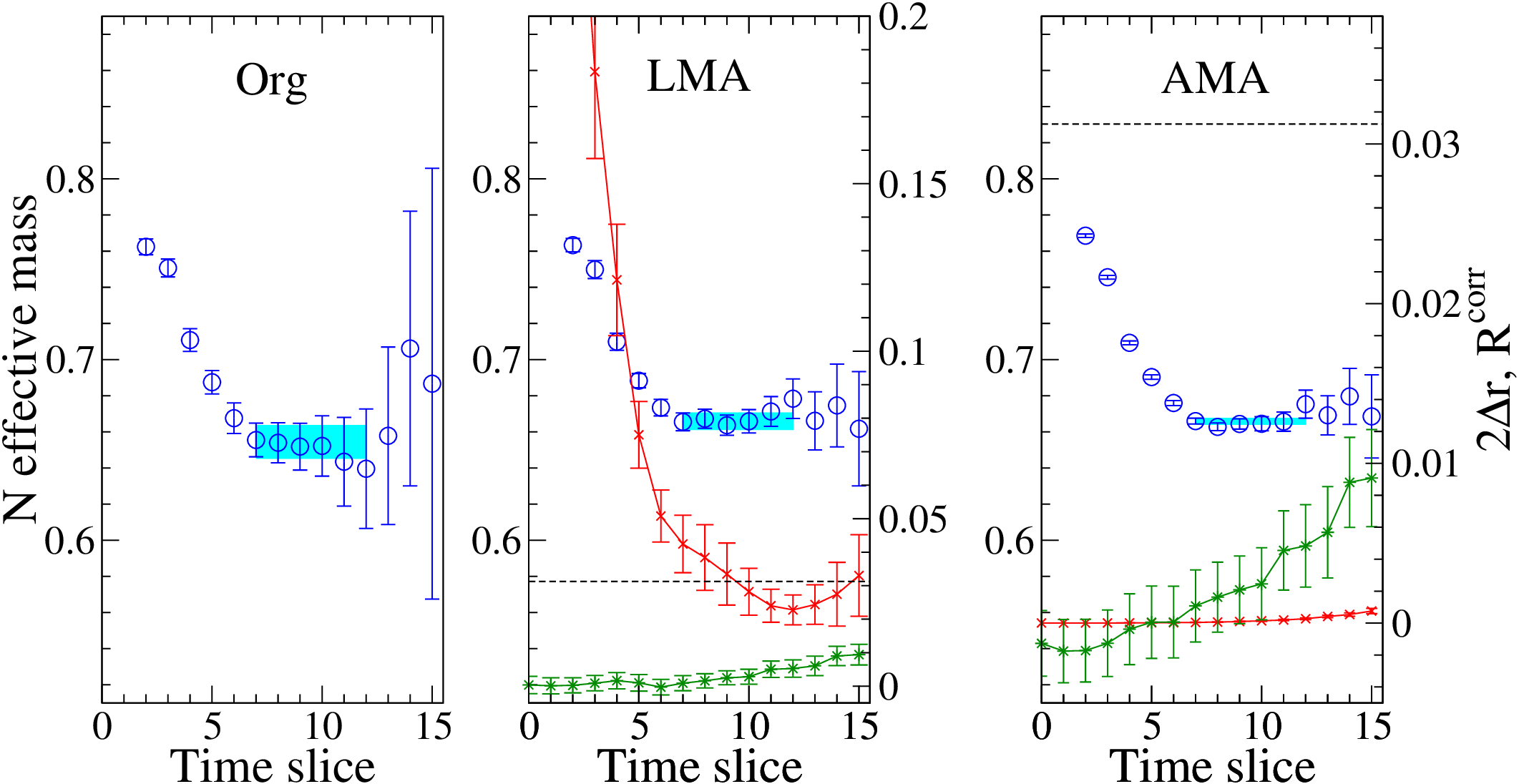}
\vskip 5mm
\includegraphics[width=140mm]{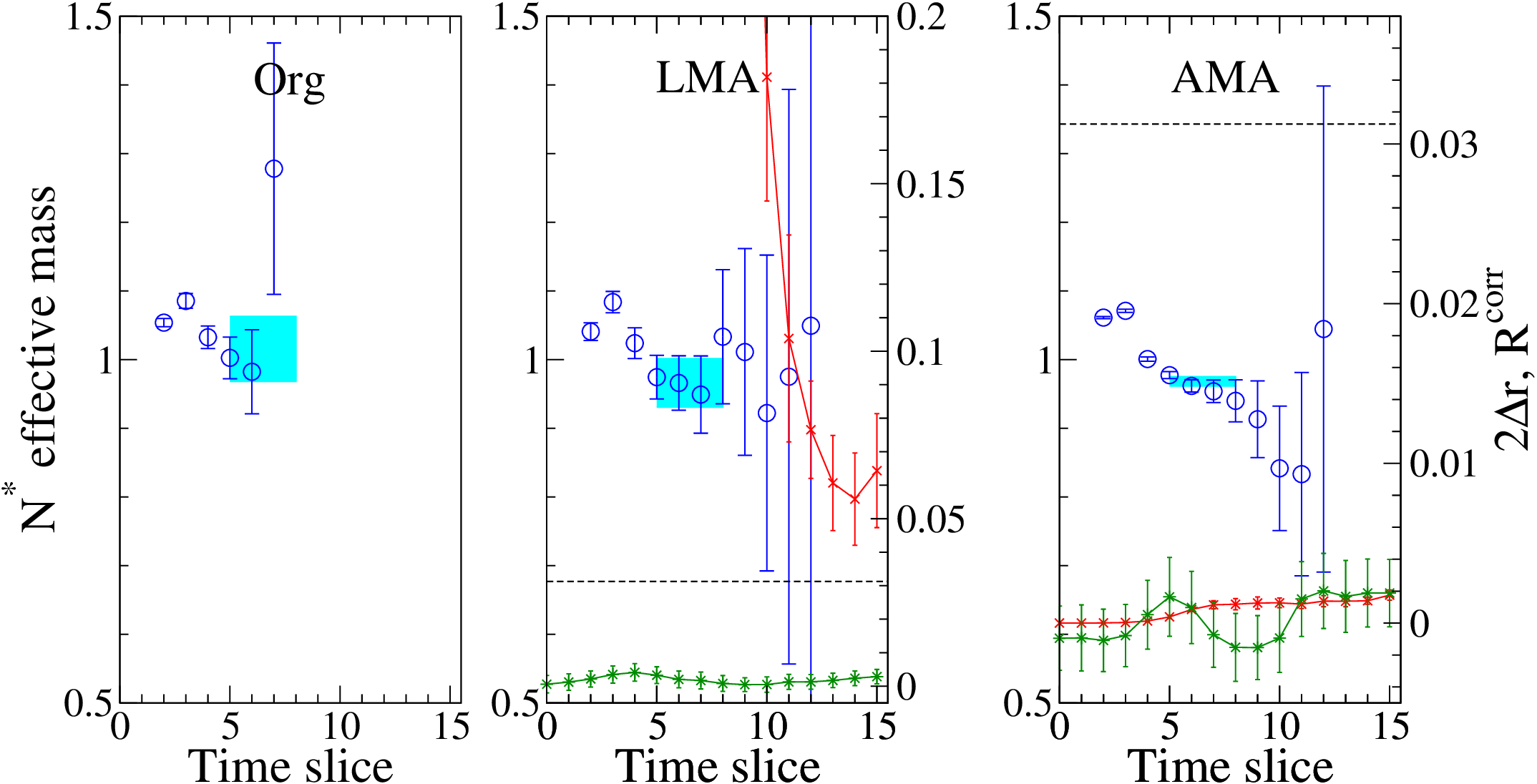}
\caption{
The effective mass plot of nucleon propagator (top) and 
its parity partner (bottom) with smeared source and 
point sink using original (left panel), LMA (middle panel) 
and AMA (right panel) at $m=0.005$. 
The cross symbols show the magnitude of 
$2\Delta r$, and star symbols denote $R^{\rm corr}$, 
as defined in Eq.~(\ref{eq:corr_r}).
The right-perpendicular axis corresponds to this value.
The dashed-line shows the value of $1/N_G$.
}\label{fig:Neffm}
\end{center}
\end{figure}

\begin{figure}
\begin{center}
\includegraphics[width=140mm]{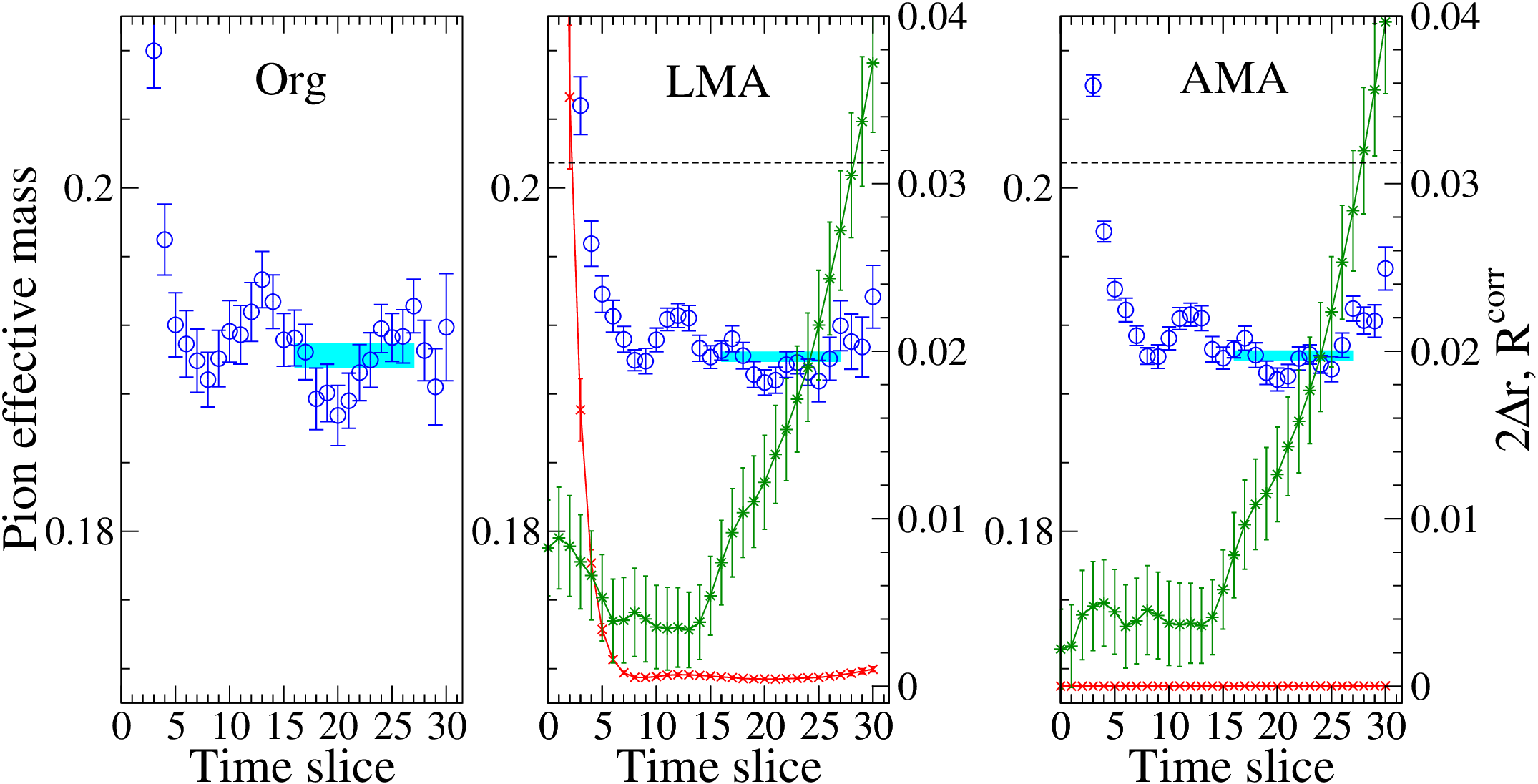}
\vskip 5mm
\includegraphics[width=140mm]{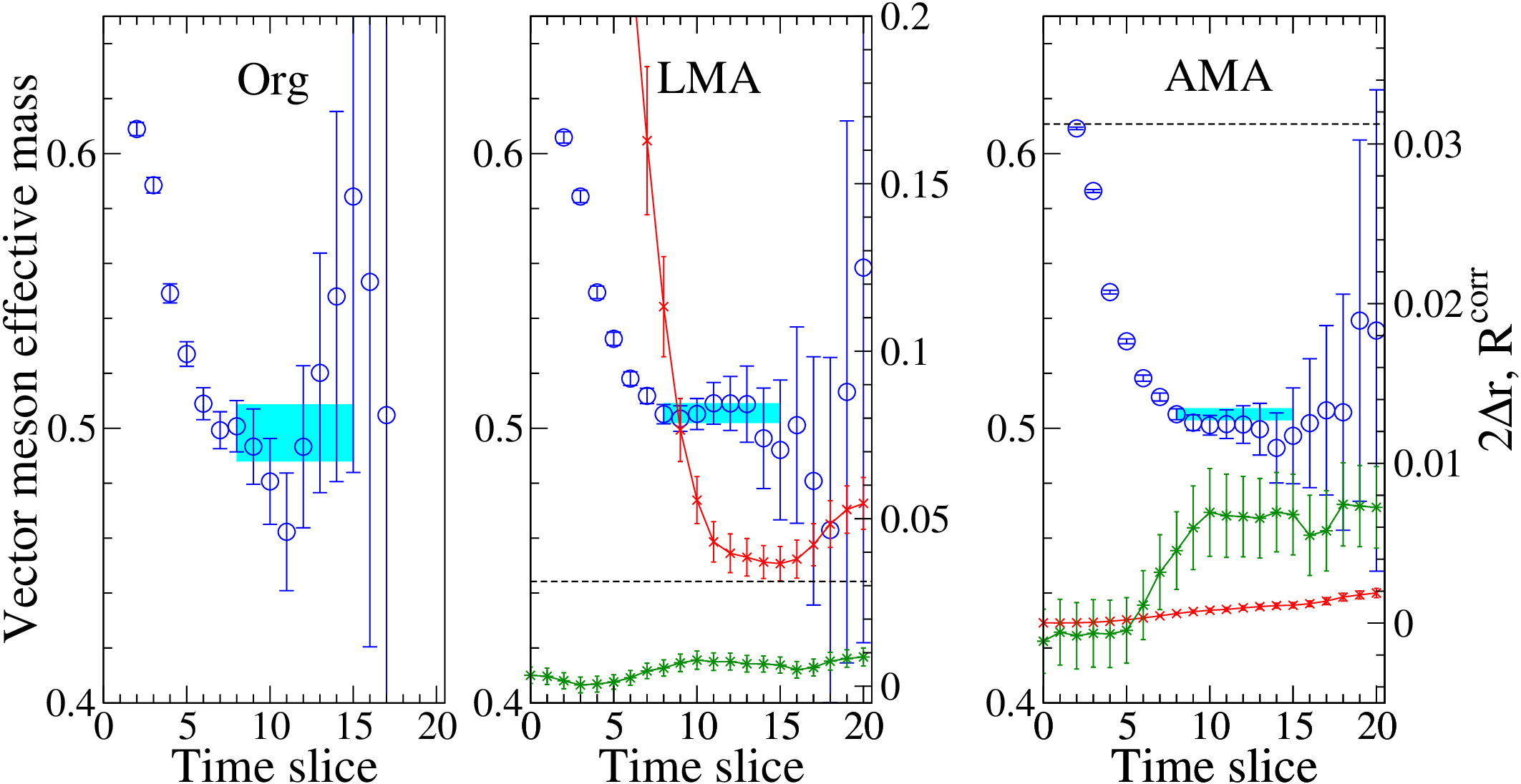}
\caption{
Similar plot to Figure \ref{fig:Neffm} of effective mass of
pion (top) and vector meson (bottom).
}\label{fig:PiVseffm}
\end{center}
\end{figure}

In Figure \ref{fig:Neffm} and \ref{fig:PiVseffm}, 
we plot the effective mass of several hadron channels together with 
$2\Delta r$ and $R^{\rm corr}$ defined in Eq.~(\ref{eq:corr_r}) 
and Eq.~(\ref{eq:r_corr}).
As previously discussed, an approximation having strong correlation 
with $\mathcal O$ has small $2\Delta r$. 
In the case of AMA the effective mass for both the nucleon and 
vector meson is improved over LMA, especially for $t$ less than 15
where $2\Delta r$ is less than 0.1\%.
On the other hand, $R^{\rm corr}$ of AMA within the fitting region 
is similar to $R^{\rm corr}$ of LMA, and it is less than 20\% of $1/N_G$
for the nucleon (and its parity partner $N^*$, which is given by the 
negative Parity projection for the nucleon two point function.
More detailed discussion and recent lattice study refers to 
for example \cite{Sasaki:2001nf,Lin:2011ti} and references therein) and the vector meson. 
Thus the two contributions in Eq.~(\ref{eq:sigma_imp}), 
$2\Delta r$ and $R^{\rm corr}$, are negligible 
compared to $1/\sqrt{N_G}$, and therefore 
the error reduction of these hadron masses is close to $1/\sqrt{N_G}$ 
in AMA (see Table \ref{tab:hadmass_m0.005}). 
However, for the pion propagator, we observe that $2\Delta r$ in AMA at 
below $t=5$ is much smaller than LMA, 
otherwise at $t>5$ both cases become similarly tiny as seen in Figure \ref{fig:PiVseffm}.
On the other hand, $R^{\rm corr}$ of the pion propagator 
is similar between LMA and AMA, with magnitude around 40\%--90\% of $1/N_G$. 
As the consequence the error reduction of AMA for 
pion propagator and pion mass is similar in magnitude with LMA 
in a region where the pion ground state dominates. 
We note that the relatively large correlation 
between different source locations for the pion propagator
may result in a slightly smaller error reduction of the pion mass
(see the ``$m_\pi$'' row in Table \ref{tab:hadmass_m0.005}).

In Tables~\ref{tab:hadmass_m0.005} and \ref{tab:hadmass_m0.01} 
we compare the fit results of hadron masses and 
scaled costs of LMA/AMA to achieve the same statistical error of the standard method. 
Here we use the chi-squared fitting with single exponential function including
the correlation in the temporal direction. 
$\chi^2$/dof is between 0.6 and 3 using the fitting range 
as shown in Tables~\ref{tab:hadmass_m0.005} and \ref{tab:hadmass_m0.01}.
The quantity $r_{\rm Cost}$ defined in Eq.~(\ref{eq:r_cost_wdefl}) and (\ref{eq:r_cost}) 
indicates the computational cost compared to the standard method,
with and without deflation, respectively. 
Comparing costs for masses of the nucleon, $N^*$, and vector mesons 
with LMA and AMA, one sees that error reduction in AMA is much larger than from LMA
at both $m=0.005$ and $m=0.01$. 
AMA has a cost reduction for those observables of 
about 5 to 20 times larger compared to the standard method and LMA.
It can be easily understood by looking at $r_{\rm Error}$ of those hadron masses in AMA
which is close to the ideal ratio ($1/\sqrt{N_G}\simeq 0.18$), and 
the construction cost of $\mathcal O^{\rm (appx)}$ is much cheaper than 
original one.
In particular, for the $N^*$, the
gain from AMA  compared to LMA is even more dramatic. 
Actually, in LMA, the $\Delta r$ term dominates the total error in 
Eq.~(\ref{eq:r_corr}), and it turns out that 
error reduction by LMA is limited to $\sqrt{2\Delta r}$ 
even if $N_G$ is increased to $N_G = V$, as is usually done.
Improvement for heavy mesons and baryons would also be interesting work.

Considering the multiple-source method with deflation, 
statistics are increased by averaging over hadron propagators with 
$N_{\rm src}$ different source locations. 
In such a case, the original cost is given by the CG cost times 
$N_{\rm src}$ plus the cost of generating eigenvectors,
\begin{equation}
  {\rm Cost}_{\rm w/defl.}\textrm{(multi-source)} 
  = \big( \textrm{\#Mult}_{\rm defl.CG(org)}\times 12 \times N_{\rm src}+\textrm{\#Mult}_{\rm Lanczos}\big) \times N_{\rm conf}.
\end{equation}
Assuming that there is no correlation between different source locations, 
we can set $N_{\rm src}=N_G$, so the reduction 
of computational cost is 
\begin{eqnarray}
  r_{\rm Cost}\textrm{(multi-source)} = \frac{\rm Cost(AMA)}{{\rm Cost}_{\rm w/defl.}\textrm{(multi-source)}}
  &\simeq& 0.49\,(m=0.005),\nonumber\\
  &\simeq& 0.33\,(m=0.01).
\end{eqnarray}
The computational cost advantage of AMA 
is cut in half compared to the case with no deflation. 
However this relative cost will decrease again if additional 
propagators are computed, for instance, for three-point functions
(see next section), or if the lattice size is increased and more source translations are used.

\begin{figure}
\begin{center}
\includegraphics[width=140mm]{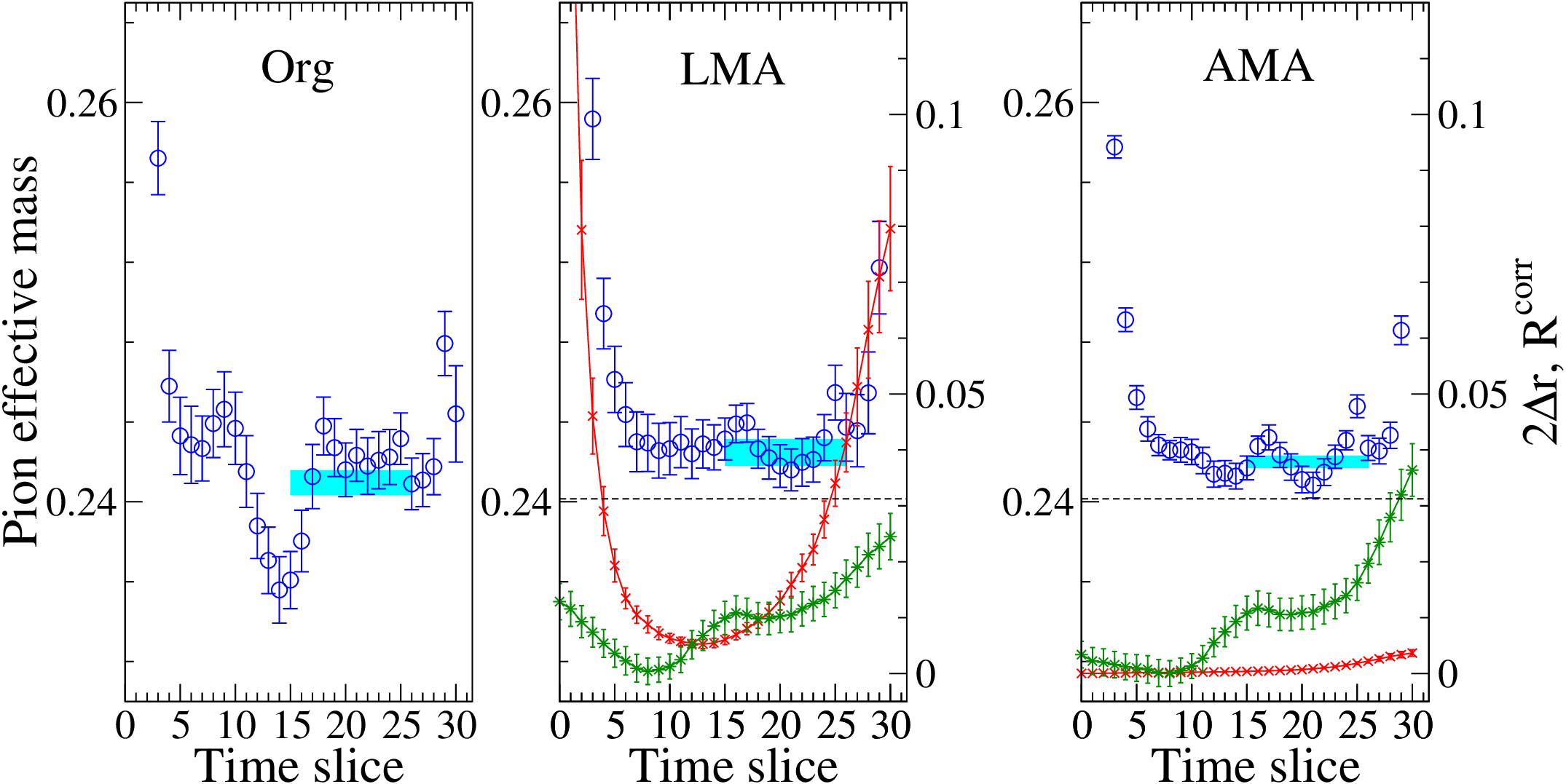}
\caption{
The effective mass plot of
pion at $m=0.01$, and $2\Delta r$ and $R^{\rm corr}$ of 
pion propagator as in Figure \ref{fig:PiVseffm}.
}\label{fig:Piseffm2}
\end{center}
\end{figure}

In the case of the pion, 
comparing $r_{\rm Error}$ in LMA between $m=0.005$ and $m=0.01$, 
we find $\Delta r$ at $m=0.01$ is much larger than at $m=0.005$.
This is due to less dominance of the low-modes and 
the use of fewer low-modes in our setup at $m=0.01$: 
the approximation is worse as seen in Figs.~\ref{fig:PiVseffm} and \ref{fig:Piseffm2}.
Using AMA, thanks to the relaxed CG, 
the approximation is improved. 
We also notice that $r_{\rm Error}$ for the pion mass is about 1.5 times larger
than for the pion propagator (see Fig.~\ref{fig:Npiv_prop} and 
Tab.~\ref{tab:hadmass_m0.005}). 
This is due to the relatively large value of $R^{\rm corr}$ for 
pion propagator above $t=16$.
This observation is confirmed if we extend the distance 
between $\mathcal O^{{\rm (appx)}\,g}$ and $\mathcal O^{{\rm (appx)}\,g'}$. 
For example, using source shifts only in the temporal direction
(source separation in the temporal direction is longer than in the spatial direction), 
$N_G=4$, $r_{\rm Error}$ of the pion mass is similar to the ideal, $1/\sqrt{N_G}=0.5$,
as shown in Tab.~\ref{tab:hadmass_m0.005_Ng4}.
It turns out that for pion the correlation $R^{\rm corr}$ is relatively 
significant in the error reduction rate.

\begin{table}
\begin{center}
\caption{
The comparison of hadron mass (nucleon with momenta, pion, vector meson
and Parity partner of nucleon) in GeV unit 
obtained by global fit of correlator (point sink and gauge-invariant Gaussian smeared source)
in AMA/LMA method with $N_G=32$. 
For reference we show the result with the correlator 
in a single source location. 
``Cost'' column shows the ratio of computational cost of AMA/LMA and 
original one after scaling to the same accuracy. 
We also compare the cost with and without the deflation method in the
original calculation using the number of low-mode presented 
in Table \ref{tab:param}.
}\label{tab:hadmass_m0.005}
\begin{tabular}{c|c|cccc|cccc}
\hline\hline
$m=0.005$\\
\hline
  & Org & LMA & $r_{\rm Error}$ & $r_{\rm Cost}^{\rm w/o\,defl}$ & $r_{\rm Cost}^{\rm w/\,defl}$ 
        & AMA & $r_{\rm Error}$ & $r_{\rm Cost}^{\rm w/o\,defl}$ & $r_{\rm Cost}^{\rm w/\,defl}$\\
\hline
Fit: $[7,12]$\\
\hline
$m_N$ & 1.1322(156) & 1.1520(78) & 0.50 & 0.48 & 0.25 & 1.1519(27) & 0.17 & 0.08 & 0.04\\ 
$E_N(n_p^2=1)$ & 1.2072(172) & 1.2349(82) & 0.48 & 0.43 & 0.23 & 1.2393(30) & 0.18 & 0.09 & 0.04\\ 
$E_N(n_p^2=2)$ & 1.3095(232) & 1.3171(96) & 0.42 & 0.33 & 0.17 & 1.3229(39) & 0.17 & 0.08 & 0.04\\ 
$E_N(n_p^2=3)$ & 1.3723(436) & 1.3941(135) & 0.31 & 0.18 & 0.10 & 1.4010(55) & 0.13 & 0.05 & 0.02\\ 
$E_N(n_p^2=4)$ & 1.5205(627) & 1.4638(192) & 0.31 & 0.18 & 0.09 & 1.4726(88) & 0.14 & 0.05 & 0.03\\ 
\hline
Fit: $[5,8]$ \\
\hline
$m_{N^*}$ & 1.757(81) & 1.671(61) & 0.75 & 1.07 & 0.56 & 1.675(11) & 0.15 & 0.06 & 0.03\\
\hline
Fit: $[16,27]$ \\
\hline
$m_{\pi}$ & 0.3291(12) & 0.3290(4) & 0.37 & 0.27 & 0.14 & 0.3291(4) & 0.36 & 0.36 & 0.19\\ 
\hline
Fit: $[8,15]$ \\
\hline
$m_{V}$ & 0.8621(176) & 0.8746(58) & 0.33 & 0.21 & 0.11 & 0.8738(34) & 0.20 & 0.11 & 0.06\\ 
\hline\hline
\end{tabular}
\end{center}
\end{table}

\begin{table}
\begin{center}
\caption{
Same as shown in Table \ref{tab:hadmass_m0.005} at 
$m=0.01$.
}\label{tab:hadmass_m0.01}
\begin{tabular}{c|c|cccc|cccc}
\hline\hline
$m=0.01$\\
\hline
  & Org & LMA & $r_{\rm Error}$ & $r_{\rm Cost}^{\rm w/o\,defl}$ & $r_{\rm Cost}^{\rm w/\,defl}$ 
        & AMA & $r_{\rm Error}$ & $r_{\rm Cost}^{\rm w/o\,defl}$ & $r_{\rm Cost}^{\rm w/\,defl}$ \\
\hline
Fit: $[7,12]$\\
\hline
$m_N$ & 1.2279(127) & 1.2234(63) & 0.50 & 0.51 & 0.25 & 1.2422(24) & 0.19 & 0.14 & 0.07\\ 
$E_N(n_p^2=1)$ & 1.2877(156) & 1.2992(76) & 0.49 & 0.49 & 0.24 & 1.3222(27) & 0.17 & 0.12 & 0.06\\ 
$E_N(n_p^2=2)$ & 1.3438(207) & 1.3682(97) & 0.47 & 0.46 & 0.22 & 1.3981(32) & 0.16 & 0.09 & 0.05\\ 
$E_N(n_p^2=3)$ & 1.3695(289) & 1.4256(145) & 0.50 & 0.52 & 0.25 & 1.4677(45) & 0.16 & 0.09 & 0.05\\ 
$E_N(n_p^2=4)$ & 1.4661(437) & 1.4944(206) & 0.47 & 0.46 & 0.22 & 1.5379(63) & 0.15 & 0.08 & 0.04\\ 
\hline
Fit: $[5,8]$ \\
\hline
$m_{N^*}$ & 1.800(49) & 1.659(69) & 1.40 & 4.02 & 1.95 & 1.787(11) & 0.23 & 0.20 & 0.10\\
\hline
Fit: $[15,26]$ \\
\hline
$m_{\pi}$ & 0.4169(10) & 0.4195(11) & 1.08 & 2.41 & 1.17 & 0.4187(4) & 0.47 & 0.83 & 0.40\\
\hline
Fit: $[8,15]$ \\
\hline
$m_{V}$ & 0.9185(124) & 0.9228(67) & 0.55 & 0.62 & 0.30 & 0.9198(29) & 0.24 & 0.22 & 0.11\\ 
\hline\hline
\end{tabular}
\end{center}
\end{table}

\begin{table}
\begin{center}
\caption{
Pion and vector meson mass as shown in Table \ref{tab:hadmass_m0.005} at 
$m=0.005$ and $N_G=4$.
}\label{tab:hadmass_m0.005_Ng4}
\begin{tabular}{c|cccc|cccc}
\hline\hline
$m=0.005$\\
\hline
  & LMA & $r_{\rm Error}$ & $r_{\rm Cost}^{\rm w/o\,defl}$ & $r_{\rm Cost}^{\rm w/\,defl}$ 
  & AMA & $r_{\rm Error}$ & $r_{\rm Cost}^{\rm w/o\,defl}$ & $r_{\rm Cost}^{\rm w/\,defl}$ \\
\hline
Fit: $[15,26]$ \\
\hline
$m_{\pi}$ & 0.3286(6) & 0.52 & 0.52 & 0.27 & 0.3287(6) & 0.51 & 0.53 & 0.28\\
\hline
Fit: $[8,15]$ \\
\hline
$m_{V}$ & 0.8840(94) & 0.54 & 0.55 & 0.29 & 0.8801(83) & 0.47 & 0.45 & 0.23\\
\hline\hline
\end{tabular}
\end{center}
\end{table}

\subsection{Nucleon form factors}

In this section we apply AMA to nucleon
three-point functions which have a more complicated structure in terms 
of quark propagators. 
We carry out the measurement of three-point functions ((nucleon)-(operator)-(nucleon)) where the operators 
are vector ($V_\mu$) or axial-vector ($A_\mu$) currents, and we 
evaluate the axial-charge and isovector 
form factors defined from the matrix elements,
\begin{eqnarray}
  \langle N_1(p_1,s) | V^a_\mu | N_0(p_0,s) \rangle
  &=& \bar u_{N_1}(p_1,s)\Big[ \gamma_\mu F^a_1(q^2) 
   + \frac{\sigma_{\mu\nu}q_\nu}{2m_N}F^a_2(q^2)\Big] u_{N_0}(p_0,s),\\
  \langle N_1(p_1,s) | A^a_\mu | N_0(p_0,s)\rangle
  &=& \bar u_{N_1}(p_1,s)\Big[ \gamma_\mu\gamma_5 F^a_A(q^2) 
   + iq_\mu\gamma_5F^a_2(q^2)\Big] u_{N_0}(p_0,s),
\end{eqnarray}
with momenta $\vec p_{0}$ and $\vec p_1$  of on-shell nucleon states $N_{0}$ and $N_1$, 
respectively, with spin $s$. The superscript ``$a$'' is an SU(2) flavor index 
referring to either isovector or isoscalar components.  
Below we study matrix elements of the isovector currents ($a=+$).
$F^{a}_{1}$ and $F^{a}_{2}$ are obtained from the Sachs form factors, 
\begin{equation}
  G_E(q^2) = F^a_1(q^2) - \frac{q^2}{4m_N^2}F^a_2(q^2),\quad
  G_M(q^2) = F^a_1(q^2) + F^a_2(q^2).
  \label{eq:Gem}
\end{equation}
The isovector form factor $F^+_A(q^2)$ at zero momentum transfer is known as the 
axial-charge of the nucleon, $g_A=F^+_A(0)$, 
which is an important quantity governing neutron $\beta$ decay. 

To obtain the form factors, we construct ratios of three-point correlation functions,
$C^N_{J_\mu}$, and nucleon two-point functions, $C_{G,L}^N$, as
\begin{eqnarray}
  R_{J_\mu}(t_1,t,t_0|p_1,p_0) = K
  \frac{C^N_{J_\mu}(\vec q,t)}{C_G^N(t_1-t_0,0)}\bigg[ 
  \frac{C_L^N(t_1-t,\vec q)C_G^N(t-t_0,0)C_L^N(t_1-t_0,0)}
       {C_L^N(t_1-t,0)C_G^N(t-t_0,\vec q)C_L^N(t_1-t_0,\vec q)}\bigg]^{1/2}
  \label{eq:ratio_mu}
\end{eqnarray}
with $K=\sqrt{2(E_N+m_N)/E_N}$, 
where $C_L^N$ is with point-sink and gauge-invariant Gaussian smeared source, 
$C_G^N$ is with gauge-invariant Gaussian smeared source and sink. 
$t_{0},\,t_1$ denote the temporal location of the initial and final states of 
nucleon which are fixed, and 
$t$ is the temporal location of the operator which moves between $t_0$ and $t_1$.
The momentum transfer is defined as $q=p_0-p_1$ 
, and in our setup 
we use $p_0=(E_N,\vec p)$ and $p_1=(m_N,0)$ with $\vec p=(p_x,p_y,p_z)=2\pi\vec n_p/L$, 
$\vec n_p^2 = 0,\cdots,4$.
In order to extract the form factors of the ground state nucleon 
from $R_{J_\mu}$ we use the spin-projection matrix $P_4=(1+\gamma_4)/2$
and $P_{5z}=P_4\gamma_5\gamma_3$, as in \cite{Yamazaki:2009zq}. For the vector case,
\begin{eqnarray}
 &&\lim_{t_1-t,t-t_0\gg 1}{\rm tr}\big[ P_{5z} (R_{V_1}+R_{V_2})\big](t_1,t,t_0|p_1,p_0) 
 = \frac{-ip_x + ip_y}{m_N}G_M(q^2),\\
 &&\lim_{t_1-t,t-t_0\gg 1}{\rm tr}\big[ P_4 R_{V_4}\big](t_1,t,t_0|p_1,p_0) 
 = \frac{E_N+m_N}{m_N}G_E(q^2),
\end{eqnarray}
and for the axial-vector,
\begin{eqnarray}
 &&\lim_{t_1-t,t-t_0\gg 1}{\rm tr}\big[ P_{5z} (R_{A_1}+R_{A_2})\big](t_1,t,t_0|p_1,p_0) 
 = -\frac{(p_x+p_y)p_z}{m_N}F_P(q^2),\\
 &&\lim_{t_1-t,t-t_0\gg 1}{\rm tr}\big[ P_{5z} R_{A_3}\big](t_1,t,t_0|p_1,p_0) 
 = \frac{1}{m_N}\big[ m_NF_A(q^2) - p_z^2F_P(q^2)\big],
\end{eqnarray}
after taking $t_1\gg t\gg t_0$ to project on the nucleon ground state.
In the above derivation we use the normalization for Dirac spinors,
$\sum_s \bar u_{N}(p,s)u_{N}(p,s) = 2m_N$.
The parameters of the gauge-invariant Gaussian smeared source-sink are the same as 
in \cite{Yamazaki:2009zq}, and $t_0=0$, $t_1=12$.
In this calculation we employ the local currents $V^a_\mu=\bar q\gamma_\mu\tau^a q$
and $A^a_\mu=\bar q\gamma_\mu\gamma_5\tau^a q$ where $\tau^a$ is flavor SU(2) generator
normalized as ${\rm tr}\,\tau^a\tau^b=\delta^{ab}$, and hence
we multiply matrix elements of the currents by the renormalization constant $Z_V=0.7178$, determined 
non-perturbatively \cite{Aoki:2010dy}.

\begin{figure}
\begin{center}
\includegraphics[width=140mm]{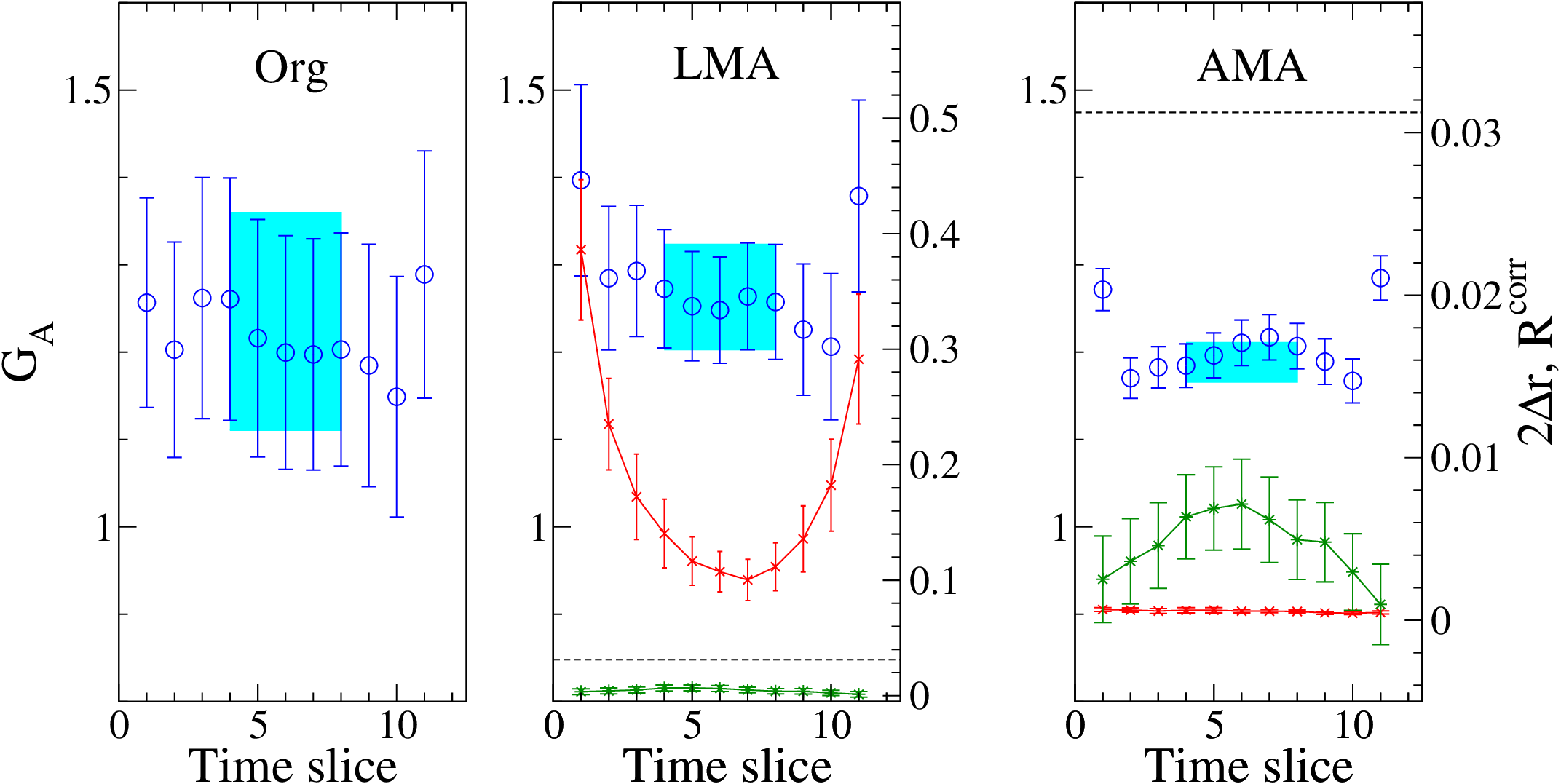}
\vskip 5mm
\includegraphics[width=140mm]{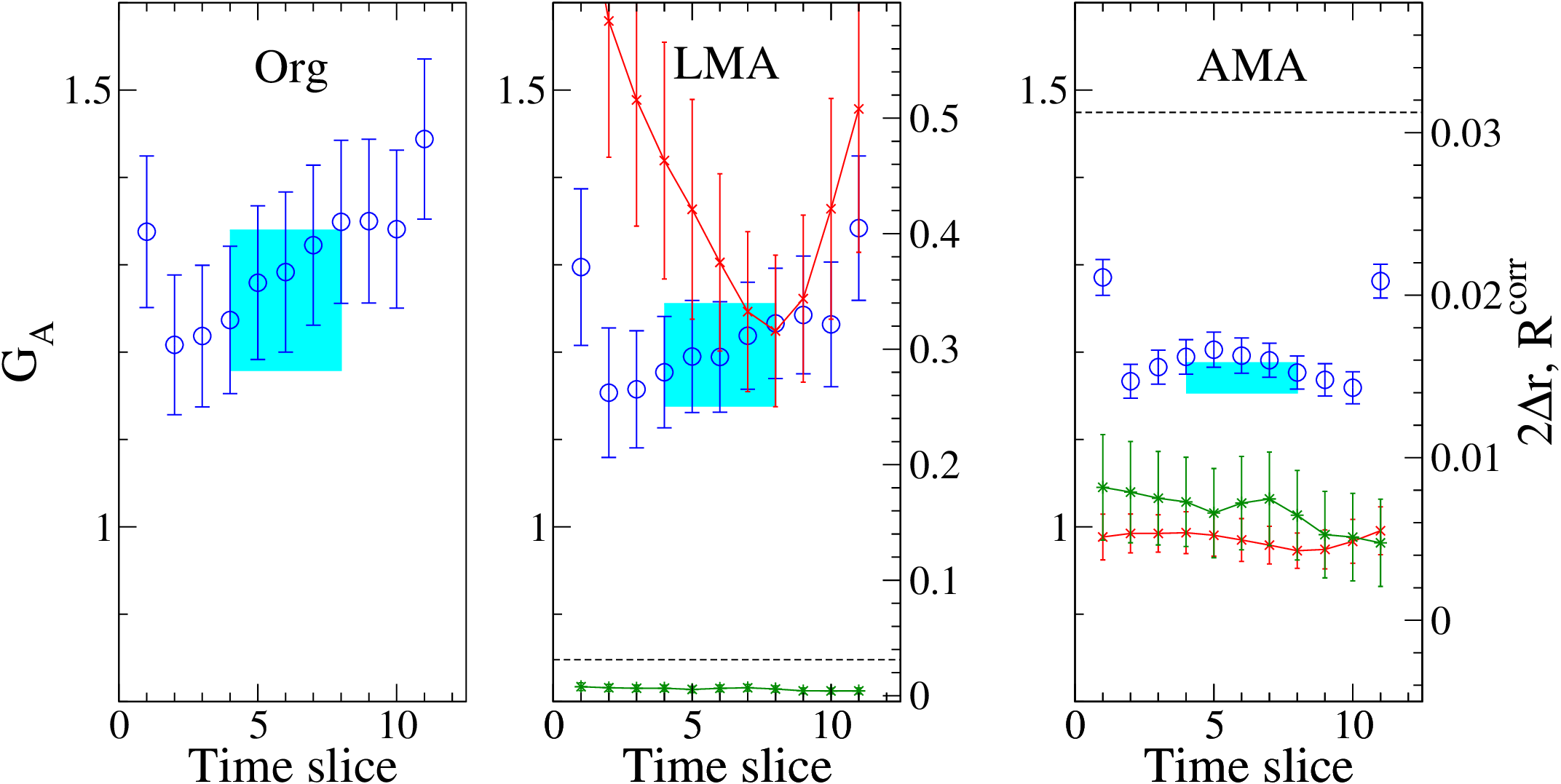}
\caption{
Time-slice dependence of axial-charge $G_A$
in $m=0.005$ (top) and $m=0.01$ (bottom) with standard method (left), 
LMA (middle) and AMA (right).
The cross symbols and star symbols denote $2\Delta r$ and $R^{\rm corr}$ 
for three-point function which is in nominator in Eq.~(\ref{eq:ratio_mu}).
The colored band is the constant fitting result in this range.
}\label{fig:ga}
\end{center}
\end{figure}

We compare the axial charge and isovector form factor at each momentum
between the standard method and LMA or AMA.
Figure \ref{fig:ga} shows $g_A$
for two different masses. 
A ground state plateau is clearly observed for $4\le t\le 8$ for both masses. 
Comparing the contribution of $\Delta r$ and $R^{\rm corr}$ in LMA and AMA, 
one sees that $\Delta r$ in AMA is much smaller, and the quality of the approximation is
significantly enhanced. 
In cost estimates of the three-point functions, we compute
``polarized'' and ``unpolarized'' matrix elements for both up-type and down-type contractions 
which is an additional cost factor of four quark propagators.
As shown in Tabs.~\ref{tab:ga_24c}~and~\ref{tab:f12_24c}, 
AMA achieves error reductions in $G_A$, $F_1^+$ and $F_2^+$
close to $1/\sqrt{N_G}$ with 5--20 times smaller computational cost
than the standard method or LMA.
Comparing the results for AMA at the two masses $m=0.005$ and $m=0.01$,
the error reduction compared to the standard method is 
significant for both, despite having fewer eigenvectors for the latter.
The cost ratios, comparing to the multi-source method with $N_{\rm src}=N_G$, are 
\begin{eqnarray}
  r_{\rm Cost}\textrm{(multi-source)} 
  &\simeq& 0.32\,(m=0.005),\nonumber\\
  &\simeq& 0.24\,(m=0.01),
\end{eqnarray}
in which we have gains greater than factors of 3 and 4 for AMA. We also note that not only have the statistical errors decreased dramatically, but the plateaus are much more readily observed for AMA.

\begin{table}
\begin{center}
\caption{
Table of axial charge $G_A$ with standard method, LMA and AMA 
in $m=0.005$ and $m=0.01$.
}\label{tab:ga_24c}
\begin{tabular}{c|c|cccc|cccc}
\hline\hline
$G_A$ & Org
 & LMA & $r_{\rm Error}$ & $r_{\rm Cost}^{\rm w/o\,defl}$ & $r_{\rm Cost}^{\rm w/\,defl}$ 
 & AMA & $r_{\rm Error}$ & $r_{\rm Cost}^{\rm w/o\,defl}$ & $r_{\rm Cost}^{\rm w/\,defl}$\\
\hline
Fit: $[4,8]$ \\
\hline
$m=0.005$ & 1.235(124) & 1.263(60) & 0.48 & 0.11 & 0.23 & 1.188(22) & 0.18 & 0.04 & 0.09\\
$m=0.01$  & 1.259(80) & 1.197(58) & 0.73 & 0.35 & 0.53 & 1.170(17) & 0.21 & 0.11 & 0.17\\
\hline\hline
\end{tabular}
\end{center}
\end{table}

\begin{table}
\begin{center}
\caption{
Table of $F_1^+$ and $F_2^+$ with standard method, LMA and AMA 
in $m=0.01$.
}\label{tab:f12_24c}
\begin{tabular}{c|c|cccc|cccc}
\hline\hline
 & Org
 & LMA & $r_{\rm Error}$ & $r_{\rm Cost}^{\rm w/o\,defl}$ & $r_{\rm Cost}^{\rm w/\,defl}$ 
 & AMA & $r_{\rm Error}$ & $r_{\rm Cost}^{\rm w/o\,defl}$ & $r_{\rm Cost}^{\rm w/\,defl}$\\
\hline
Fit: $[4,8]$ \\
\hline
$F_1^+(n_p^2=1)$ & 0.849(53) & 0.860(52) & 0.99 & 0.64 & 0.97 & 0.799(10) & 0.20 & 0.10 & 0.15\\ 
$F_1^+(n_p^2=2)$ & 0.695(50) & 0.730(47) & 0.95 & 0.60 & 0.91 & 0.678(10) & 0.20 & 0.10 & 0.15\\ 
$F_1^+(n_p^2=3)$ & 0.493(57) & 0.618(47) & 0.82 & 0.45 & 0.68 & 0.583(11) & 0.21 & 0.10 & 0.16\\ 
$F_1^+(n_p^2=4)$ & 0.406(50) & 0.524(49) & 0.97 & 0.62 & 0.94 & 0.555(17) & 0.35 & 0.30 & 0.45\\ 
\hline
$F_2^+(n_p^2=1)$ & 2.61(26) & 2.35(17) & 0.66 & 0.28 & 0.43 & 2.37(5) & 0.19 & 0.09 & 0.13\\ 
$F_2^+(n_p^2=2)$ & 1.88(22) & 1.91(14) & 0.66 & 0.29 & 0.44 & 1.85(4) & 0.19 & 0.09 & 0.13\\ 
$F_2^+(n_p^2=3)$ & 1.52(16) & 1.62(13) & 0.82 & 0.44 & 0.67 & 1.52(4) & 0.25 & 0.15 & 0.23\\ 
$F_2^+(n_p^2=4)$ & 1.12(15) & 1.17(13) & 0.86 & 0.49 & 0.74 & 1.32(5) & 0.35 & 0.29 & 0.44\\ 
\hline\hline
\end{tabular}
\end{center}
\end{table}

\section{Future extension}
\label{sec:other}

This paper has shown numerical tests of AMA
using the relaxed CG as
the approximation, but there are many other 
examples of $\mathcal O^{(\rm appx)}$. 
One idea is to employ improved DWF actions, {\it e.g.} M{\"o}bius-type 
\cite{Brower:2012vk} or Borici-type 
\cite{Borici:1999zw,Borici:2004pn}, 
which are extensions of DWF allowing smaller $L_s$
without enhancing chiral symmetry breaking,
in addition to the relaxed CG solver.
Such improvements have other benefits like  
the reduction of memory or disk-storage size of eigenvector data stored on disk.

\begin{figure}
\begin{center}
\includegraphics[width=75mm]{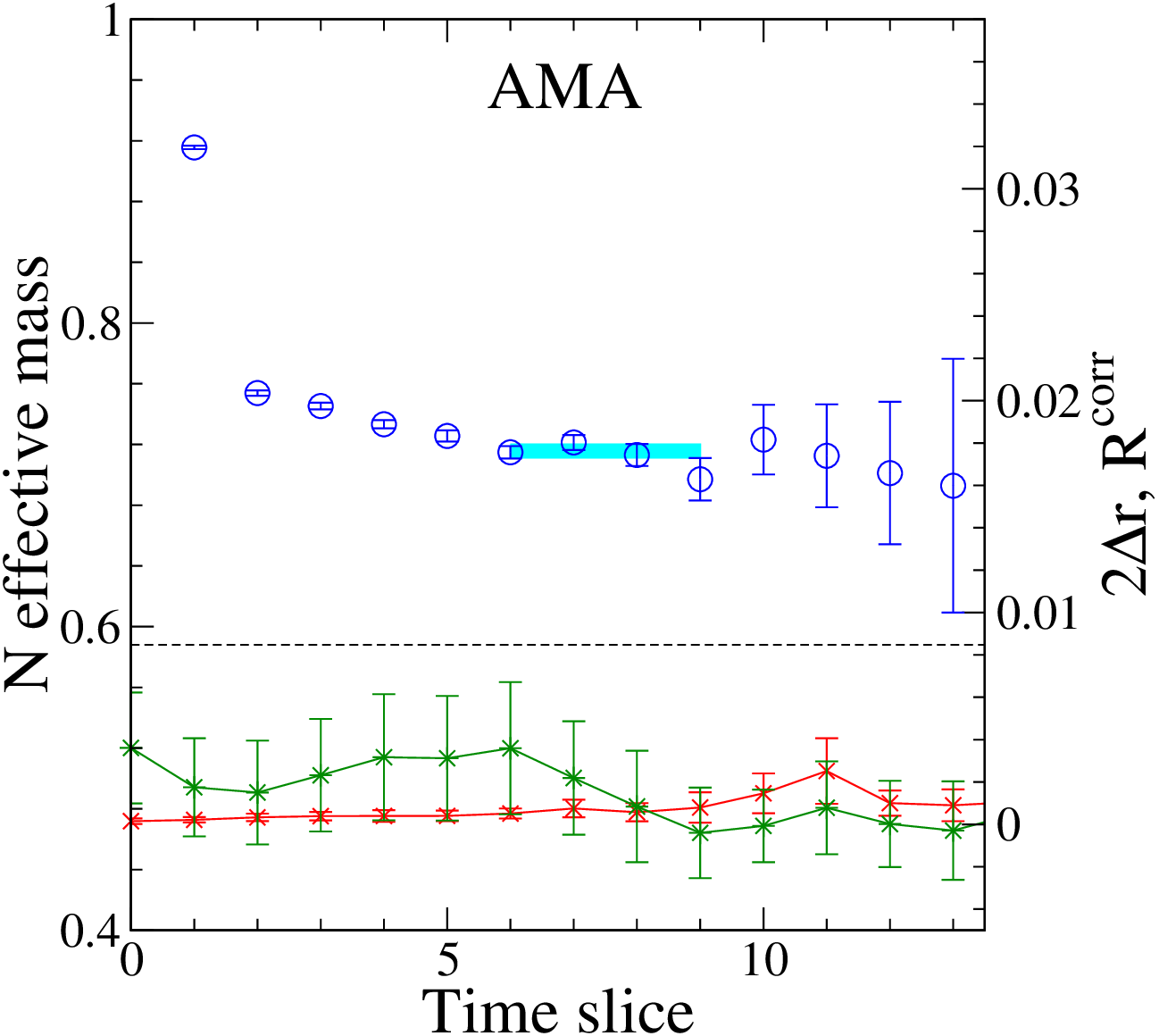}
\hspace{3mm}
\includegraphics[width=77mm]{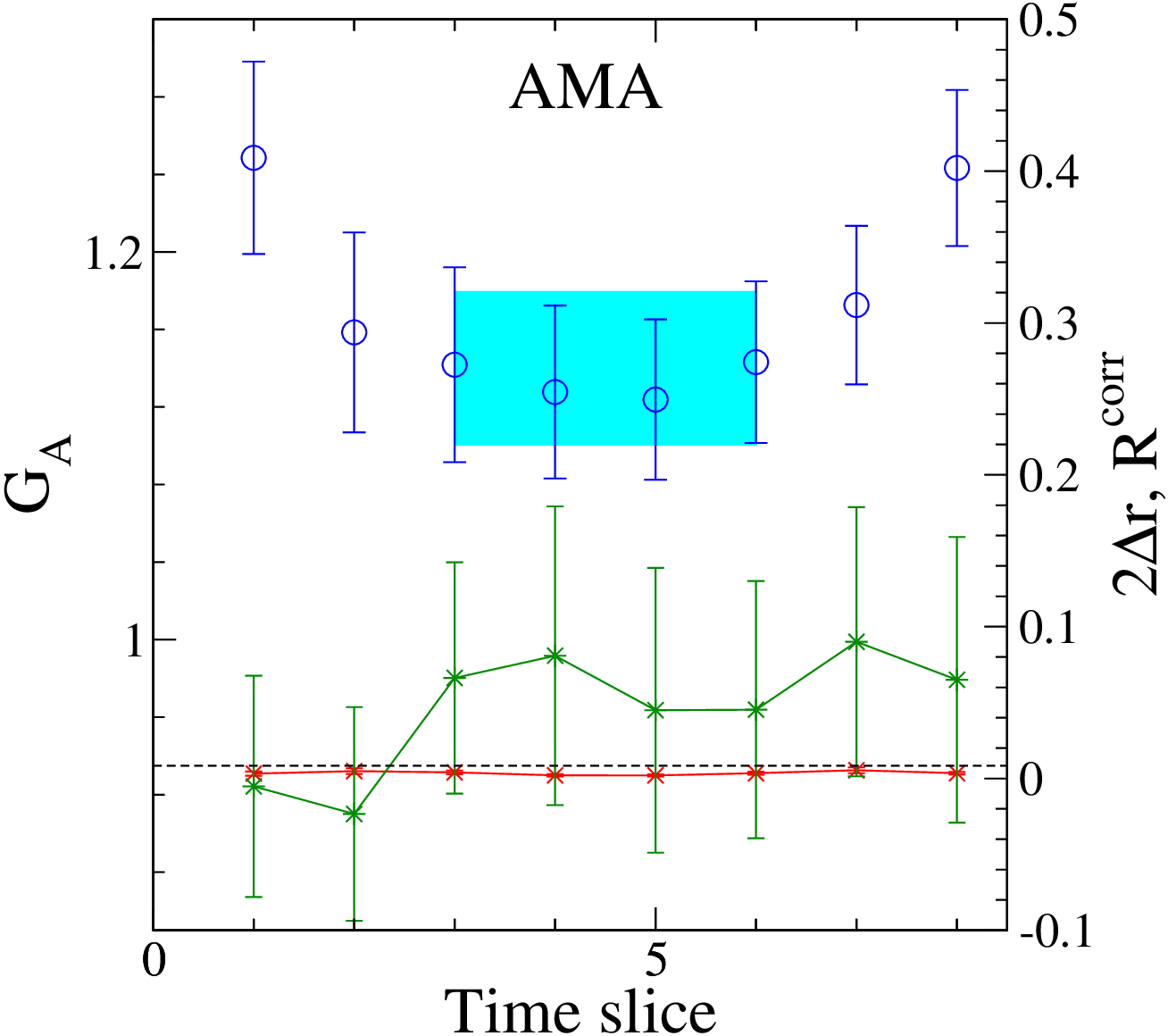}
\caption{
The nucleon effective mass plot and 
the axial charge, as shown in Figure \ref{fig:ga}, at $m=0.001$ with 
Iwasaki+DSDR action in $32^3\times 64$ lattice. 
}\label{fig:ga_ID}
\end{center}
\end{figure}

We test the above strategy on another DWF ensemble generated 
by the RBC/UKQCD collaboration~\cite{Arthur:2012opa},
with larger lattice size ($32^3\times 64$) and $L_s=32$, 
and smaller pion mass, $m_\pi\approx 170$ MeV. 
For the approximation we take a M{\"o}bius-type DWF Dirac operator with $L_s=16$.
We use 1000 low-modes, computed with a 200 degree Chebychev polynomial, and then only
2 restarts of the Lanczos procedure are needed.
In this case, the computational cost ratio reads
\begin{eqnarray}
  {\rm Cost(AMA)} &=& \big(\textrm{\#Mult}_{\rm Lanczos} \times 0.6
  + \big(\textrm{\#Mult}_{\rm defl.CG(org)} \nonumber\\
  &+& \textrm{\#Mult}_{\rm defl.CG(AMA)}\times 0.6N_G\big)
  \times 12 \big) \times N_{\rm conf}.
\end{eqnarray}
where the factor 0.6 arises from the fact that there is an additional 20\% cost for 
the multiplication with the M{\"o}bius-type Dirac operator compared to a DWF 
operator with same $L_s$ length together with the having 
of the cost due to using $L_s/2$ for the M{\"o}bius-type Dirac operator, 
{\it i.e.} 1.2/2 = 0.6. 
The axial charge is shown in Fig.~\ref{fig:ga_ID}. 
One sees that there is a clear plateau between 3 and 6, where 
we set the source and sink operator at time-slice 0 and 9 respectively, 
and around the plateau the correlation $\Delta r$ has a similar order 
as for the $m=0.01$, $24^3\times 64$ case discussed in the last section.
In table \ref{tab:mass_32cID} and \ref{tab:ga_32cID}
we summarize hadron masses and the axial charge 
for both the standard method and AMA. 
From those tables, the ratio of errors is close to 
the ideal one, $1/\sqrt{112}\simeq 0.094$, 
and thus $\mathcal O^{\rm (appx)}$ is still a good approximation
to the original even though we use M{\"o}bius-type DWF. 
AMA reduces the computational cost by
10 to 30 times in this case. 

\begin{table}
\begin{center}
\caption{Result of hadron mass in DSDR lattice in $m=0.001$.}
\label{tab:mass_32cID}
\begin{tabular}{c|c|ccccc}
\hline\hline
 & Org & AMA & $r_{\rm Error}$ & $r_{\rm Cost}^{\rm w/o\,defl}$ \\
\hline
Fit: $[6,9]$ \\
\hline
$m_N$ & 0.9625(538) & 0.9822(57) & 0.11 & 0.04 \\ 
$E_N(n_p^2=1)$ & 0.9759(524) & 1.0201(59) & 0.11 & 0.04 \\ 
$E_N(n_p^2=2)$ & 1.0090(515) & 1.0568(65) & 0.13 & 0.06 \\ 
$E_N(n_p^2=3)$ & 1.0466(509) & 1.0900(74) & 0.15 & 0.08 \\ 
$E_N(n_p^2=4)$ & 1.0035(544) & 1.1268(84) & 0.16 & 0.08 \\ 
\hline
Fit: $[4,7]$ \\
\hline
$m_N$ & 1.445(258) & 1.430(24) & 0.09 & 0.03 \\ 
\hline
Fit: $[8,21]$ \\
\hline
$m_{\pi}$ & 0.1694(21) & 0.1712(3) & 0.18 & 0.11 \\ 
\hline
Fit: $[6,10]$ \\
\hline
$m_{V}$ & 0.8502(821) & 0.7414(77) & 0.09 & 0.03 \\
\hline\hline
\end{tabular}
\end{center}
\end{table}

\begin{table}
\begin{center}
\caption{Result of $G_A$ in DSDR lattice in $m=0.001$.}
\label{tab:ga_32cID}
\begin{tabular}{c|c|ccccc}
\hline\hline
 & Org & AMA & $r_{\rm Error}$ & $r_{\rm Cost}^{\rm w/o\,defl}$ \\
\hline
Fit: $[3,6]$ \\
\hline
$G_A$ & 1.401(275) & 1.135(42) & 0.15 & 0.05 \\
\hline\hline
\end{tabular}
\end{center}
\end{table}

Still other approximations are possible. 
For instance, the inexactly deflated CG, using the EigCG algorithm 
\cite{Abdel-Rehim:2013xgv} with low-precision, is adopted
as $\mathcal O^{\rm (appx)}$.
This uses low-precision eigenmodes as well as deflation, 
and will be beneficial for
long-distance observables corresponding to pion and Kaon physics.
Especially for large lattice sizes, since there are many available source locations, 
it is possible to reduce the size of gauge ensembles while still maintaining statistical precision. 
Furthermore we also note that in \cite{Bali:2009hu}
the hopping parameter expansion for the inverse of the Dirac matrix is  
used as the approximation $\mathcal O^{\rm (appx)}$.
These are a few of the new directions to pursue high precision 
calculations without additional computational cost in a Monte-Carlo simulation
(however, a careful analysis of autocorrelation times is necessary).

\section{Discussion and summary}

\begin{figure}
\begin{center}
\includegraphics[width=120mm]{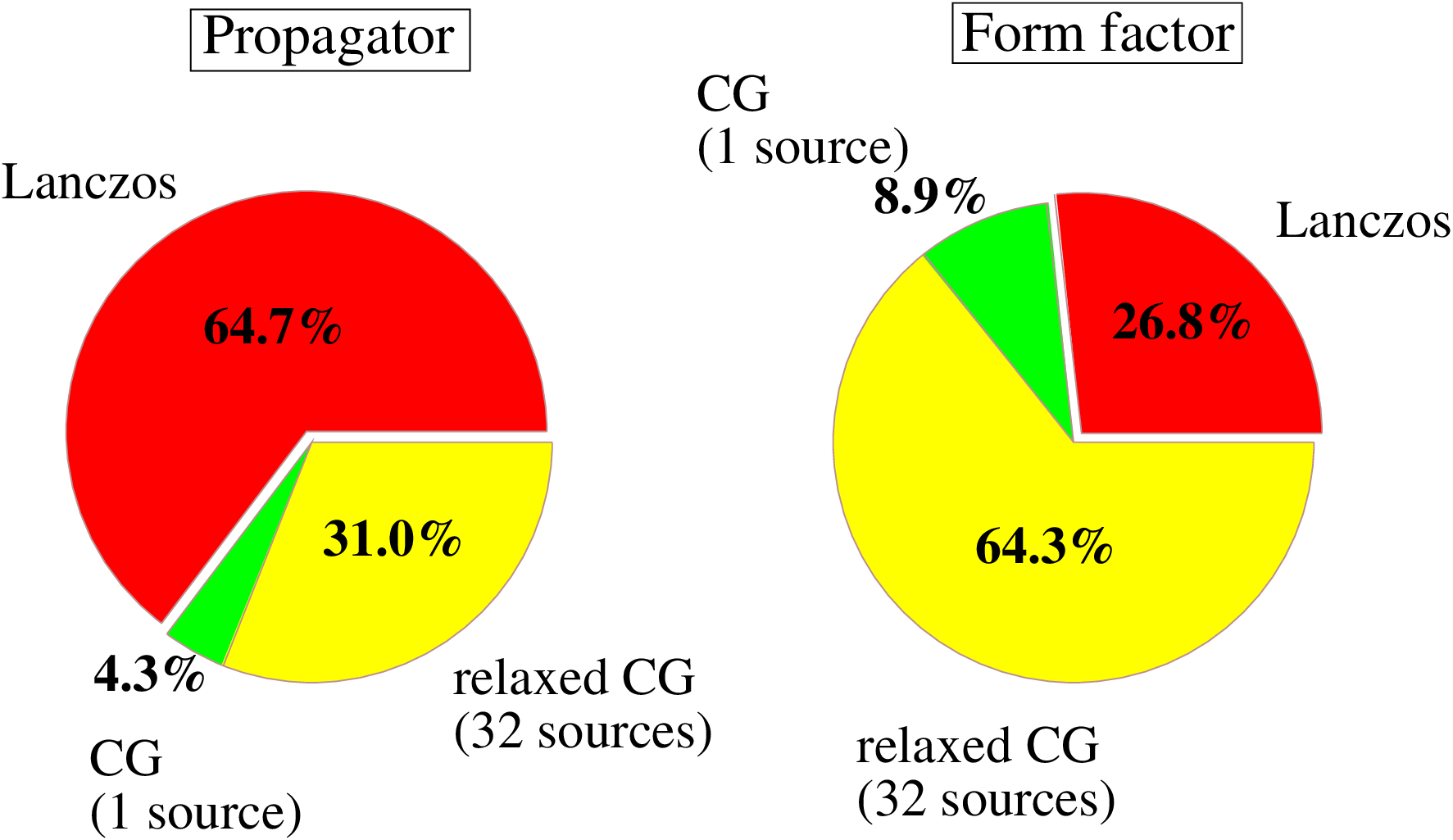}
\caption{
The rate of computational cost of AMA for hadron propagator (left)
and three-point function of form factor (right) at $m=0.005$.
This is in the case of 400 eigenmodes computation
and use of 32 source locations for relaxed CG
($\varepsilon = 0.003$) in AMA. 
}\label{fig:pichart_m0.005}
\end{center}
\end{figure}

\begin{figure}
\begin{center}
\includegraphics[width=130mm]{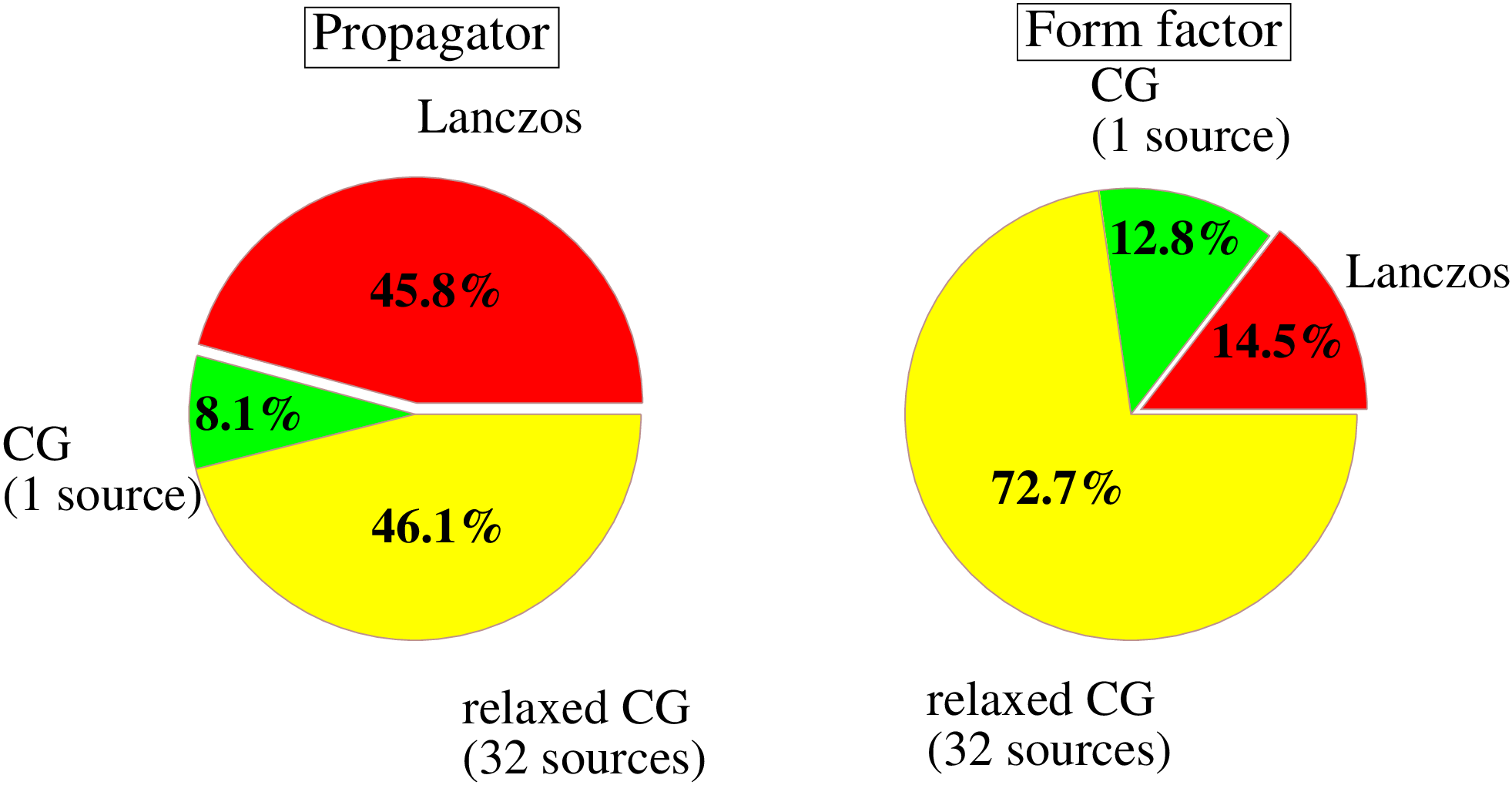}
\caption{
Same figure as Figure \ref{fig:pichart_m0.005} at $m=0.01$. 
This is in the case of 180 eigenmodes computation
and use of 32 source locations for relaxed CG
($\varepsilon = 0.003$) in AMA. 
}\label{fig:pichart_m0.01}
\end{center}
\end{figure}

\begin{figure}
\begin{center}
\includegraphics[width=130mm]{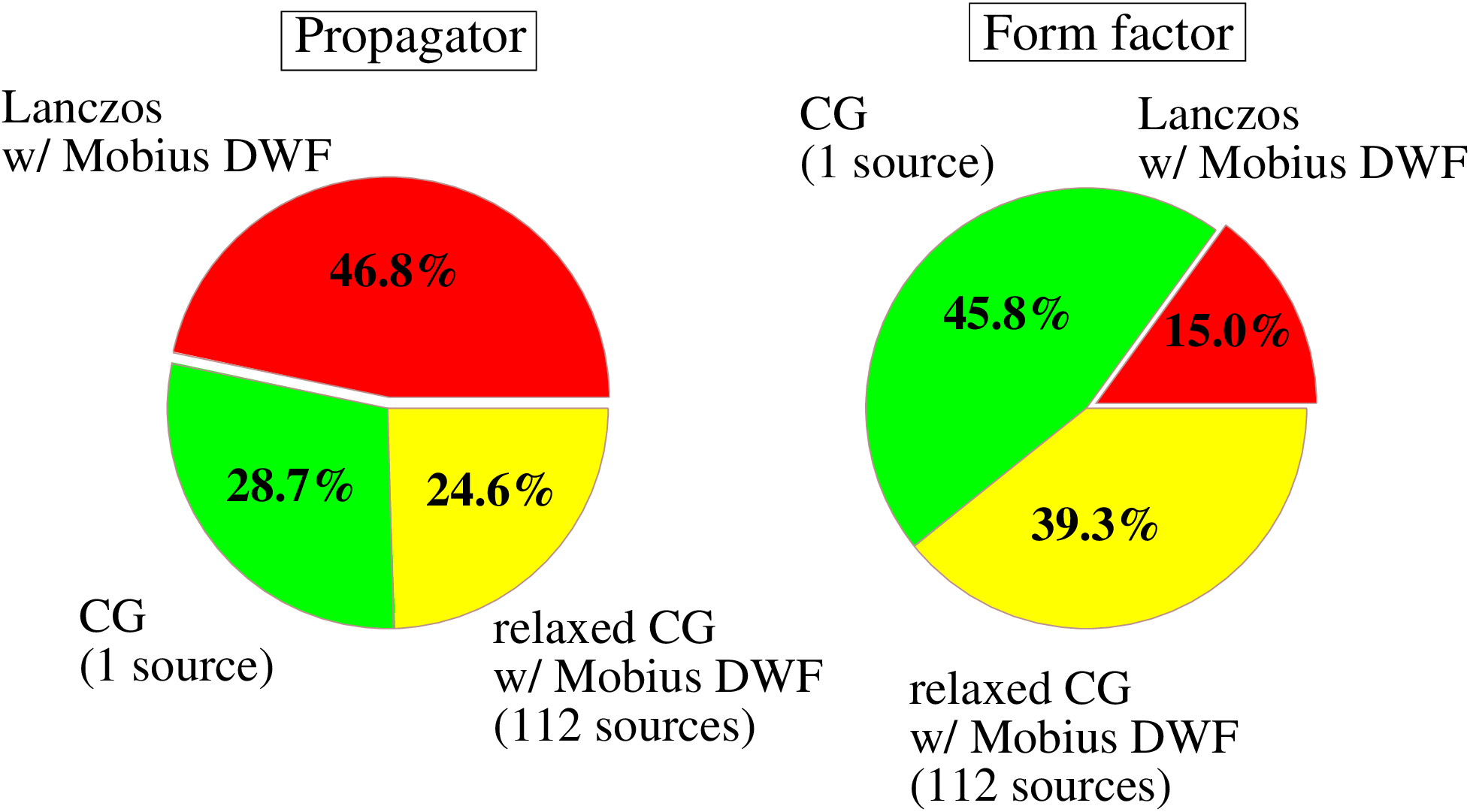}
\caption{
Same figure as Figure \ref{fig:pichart_m0.005} for 
$32^3\times 64\times32$ DSDR lattice. 
This is in the case of 1000 eigenmodes computation
and use of 112 source locations for relaxed CG
with M{\"o}bius DWF kernel in AMA. 
}\label{fig:pichart_m0.001}
\end{center}
\end{figure}

\begin{figure}
\begin{center}
\includegraphics[width=130mm]{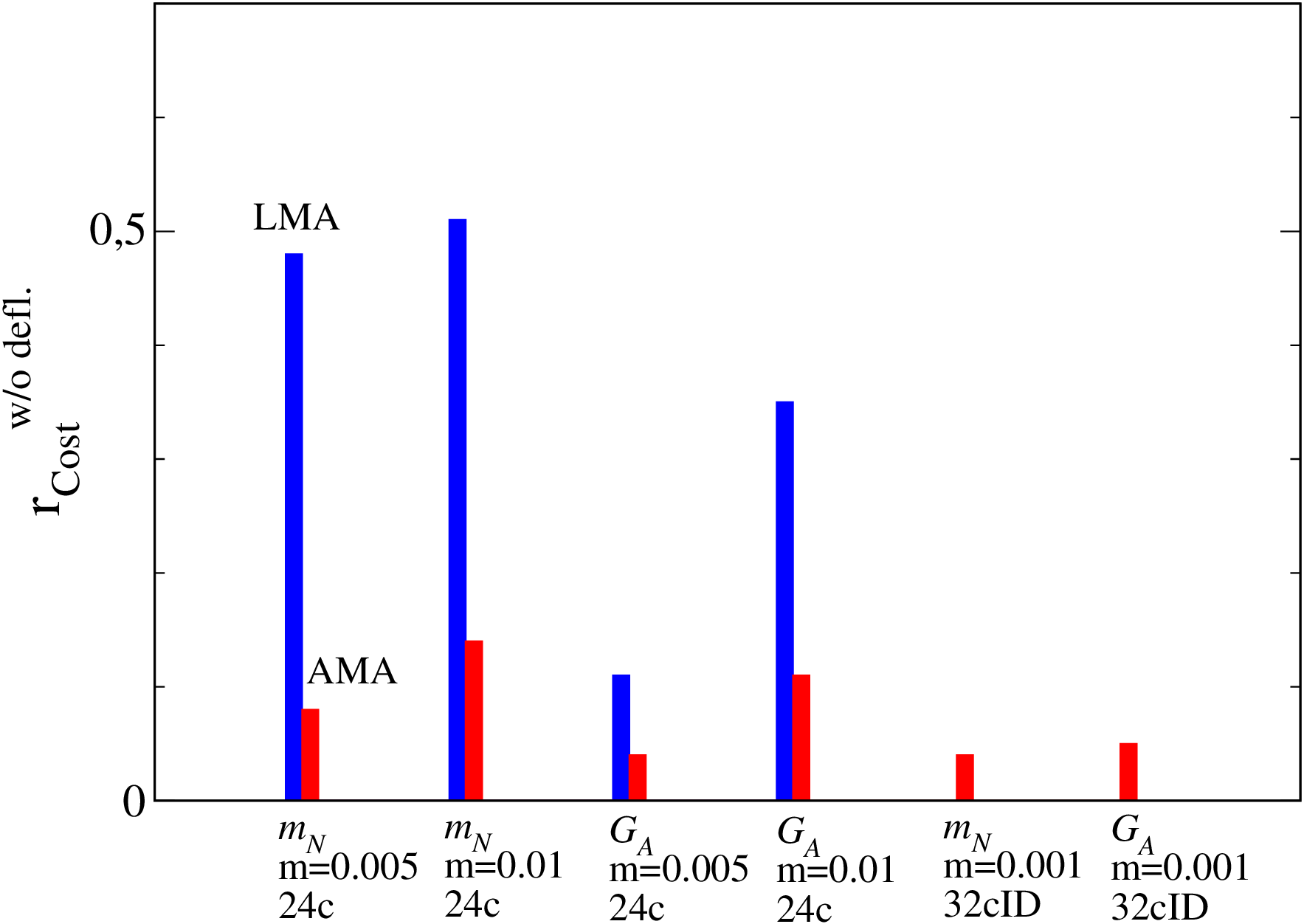}
\caption{
$r_{\rm Cost}^{\rm w/o defl.}$ of nucleon mass $m_N$ and axial charge $G_A$
for LMA (blue bar) and AMA (red bar).
}\label{fig:bar_cost}
\end{center}
\end{figure}

As shown in the previous section, 
all-mode averaging (AMA) is a powerful tool for the precise measurement of 
observables obtained from correlation functions in Monte-Carlo simulations. 
Defining the improved estimator $\mathcal O^{\rm (imp)}$
using the approximation $\mathcal O^{\rm (appx)}$, 
which has the same covariance properties as the original $\mathcal O$
but a much smaller construction cost,
$\mathcal O^{\rm (imp)}$ has smaller statistical errors without additional computational cost. 
In this paper we employ the relaxed CG with deflation to produce the approximation. 
Since the computational cost of the approximation using the  
relaxed CG is much less than the original one, 
the observables needing many quark propagators with CG solve of the 
Dirac matrix benefit accordingly from the AMA method. 
Figures \ref{fig:pichart_m0.005}, \ref{fig:pichart_m0.01} and 
\ref{fig:pichart_m0.001}
show the ratio of computational costs for AMA. 
One sees that, compared to the propagator, 
the cost of the CG solves for the nucleon form factor dominates the total cost. 
This is because 4 extra CG solves are necessary to construct 
the three-point functions. 
Figure \ref{fig:bar_cost} shows the summary of 
reduction rate of computational cost for LMA and AMA
as in Table \ref{tab:hadmass_m0.005}, \ref{tab:hadmass_m0.01}, \ref{tab:ga_24c}
\ref{tab:mass_32cID} and \ref{tab:ga_32cID}. 
The computational cost of $G_A$ in AMA is more reduced rather than the two-point function,
and also AMA has an advantage of more than 7 times speed-up for computation of 
two- and three-point function compared to traditional method.
We also notice that, for 32$^3\times 64$ lattice size and DSDR gauge action
(``32cID''), there is more than 10 times reduction 
of $r_{\rm Cost}^{\rm w/o\,defl.}$ by employing the
M{\"o}bius operator in the approximation. 
There are also realistic DWF simulations at the physical quark mass
point with 5.5 fm volume with two lattice spacings, which employed
AMA \cite{Blum:2014wsa}.
It turns out that AMA also works well for an approximation 
which is made from a different action than the original one. 
As shown in Fig.~\ref{fig:pichart_m0.001}, 
the computational cost of a precise CG solve with DWF
is still large, in fact 29\% for the propagator 
and 46\% for the form factor, since we did not use deflation method in the original one. 
Further cost reduction by applying the modified deflation 
method in CG with M{\"o}bius DWF eigenmodes is currently under way \cite{Yin:2011np}. 

\begin{figure}
\begin{center}
\includegraphics[width=120mm]{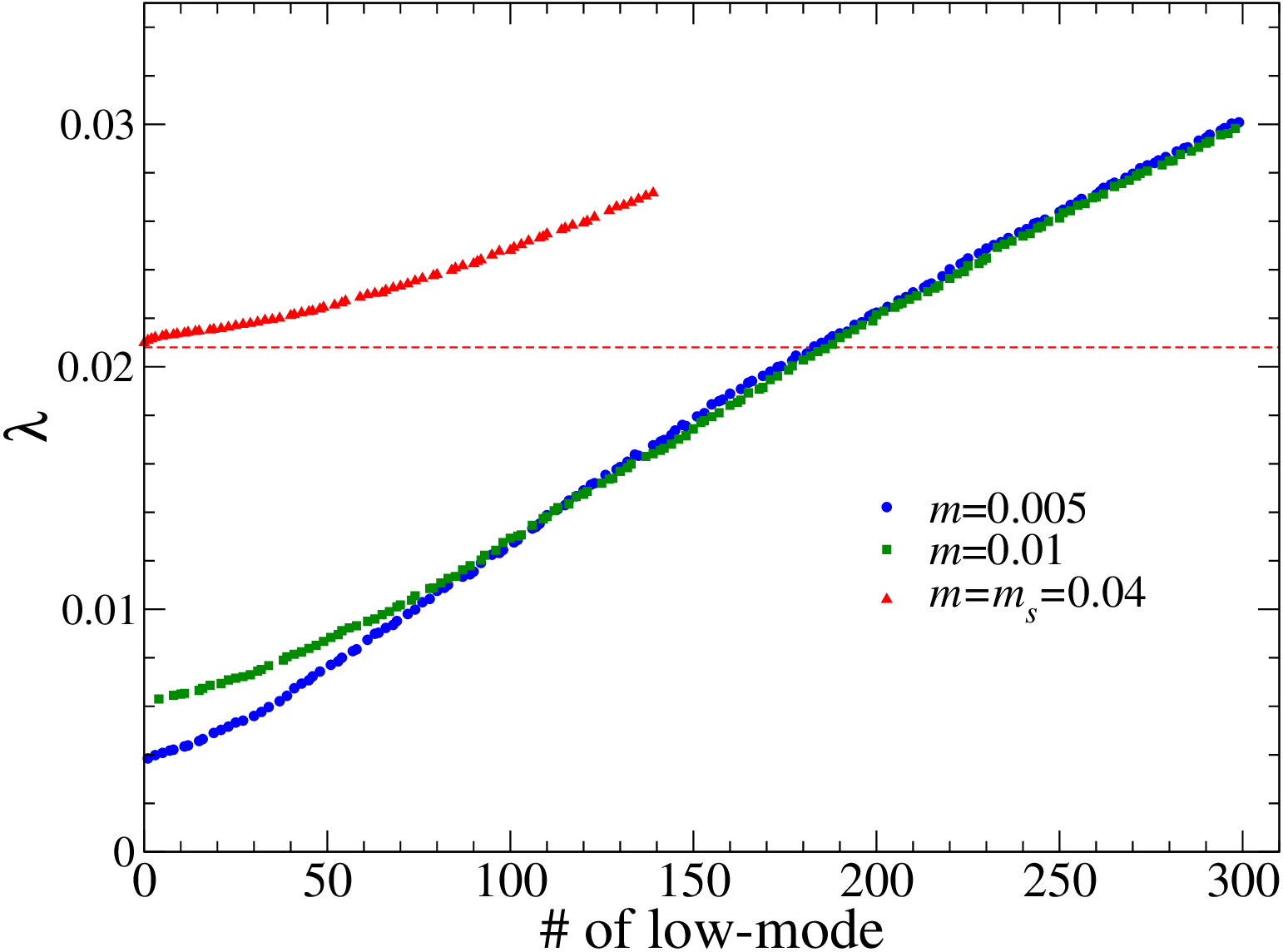}
\caption{
Distribution of positive low-lying eigenvalue at 
light quark mass $m=0.005,\,0.01$ and strange quark mass $m=m_s=0.04$.
Dashed line shows the lowest eigenvalue for strange quark mass. 
}\label{fig:eigen_dist}
\end{center}
\end{figure}

We comment on the relation of the approximation with the low-mode distribution of the Dirac operator. 
As in Eqs.~(\ref{eq:all-mode})~and~(\ref{eq:all-mode2}), 
the deflation with low-modes increases the quality of the 
approximation since these are treated exactly in the inverse of the Dirac operator. 
However in this case there appears the additional computational cost of the eigenvectors. 
So that in AMA we need to find the appropriate value of $N_\lambda$ 
by considering a balance between additional eigenmode cost and benefit for deflation. 
In the DWF case, the benefit of deflation in strange quark mass regime is much less  
than in light quark mass regime. 
As shown in Figure \ref{fig:eigen_dist}, 
one sees that the lowest eigenvalue of the strange quark Dirac operator has 
similar magnitude as in the $N_\lambda=180$ point in both $m=0.005$ and $m=0.01$.
It turns out that the approximation for the strange quark without deflation 
has a similar gain as in the light quark mass with $N_\lambda =$180. 
We know that AMA with $N_\lambda=180$ in $m=0.01$
has a certain cost reduction for two- and three-point functions, and 
thus, at the strange quark mass, AMA without low-mode deflation also 
has an advantage.

AMA is an example of a new class of covariant approximation averaging (CAA)
which reduces the statistical error
on correlation functions in Monte-Carlo simulations in an efficient way. 
Although AMA is similar to low-mode averaging (LMA), 
we have shown that it works not only for low-mode dominated observables 
(associated with the pion)
but also for a broad range of observables involving baryons and other mesons
by taking account of contributions from all modes of the Dirac operator. 
In AMA we have used the conjugate gradient inverter with a relaxed stopping criterion
as the approximation, and numerically tested this method
in lattice QCD with $N_f=2+1$ dynamical domain-wall fermions (DWF) 
on lattice sizes of $24^3\times 64$ and $L_s=16$ and inverse spacing $a^{-1}=1.73$ GeV.
Our tests correspond to pions with masses in the range
300 to 500 MeV.
Using AMA, we have shown reductions of computational cost of more than 5 times compared to the 
standard method 
for nucleon and vector meson masses, the axial charge and isovector form factors of the nucleon.
These results suggest interesting applications to
observables having long-standing hurdles of large statistical noise to precise measurements, 
{\it e.g.} the neutron electric dipole moment, muon anomalous magnetic moment,
and proton decay matrix elements \cite{Aoki:2013yxa}. 
The application of AMA to all of these is now under way.

\begin{acknowledgments}
We thank Norman H. Christ for giving an idea of randomly shifted source method without 
covariant symmetry presented in appendix \ref{sec:notCAA}.
We also thank Yasumichi Aoki, Peter Boyle, Tomoni Ishikawa, Meifeng Lin, 
Robert Mawhinney, Amarjit Soni, Oliver Witzel and fellow members of RBC/UKQCD
collaboration for useful discussion and suggestion.
Numerical calculations were performed using the RICC at RIKEN and the Ds cluster at FNAL. 
This work was supported by the Japanese Ministry of Education Grant-in-Aid, 
Nos. 22540301 (TI), 23105714 (ES), 23105715 (TI) and U.S. DOE grants DE-AC02-98CH10886 (TI) 
and DE-FG02-13ER41989 (TB).
We are grateful to BNL, the RIKEN BNL Research Center, and USQCD for providing resources 
necessary for completion of this work.
\end{acknowledgments}

\appendix

\section{Standard deviation of the improved estimator}\label{sec:sigma}
The standard deviation of the improved estimator in (\ref{eq:imp}) 
is given as
\begin{equation}
  \sigma^{\rm (imp)} = \sqrt{\langle (\Delta \mathcal O^{\rm (imp)})^2\rangle}.
\end{equation}
Here we express the correlation between $\mathcal O$, 
$\mathcal O^{\rm (appx)}$ and $\mathcal O^{{\rm (appx)}\,g}$ as
\begin{eqnarray}
 r_g &=& \frac{\langle\Delta\mathcal O\Delta\mathcal O^{{\rm(appx)}\,g}\rangle}
       {\sigma\sigma_g^{\rm (appx)}},\label{eq:rg}\\
 r_g^{\rm corr} 
     &=& \frac{\langle\Delta\mathcal O^{\rm (appx)}\Delta\mathcal O^{{\rm (appx)}\,g}\rangle}
       {\sigma^{\rm (appx)}\sigma_g^{\rm (appx)}},\label{eq:rgcorr}\\
 r_{gg'}^{\rm corr} 
     &=& \frac{\langle\Delta\mathcal O^{{\rm (appx)}\,g}\Delta\mathcal O^{{\rm (appx)}\,g'}\rangle}
       {\sigma_g^{\rm (appx)}\sigma_{g'}^{\rm (appx)}},\label{eq:rggcorr}
\end{eqnarray} 
where, if $g$ is the unit transformation $I$, we have $r_{I} = r$ 
and $r^{\rm corr}_{Ig'} = r_{g'}^{\rm corr}$.
Substituting (\ref{eq:rg}), (\ref{eq:rgcorr}) and (\ref{eq:rggcorr})
into (\ref{eq:appx}) and (\ref{eq:imp}), we have 
\begin{eqnarray}
  \sigma^{\rm (imp)} 
&=& \Big[
  \sigma^2 - 2r\sigma^{\rm (appx)}\sigma + \sigma^{\rm (appx)\,2}
  + \frac{2}{N_G}\sum_{g}\sigma_g^{\rm (appx)}(r_g\sigma - r_g^{\rm corr}\sigma^{\rm (appx)})
  \nonumber\\
&+& \frac{1}{N_G^2}\Big( \sum_{g}\sigma_g^{\rm appx\,2} 
  + \sum_{g\ne g'}\sigma_g^{\rm (appx)}\sigma_{g'}^{\rm (appx)}r_{gg'}^{\rm corr}
  \Big)\Big]^{1/2}.
\end{eqnarray}
Assuming that the standard deviation of $\mathcal O$ is equivalent with
$\mathcal O^{{\rm (appx)}}$, 
\begin{eqnarray}
&&\sigma^{\rm (appx)}\simeq \sigma^{\rm (appx)}_g \simeq \sigma,\label{eq:sigma_appx}
\end{eqnarray}
we have
\begin{equation}
  \sigma^{\rm (imp)} \simeq \sigma\Big[2(1-r)
  + \frac{2}{N_G}\sum_{g}(r_g - r^{\rm corr}_g)
  + \frac{1}{N^2_G}\Big(N_G + \sum_{g\ne g'}r^{\rm corr}_{gg'}\Big)\Big]^{1/2}.
\end{equation}
Furthermore if the correlation between
$\mathcal O^{\rm (appx)}$ and $\mathcal O^{{\rm (appx)}\,g}$ is negligibly small, 
\begin{equation}
  r_{g}^{\rm corr}\simeq 0,\quad r_{gg'}^{\rm corr}\simeq 0,\quad
  r_{g} \simeq 0,
\end{equation}
(the last one assumes the correlation between 
$\mathcal O^{{\rm (appx)}\,g}$ and $\mathcal O$ is small), 
we have 
\begin{equation}
  \sigma^{\rm (imp)} \simeq \sigma\sqrt{2(1-r)+\frac{2r-1}{N_G}}.
\end{equation}

\section{Note on possible bias due to round-off error}
\label{sec:bias}
In this section, we address the possible appearance of bias due to 
the round-off error for finite machine precision.
Although AMA estimator does not have any bias 
if the exact arithmetic is carried out, 
it is important to notice whether or not a significant breaking of covariant symmetry by round-off error appears. 
We strongly advise that, in practice, one should explicitly check that  
the size of the bias is negligible on a few configuration as is done below (Fig.~\ref{fig:ama_test}), 
or follow the method in Appendix \ref{sec:notCAA} to remove the bias completely.

There are two possible sources.
One is, only when a fixed norm of the residual vector in the CG is used as the stopping condition 
in the approximation part of the improved estimator,
the difference of CG iteration rarely occurring 
in a verge of stopping condition because of inexact arithmetic 
of residual vector-norm computation. 
Second is round-off error accumulating in iterative solver algorithm 
at arithmetic step of multiplication of vector-vector and vector-matrix. 
In our numerical study, however, we show there does not appear it even in sub-\% precision. 

Here the bias is defined as the violation 
of the equivalence Eq.~(\ref{eq:latsym}),
\begin{equation}
  \langle \mathcal O^g[U]\rangle = \langle \mathcal O[U^g]\rangle + \delta_{\mathcal O},
  \label{eq:bias}
\end{equation}
where $\delta_{\mathcal O}\ne 0$ indicates the amount of systematic error. 
This is a consequence of the breaking of covariance
in Eq.~(\ref{eq:cov1}), 
\begin{equation}
  \mathcal O^g[U] \ne \mathcal O[U^g].
  \label{eq:vio_cov}
\end{equation}
This breaking may not be negligible when a very crude approximation is employed, or accumulation of machine-epsilon 
is somehow enhanced under weak circumstances for round-off effect.

\subsection{Threshold error in fixed stopping condition for residual vector}
\label{sec:Threshold}
\begin{figure}
\begin{center}
\includegraphics[width=100mm]{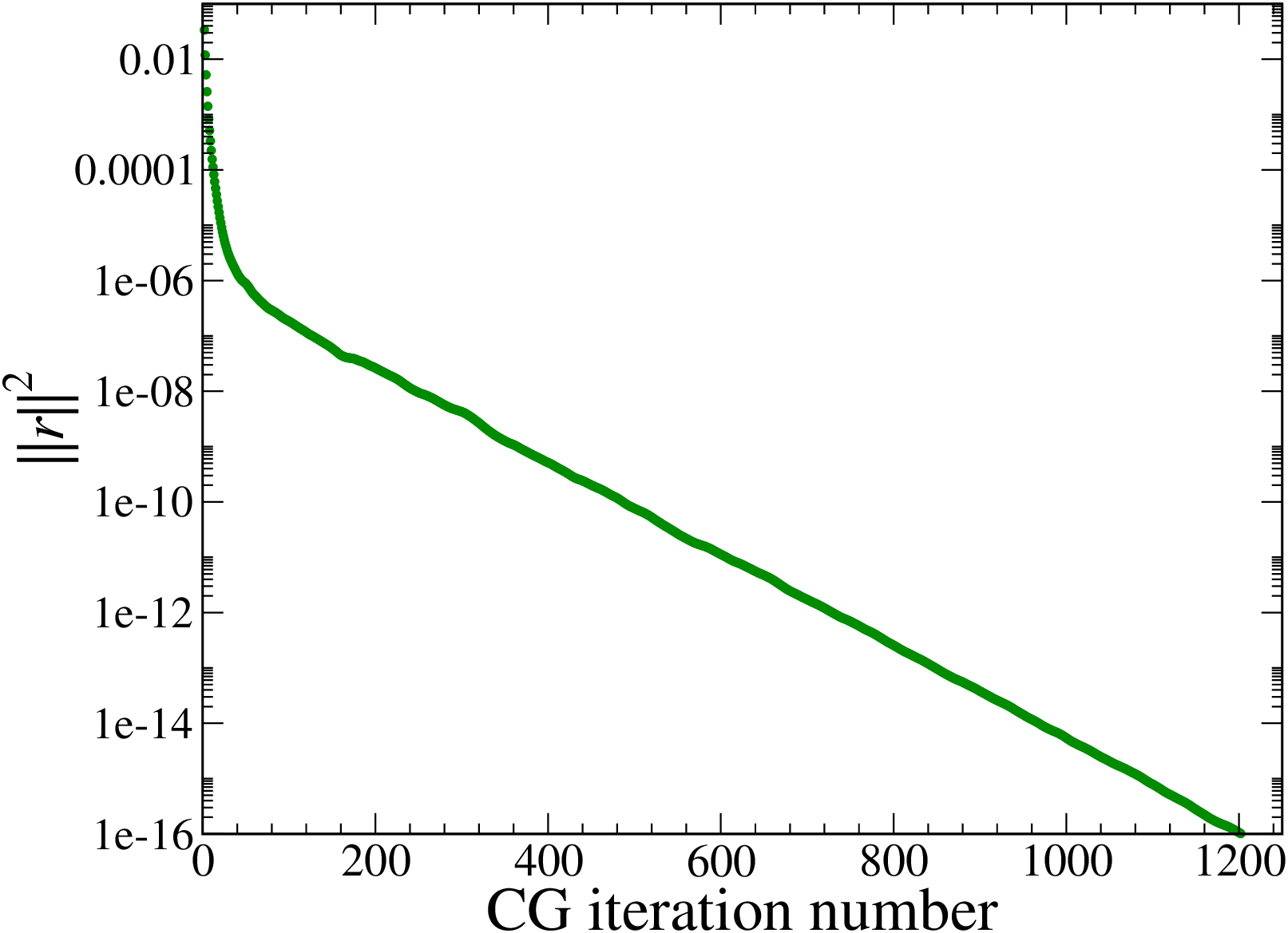}
\caption{
The relation between the squared norm of residual vector and 
CG iteration number. 
}
\label{fig:convert}
\end{center}
\end{figure}

In the following, we show the first example of bias effect and numerical check. 
This is only the most obvious place where small differences due to the finite precision matters.
When we use the CG for the construction of $f_\varepsilon$ 
in the second term of Eq.~(\ref{eq:all-mode}),
the accuracy of $f$ is measured by using the 
residual vector $r$ defined as the difference between
the source vector and matrix $H$ times the approximation vector $f$, $r=b-Hf$.
Its norm corresponds to the accuracy of $f$, $f_\varepsilon$, and 
it is given as the sum over lattice sites, 
\begin{equation}
  ||r||^2 = \sum_{x}r^\dag(x) r(x) = r^\dag(x_1)r(x_1) + r^\dag(x_2)r(x_2) 
          + \cdots + r^\dag(x_V)r(x_V).
\end{equation}
We notice that the above norm 
is slightly different from the one resulting if the right hand side 
of Eq.~(\ref{eq:vio_cov}) is computed instead,
due to the order of arithmetic, 
\begin{eqnarray}
  ||r^g||^2 &=& \sum_{x=x^g} r^\dag(x) r(x) \nonumber\\
  &=& r^\dag(x_1+\delta)r(x_1+\delta) + r^\dag(x_2+\delta)r(x_2+\delta) + \cdots
  + r^\dag(x_V+\delta)r(x_V+\delta) \nonumber\\
  &\ne& ||r||^2, 
\end{eqnarray}
where $g$ denotes the transformation for $x^g= x+\delta$
with constant shift-vector $\delta$. 
When the stopping condition $\varepsilon$ 
used in AMA falls between $||r^g||$ and $||r||$, 
the number of CG iterations is different, 
\begin{equation}
  N_{\rm CG}(||r||) \ne N_{\rm CG}(||r^g||), 
  \label{eq:CGiter}
\end{equation}
which leads to the breaking of Eq.~(\ref{eq:vio_cov}).
This discrepancy affects Eq.~(\ref{eq:all-mode2}),
\begin{equation}
  f_\varepsilon(H[U](x,y)) = \sum_{k=1}^{N_{\rm CG}(||r||)} c_k[U] (H[U])^k(x,y),
\end{equation}
where $N_{\rm CG}(||r||)$, the number of CG iterations when the fixed stopping condition 
of norm of residual vector is used.  
$c_k[U]$ is a coefficient implicitly determined by the CG procedure. 
Because of Eq.~(\ref{eq:CGiter}), 
the discrepancy of the CG part under the transformation $g$ arises as 
\begin{eqnarray}
  f_\varepsilon(H[U](x^g,y^g)) 
  &=& \sum_{k=1}^{N_{\rm CG}(||r||)} c_k[U] H^k[U](x^g,y^g)\nonumber\\
  &=& \sum_{k=1}^{N_{\rm CG}(||r^g||)} c_k[U^g] H^k[U^g](x,y) + \Delta_f
  = f_\varepsilon(H[U^g](x,y)) + \Delta_f,
\end{eqnarray}
(here we assume that $c_k[U] H^k[U](x^g,y^g) = c_k[U^g] H^k[U^g](x,y)$
within machine precision).
$\Delta_f$ does not vanish when 
accidentally different number of iteration by round-off error appears as in Eq.~(\ref{eq:CGiter}). 
Therefore there is no guarantee of cancellation between 
$\langle\mathcal O^{\rm AMA}\rangle$ and $\langle\mathcal O_G^{\rm AMA}\rangle$. 
This breaking may be significant if a very low precision for the stopping condition is chosen, 
where $f_\varepsilon$ rapidly changes for the initial CG iterations.
For example, as seen from Figure~\ref{fig:convert}, 
when the CG iteration number is changed from $N_{\rm CG}=20$ to 21, 
the accuracy of solution vector changes by the order $||r(x)||\simeq 10^{-3}$. 
On the other hand, in the region of $N_{\rm CG}=1200$, 
even if $N_{\rm CG}$ is changed from 1200 to 1201, 
the accuracy of solution vector is still 
less than $||r(x)||\simeq 10^{-9}$, 
and it turns out that the effect of different $N_{\rm CG}$ of relaxed CG in $\mathcal O^{\rm (appx)}$
is more significant than $N_{\rm CG}$ of exact CG in $\mathcal O$
(and also such bias is totally suppressed within machine precision for $\mathcal O$).
Obviously this kind of bias does not appear when 
$f_\varepsilon$ is constructed by a fixed CG iteration number
instead of fixed norm of residual vector as the stopping condition. 

\begin{algorithm}[t]
\caption{CG algorithm for solving $A x_{\rm CG} = b$ with positive Hermitte matrix $A$}
\label{alg:CG}
\begin{algorithmic}[1]
 \IF{ $k := 0$ } 
 \STATE $x_0 :=0$ \label{CG x0}
 \STATE $r_0 := b-A x_0$, $p_0 := r_0$ \label{CG r0}
 \ENDIF
\WHILE{ $||r_k|| > \epsilon$ } 
\STATE $\alpha_k := { (r_k,r_k)\over (p_k, Ap_k)}$ 
\STATE $x_{k+1} := x_k + \alpha_k p_k$, $r_{k+1} := r_k - \alpha_k Ap_k$ \label{CG rk}
\STATE $\beta_k :={(r_{k+1},r_{k+1})\over (r_{k},r_{k})} $
\STATE $p_{k+1} := r_{k+1} + \beta_k p_k$ \label{CG pk}
\STATE $k := k+1$
\ENDWHILE
\end{algorithmic}
\end{algorithm}


\begin{figure}
\begin{center}
\includegraphics[width=100mm]{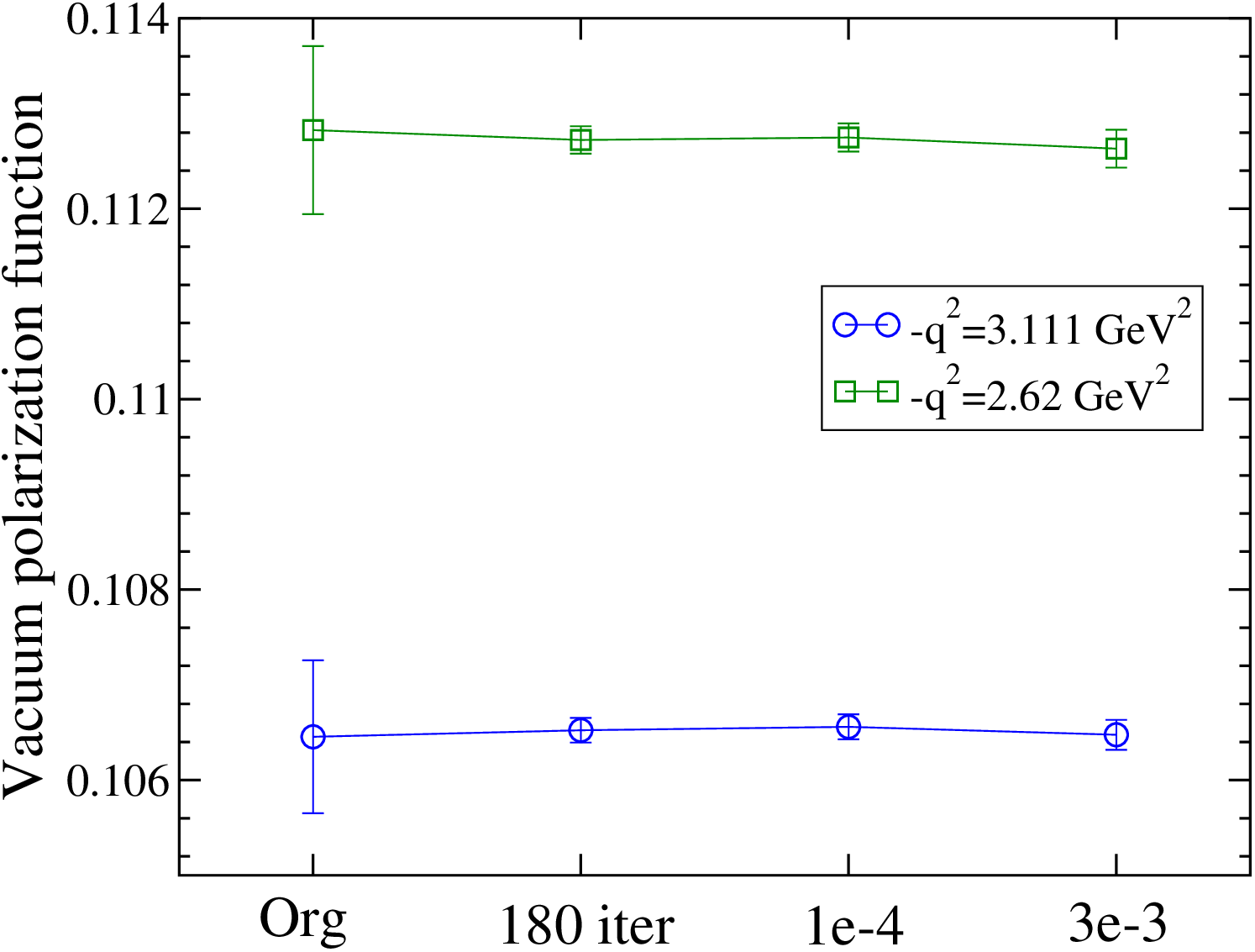}
\caption{
The vacuum polarization function of vector type using same gauge configurations 
at $m=0.005$. The number of configurations are 51. 
``Org'' denotes the results without AMA, 
``1e-4'', ``3e-3'' and ``180 iter'' denote the AMA results using $10^{-4}$, $3\times10^{-3}$ stopping 
criteria of norm of residual vector and 180 CG iteration respectively. 
The different symbols are results in different $-q^2$ point of vacuum polarization function.
}\label{fig:VP_test}
\end{center}
\end{figure}

In Figure~\ref{fig:VP_test} we numerically compare the result of the vacuum polarization function (VPF)
with two procedures of AMA used in $3\times 10^{-3}$ and $10^{-4}$ stopping condition 
for the norm of residual vector and 180 CG iterations. 
The VPF is extracted from the conserved vector and local vector current 
correlator following \cite{Shintani:2008ga,Boyle:2011hu,Aubin:2006xv}.
One sees that the resulting values of the VPF from two different stopping conditions are consistent
within statistical error whose accuracy is at the sub-percent level.
This result supports that the systematic error of arithmetic bias addressed in this section 
is not visible in the practical calculations.
Note that the mechanism that enhances the size of the bias due to the threshold effect of the residual 
vector norm mentioned above is avoided when using fixed CG iteration number. 

\subsection{Accumulated round-off error}
\label{sec:AccRoundoff}
The round-off error due to inexact arithmetic in an iterative solver could potentially destroy 
the covariance that is crucial for AMA and introduce bias. Below we show in a realistic case 
that the round-off error is innocuous. 
CAA conceptually relies on preserving the covariant symmetry in each
iteration, {\it e.g.} from step 6 to step 9 in Algorithm~\ref{alg:CG}. 
After many vector-vector and matrix-vector multiplies to determine the residual and search vectors, 
the accumulation of round-off error due to the different order of arithmetic may spoil 
the exact covariant symmetry.
The extent to which the symmetry is violated, of course, depends too on the details of the algorithm
\footnote{
For example, the BiCG-type algorithm which is much less stable than CG 
may be more susceptible to accumulated effects of round-off.
We thank T. Doi for pointing this out to us after making a test with Wilson-clover fermions.
}.

\begin{figure}
\begin{center}
\includegraphics[width=100mm]{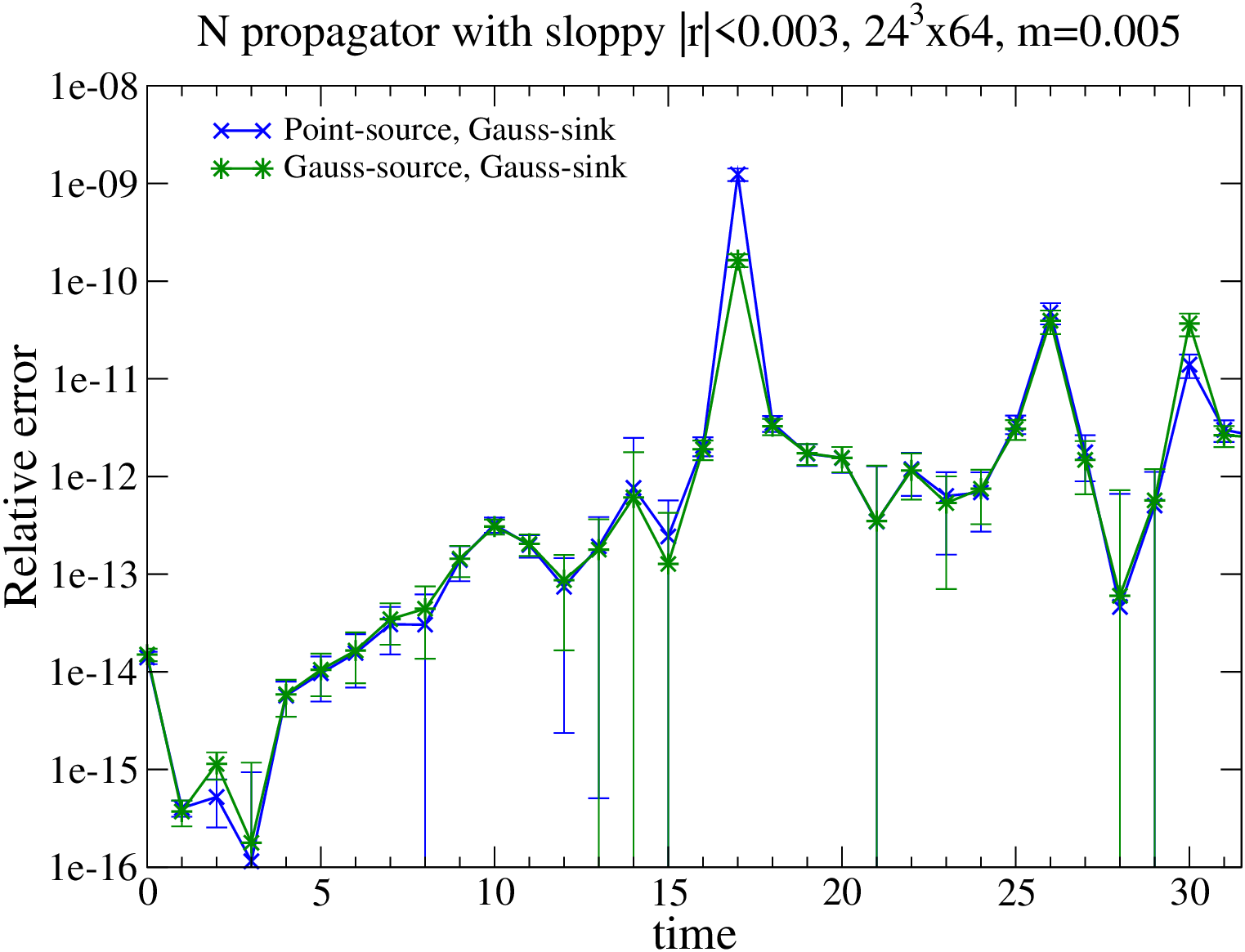}
\caption{
Relative error of $\delta_c$ for nucleon propagator 
with Gaussian-source and point sink (cross) and 
Gaussian-source and Gaussian-sink (star) as a function of time-slice. 
This is averaged one using 16 source locations on one reference configuration with low-mode deflation 
using 400 low-modes at $m=0.005$ in $24^3\times 64$ lattice.
}\label{fig:ama_test}
\end{center}
\end{figure}

To check the preservation of covariance in the AMA approximation,
we compare nucleon two-point correlation functions with those computed 
after translating the position of both the nucleon source and the gauge links. 
If the floating point arithmetic were exact, the nucleon correlation functions would have be identical 
which means the bias in AMA is zero. The bias caused by the finite precision arithmetic is quantified as 
\begin{equation}
  \delta_c = \mathcal O^{ {\rm (appx)}\,g }[U^{\bar g}] - \mathcal O^{ {\rm (appx)} }[U], 
\end{equation}
where $g$ denotes the transformation, and $\bar g$ denotes the 
inverse transformation of $g$. 
In our test the source position and link variables are shifted using 16 different translations, 
$(12,0,0,0)$, $(0,12,0,0)$, $\dots, (12,12,12,32)$ on one configuration. 
The only difference with the original unshifted calculation is the order of arithmetic in the Lanczos and 
CG algorithm according to the shift of the gauge configuration and fermion source point.
In Figure~\ref{fig:ama_test}, one sees that the effect of round-off error on the covariant symmetry, 
when using the low-mode deflation with 400 low-lying eigenmodes as used in the present work, 
is $O(10^{-9})$ (and much smaller in the part of the correlation function that is statistically well-resolved) and does not depend on smeared or local source type. 
Thus the approximation with sloppy CG 
using 0.003 residual stopping condition is not 
significantly affected by accumulative round-off errors, and hence systematic bias. 
In fact, even if such round-off error did introduce a bias due to the
relative order of arithmetic, it can be removed by the technique explained 
in the next section which does not rely on covariance.

\section{Error reduction technique without covariant symmetry}
\label{sec:notCAA}
In this section we introduce the another estimator
in which the random transformation $g_r\in G_r$ is adopted for 
$\mathcal O^{\rm (appx)}$ instead of covariance. 
Employing $g_r$, which is assumed as the element of group $G_r$, 
into Eq.~(\ref{eq:imp}), the improved estimator is defined as
\begin{eqnarray}
  \mathcal O^{{\rm (imp)}\,g_r}
  &=& \mathcal O^{g_r} - \mathcal O^{{\rm (appx)}\,g_r}
      + \mathcal O_{G}^{{\rm (appx)}\,g_r},
  \label{eq:imp_other}
  \\
  \mathcal O^{{\rm (appx)}\,g_r}_{G}
  &=& \frac{1}{N_G}\sum_{g\in G}\mathcal O^{{\rm (appx)}\,g\circ g_r}.
\label{eq:imp_other_g}
\end{eqnarray}
The second equation has the 
multi-transformation $g\circ g_r$ with $g$ and $g_r$ for $\mathcal O^{\rm (appx)}$. 
Here we also assume $G$ as the subset of $G_r$

We prove that this estimator does not have any bias provided the numerical procedure of $\mathcal O^{\rm (appx)}$ 
is deterministic and reproducible, these are the calculation is bit-by-bit same for the same input parameters 
(gauge configuration, source location, stopping criteria etc). 
We note that our program is always checked to reproduce bit-by-bit same results for same input.
Since the biasless estimator should satisfy the equivalence 
of expectation value as
\begin{equation}
  \Big\langle \mathcal O \Big\rangle = \Big\langle \mathcal O^{g_r} \Big\rangle =
  \Big\langle \mathcal O^{{\rm (imp)}\,g_r} \Big\rangle,
\end{equation}
(here we consider $\mathcal O$ is covariant under $g_r$)
thus, from Eq.~(\ref{eq:imp_other}) and using the 
transformation of the link variable with $g_r$, we show
\begin{equation}
  \Big\langle \mathcal O^{{\rm (appx)}\,g_r} \Big\rangle 
  = \Big\langle\mathcal O_{G}^{{\rm (appx)}\,g_r}\Big\rangle,
  \label{eq:equive}
\end{equation}
even if $\mathcal O^{\rm (appx)}$ does not follow from a covariant symmetry.
In the above, the expectation value is defined as the group integral of 
link variables and the summation over $g_r\in G_r$. 

The left-hand-side of Eq.~(\ref{eq:equive}) is described as, 
\begin{equation}
  \Big\langle \mathcal O^{{\rm (appx)}\,g_r} \Big\rangle
  = \frac{1}{Z}\sum_{g_r\in G_r} P(g_r)\int dU\,e^{-S[U]}
    \mathcal O^{{\rm (appx)}\,g_r}[U],
\label{eq:lhs}
\end{equation}
where $S[U]$ denotes the QCD action, and $P(g_r)$ denotes the 
distribution function of $g_r\in G_r$ normalized to unity.
$Z$ is the partition function. 
On the other hand the right-hand-side of Eq.~(\ref{eq:equive}) can be written as 
\begin{equation}
  \Big\langle\mathcal O_{G}^{{\rm (appx)}\,g_r}\Big\rangle
  = \frac{1}{Z}\sum_{g_r\in G_r} P(g_r)\frac{1}{N_G}\sum_{g\in G}\int dU\,e^{-S[U]}
  \mathcal O^{{\rm (appx)}\,g\circ g_r}[U].
\label{eq:rhs}
\end{equation}
Here we consider that the multiplication of $g_r\in G_r$ with $g\in G$
is also an element of $G_r$, {\it i.e.} $g\circ g_r\in G_r$, and
the distribution function of $g\circ g_r$ is the same
function of $g_r\in G_r$, {\it i.e.} $P(g\circ g_r)=P(g_r)$, 
when $G \subseteq G_r$.
In this case, Eq.~(\ref{eq:rhs}) can be expressed as 
a single sum over $g_r\in G_r$, 
and so its equation is equivalent to Eq.~(\ref{eq:lhs}). 
We notice that in this derivation it is unnecessary to use the covariance 
of $\mathcal O^{\rm (appx)}$.
Practically $g_r$ is chosen randomly
in each configuration, for instance a random shift of source location
for $\mathcal O$ and $\mathcal O^{\rm (appx)}$.
Hence, 
to avoid any bias due to the arithmetic error explained in Section \ref{sec:bias}, 
$\mathcal O^{{\rm (imp)}\,g_r}$ instead of $\mathcal O^{{\rm (imp)}}$ is appropriate
when the CG stopping condition is chosen as the fixed norm of residual vector.
Note that, in Eq.~(\ref{eq:imp_other}), 
$g_r$ is only performed for
each functional; the link variables are not transformed.
When the link variable is transformed instead of $\mathcal O^{\rm (appx)}$, 
the bias-less of $\mathcal O^{{\rm (imp)}\,g_r}$ is only guaranteed 
for $\mathcal O^{\rm (appx)}$ by the covariance under $G$ and $G_r$. 

\section{Implicitly restarted Lanczos algorithm with polynomial acceleration}
\label{sec:Lanczos}
Suppose that $A\in \mathbb C^{N\times N}$ is the Hermitian, positive definite, matrix.
Introducing the tridiagonal matrix $T\in \mathbb C^{m\times m}$
whose diagonal and off-diagonal components are $\alpha_{i=1,\cdots,m}$ and 
$\beta_{i=1,\cdots,m-1}$, respectively, the relation
\begin{equation}
  AV = VT + r_me_m^\dag,
\end{equation}
provides $T$ and the orthogonal matrix $V \in\mathbb C^{N\times m}$ 
recursively as shown in 
Algorithm~\ref{alg:lanczos}.
In the above equation $e_m$ denotes the unit vector with non-zero value
in the $m$-th component.
If $V^\dag r_m\simeq 0$, the $k(\le m)$-th eigenvector $\psi_k$ and eigenvalue ($\lambda_k$)
of matrix A are given by the multiplication of the unitary matrix 
obtained by the diagonalization for tridiagonal matrix,
$T = U^\dag \Lambda U$, as $UV = \{\psi_k\},\,\Lambda = {\rm diag}(\lambda_k)$.

The restarted Lanczos algorithm is based on the concept to recycle the 
the final vector $v_m$ in the Lanczos iteration as the new initial vector $v^{\rm new}$ 
in order to avoid the storage constraints. 
Suppose that $m$ is divided into $k$ {\it wanted} eigenvectors 
$\{v_{1},\cdots,v_{k}\}$ which is 
the desired region of the eigenvalue distribution, and 
$p$ {\it unwanted} vectors $\{v_{k+1},\cdots,v_{k+p}\}$
which are recomputed in every step of the Lanczos iteration after restarting. 
After running $m\equiv k+p$ Lanczos steps, we restart the Lanczos process with 
initial vector and $\beta$ value,
\begin{equation}
 v^{\rm new}_{k+1} = v_m, \quad \beta'_{k}=\beta_m,
 \label{eq:new_v}
\end{equation}
and thus the orthogonal matrix $V$ is constructed by 
\begin{equation}
  V = \{v_1,\cdots,v_k\}\cup\{v^{\rm new}_{k+1},\cdots,v^{\rm new}_{m}\} 
  \subset \{v_1,\cdots,v_{m+p}\}.
  \label{eq:v_region}
\end{equation}
Effectively after the restarted Lanczos step we obtain 
vectors $v_i$ spanning the Krylov space $\mathcal K_{m+p}(A,v_1)$.
The last equation in (\ref{eq:v_region}) may be broken 
due to round-off errors, leading to loss of 
orthogonality in the restarted process,
since it does not take account of 
reorthogonalization with previous {\it unwanted} vectors $\{v_{k+1},\cdots,v_{m}\}$.
Such an effect, however, depends on the choice of $p$, and 
in the actual lattice QCD simulation, less than 5 time restarted 
Lanczos process has no matter of orthogonality loss. 

Usually we implement the filtering technique using 
QR factorization and shifting the resulting tridiagonal matrix.
In this algorithm we employ the approximate {\it unwanted} eigenvalues 
as shift parameters $\mu_i=\tilde\lambda_{i=k+1,\cdots,m}$ and
obtain the orthogonal matrix $Q=\prod_{i=1}^pQ_i$ from
the QR factorization process (see Algorithm \ref{alg:QR}).

\begin{algorithm}[t]
\caption{QR factrization process}
\label{alg:QR}
\begin{algorithmic}[1]
\STATE Let set $T_1 = T$ and $i=1$
\WHILE{$i=p$}
\STATE $T_i-\mu_i = Q_iR_i$
\STATE $R_iQ_i + \mu_i= T_{i+1}$
\STATE $i = i + 1$
\ENDWHILE
\end{algorithmic}
\end{algorithm}
$V_+=VQ$ and $T_p$ are also satisfied with the Lanczos recursion relation
\begin{equation}
  (AV_+)_{ij} = (V_+ T_p)_{ij} + (r_m)_i Q_{mj} ,\quad V_+ = VQ,
\end{equation}
and thus the new initial vector $v^{\rm new}_{k+1}$ alternative to Eq.~(\ref{eq:new_v})
consists of
\begin{equation}
  r^{\rm new}_{k+1} = v_{k+1}^+  + Q_{mk}r_m,\quad 
  \beta_{k+1}^{\rm new} = ||r_{k+1}^{\rm new}||,\quad
  v^{\rm new}_{k+1} = r^{\rm new}_{k+1} /\beta_{k+1}^{\rm new},
\end{equation}
with rotated vector $v_{i}^+ = \sum_{l=1}^m Q_{li}v_l$ $[i=1,\cdots,k+1]$.
In the above we use the relation of $T_{i+1}=Q_i^\dag T_iQ_i$ and $Q_{m,i}=0$ $[i=1,\cdots,k-1]$. 
Therefore we can restart the Lanczos step from $k+1$ to $k+p$ 
following Algorithm~\ref{alg:lanczos}, 
and we generate the new orthogonal matrix:
\begin{equation}
  V^{\rm new} = \{v_1^+,\cdots,v^+_k\}\cup\{v^{\rm new}_{k+1},\cdots,v^{\rm new}_{m}\}
\end{equation}
which is also spans the Krylov space $\mathcal K_{m+p}(A,v_1)$.
Note that via QR factorization the new {\it wanted} vector $v_{1,\cdots,k}^+$ 
is automatically multiplied by the filtering polynomial function
\begin{equation}
  f_p(A) = \prod_{i=k+1}^{m}(A-\tilde\lambda_i),
\end{equation}
and thus
\begin{equation}
  v^+_i \propto f_p(A)v_i
  \label{eq:filter}
\end{equation}
which is known from the relation of $V_+e_1 = VQe_1 \propto f_p(A)v_1$.
The filtering polynomial function may suppress the {\it unwanted} vectors. 
Fulfilling the {\it unwanted} eigenvalue 
constraints on $|f(\lambda_{i=k+1,\cdots,m})|<|f(\lambda_k)|$, 
the polynomial function of Eq.~(\ref{eq:filter}) works as a filter of 
{\it unwanted} eigenmodes from spectrum of $A$
\cite{Sorensen:1992:IAP:131879.131903,calvetti1994implicitly}. 

\begin{figure}
\begin{center}
\includegraphics[width=100mm]{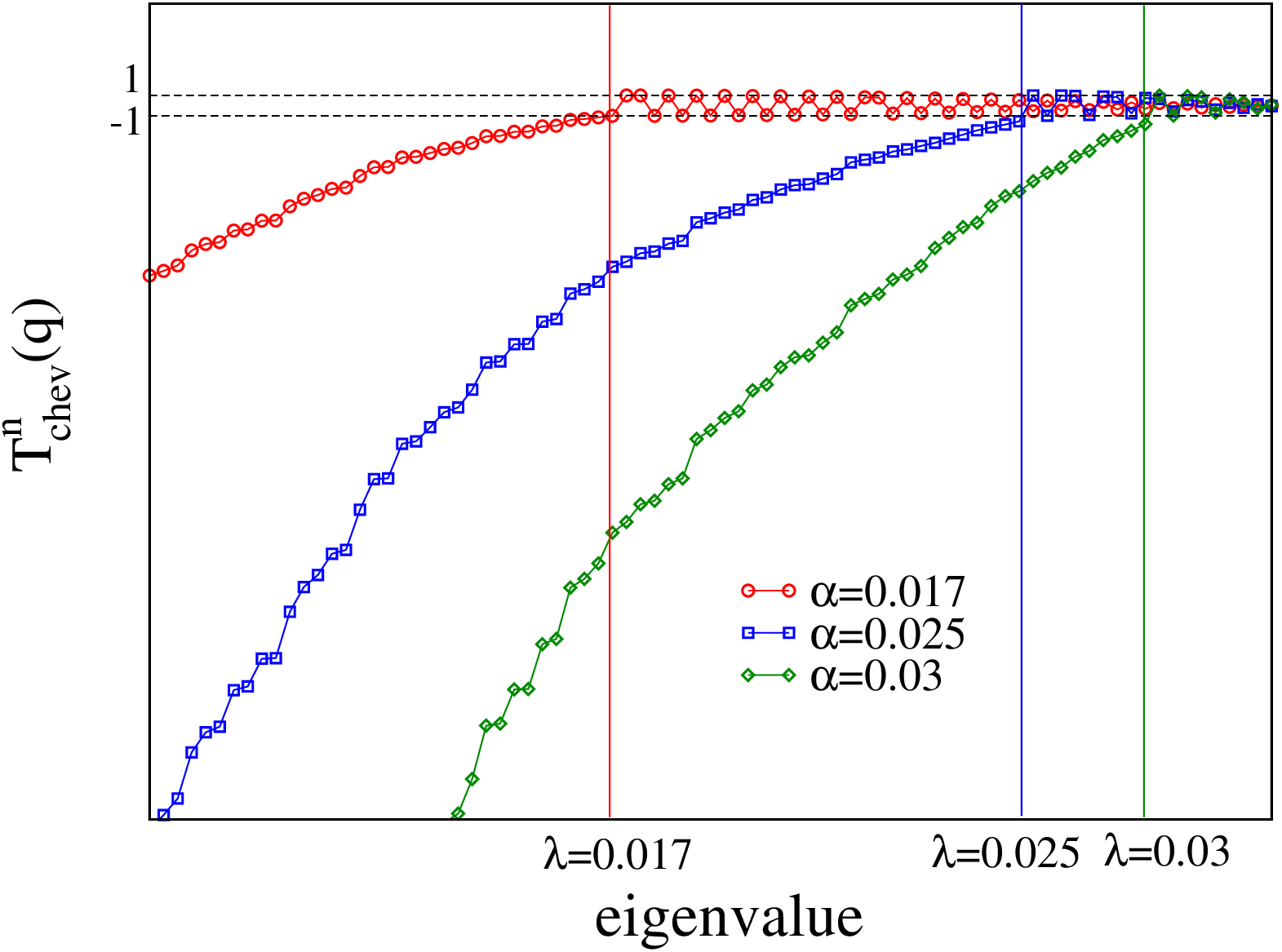}
\caption{
The sketch of Chebychev polynomial $T^n_{\rm chev}(q)$ as a function of eigenvalue. 
Different symbols illustrate the $T^n_{\rm chev}(q)$ with several $\alpha$. 
}\label{fig:Tn}
\end{center}
\end{figure}

The restarted Lanczos algorithm combined with polynomial acceleration \cite{Saad:1984}
emphasizes the low-lying 
{\it wanted} eigenvectors in the Krylov space and suppresses the {\it unwanted} vector 
via the filtering function. 
Let us consider the computation of the low-modes of Hermitian matrix $H$ 
whose maximum absolute eigenvalue is already known as $\lambda_{\rm max}$.
The Chebychev polynomial function $T_{\rm chev}$ can be used to easily control 
the eigenvalue distribution of $H$ by enhancing the wanted small eigenvalue region 
$( \lambda < \alpha)$ and suppressing the unwanted region. 
By applying $T_{\rm chev}$ with the following argument function
 \begin{eqnarray}
  q(H) = \frac{2H^2-\alpha^2-\beta^2}{\beta^2-\alpha^2},
  \label{eq:qH}
\end{eqnarray}
we have that
\begin{eqnarray}
\begin{array}{ll}
|T_{\rm chev}^n(q(\lambda))|\gg 1,& \lambda^2\not\in[\alpha^2,\beta^2],\\
T_{\rm chev}^n(q(\lambda))\in [-1,1],& \lambda^2\in[\alpha^2,\beta^2],\\
\end{array}
  \label{eq:Tn}
\end{eqnarray}
where we set  $\alpha$ slightly larger than the maximum {\it wanted}
eigenvalue, and $\beta^2\ge\lambda^2_{\rm max}$
(see Figure \ref{fig:Tn}).
$T^n_{\rm chev}(q(H))$, constructed by a recursion relation, 
$T_{\rm chev}^n(x) = 2x T_{\rm chev}^{n-1}(x) - T_{\rm chev}^{n-2}(x)$,
has the same eigenvectors as $H$ 
and the highest eigenvalue of $T^n_{\rm chev}(q(H))$
corresponds to the lowest eigenvalue of $H$.
The degree $n$ of $T^n_{\rm chev}$, which is also the number of its zeroes 
in $[-1,1]$, depends on the magnitude of 
the highest eigenvalue and the hierarchy of magnitudes for the {\it wanted} eigenvalues.
Recalling the restarted Lanczos process, if we set $\alpha$ close to 
the lowest point in the eigenvalue region $\lambda_{i=k+1,\cdots,m}$,  
the filtering function in Eq.~(\ref{eq:filter}) strongly
suppresses the {\it unwanted} eigenvalue region.

We easily extend the polynomial acceleration techniques 
to focus on an arbitrary range of {\it wanted} 
eigenvalues by introducing the shift parameter $\mu$ into Eq.~(\ref{eq:qH}),
\begin{equation}
  q(H,\mu) = \frac{2(H-\mu)^2-\alpha^2-(\beta+|\mu|)^2}{(\beta+|\mu|)^2-\alpha^2},
  \label{eq:qHmu}
\end{equation}
in which this argument function enhances the spectrum in the range 
$\lambda = ( \mu-\alpha, \mu+\alpha )$.

\begin{algorithm}[t]
\caption{Lanczos algorithm}
\label{alg:lanczos}
\begin{algorithmic}[1]
\STATE Set $v_1$ to the unit vector, $\beta_0=0$ and $k=0$;
\WHILE{ $k = m$ }
\STATE $\alpha_k = (v_k,Av_k)$;
\STATE $r_k = (A-\alpha_k)v_k - \beta_{k-1} v_{k-1}$;
\STATE $\beta_k = ||r_k||$;
\STATE $v_{k+1} = r_k/\beta_k$;
\STATE Gram-Schmit reorthogonalization for $v_1,\cdots, v_{k+1}$ if we needed;
\STATE $k = k + 1$;
\ENDWHILE
\end{algorithmic}
\end{algorithm}

\section{4D even-odd preconditioning in domain-wall fermions}
\label{sec:4Deo}

In this section we explicitly present the definition of domain-wall fermion (DWF)
4D even-odd preconditioning (see \cite{Pochinsky} and \cite{Brower:2012vk} and references therein)
which is used not only in the preconditioning of the CG solver, 
but also in the computation of eigenvectors and eigenvalues in the Lanczos algorithm.
Instead of DWF 5D even-odd preconditioning as has been used in \cite{Aoki:2010dy}, 
the DWF operator can be expressed as the even-odd hopping matrix
in 4D space-time in which the Wilson-fermion kernel of DWF is in the off-diagonal 
blocks and 5D hopping term is in diagonal blocks of the following matrix,
\begin{eqnarray}
  D_{DW}((x,s),(y,t)) 
 &=& (5-M_5)\Big[ \delta_{x,y}W_5(s,t) - KW_4(x,y)\delta_{s,t}\Big]\nonumber\\
 &=& (2K)^{-1}\left(
     \begin{array}{cc} I_{ee}W_5 & -KW_{4\,eo} \\ -KW_{4\,oe} & I_{oo}W_5
     \end{array}\right)
\end{eqnarray}
in which we use
\begin{eqnarray}
  K &=& \frac{1}{2(5-M_5)},\\
  W_4(x,y) &=& \sum_\mu\Big[ (1+\gamma_\mu)U^\dag(x-\hat\mu)\delta_{x-\hat\mu,y}
                          + (1-\gamma_\mu)U(x)\delta_{x+\hat\mu,y} \Big],\\
  W_5(s,t) &=& 1-2K\Big( P_R\delta_{s,t+1} + P_L\delta_{s+1,t} 
               - mP_R\delta_{s,1}\delta_{t,L_s}- mP_L\delta_{s,L_s}\delta_{t,1}\Big),
\end{eqnarray}
with SU(3) link variable $U_\mu(x)$ and Dirac $\gamma$-matrix.
Here we suppress color and spin indices in the DWF operator.
Even- or odd-ness  of a site of Euclidean space-time is given 
as $\mod(\sum_{\mu=1}^4 x_\mu,2) =$ 0 or 1. 
$M_5$ is the so-called domain wall height.

The inverse of the DWF operator in even-odd representation is expressed through the
Schur decomposition as,
\begin{eqnarray}
  D_{DW}^{-1} &=& (2K)^{-1}
  \left( \begin{array}{cc}
    I_{ee} & 0 \\ KW_5^{-1}W_{4\,eo} & I_{oo}W_5^{-1}
  \end{array} \right)
  \left( \begin{array}{cc}
    D_{ee}^{-1} & 0 \\ 0 & I_{oo} \\
  \end{array}\right)
  \left( \begin{array}{cc}
    I_{ee} & KW_{4\,eo}W_5^{-1} \\ 0 & I_{oo} \\
  \end{array}\right),\\
  D_{ee} &=& I_{ee}W_5 - K^2W_{4\,eo}W_5^{-1}W_{4\,oe},
  \label{eq:D_ee}
\end{eqnarray}
in which the inverse of $W_5$ can be represented explicitly,
\begin{eqnarray}
  &&W_5^{-1}(s,t) = A(s,t)P_R + B(s,t)P_L,
  \label{eq:W5inv}\\
  &&A(s,t) = \delta_{st} - \frac{1}{1+m\kappa^{L_s}}\left(
  \begin{array}{cccccc}
  m\kappa^{L_s} & m\kappa^{L_s-1} & m\kappa^{L_s-2} & \cdots & m\kappa \\
  -\kappa & m\kappa^{L_s} & m\kappa^{L_s-1} & \cdots & m\kappa^2\\
  -\kappa^2 & -\kappa & m\kappa^{L_s} & \cdots & m\kappa^3\\
  \vdots & \vdots & \vdots & & \vdots \\
  -\kappa^{L_s-1} & -\kappa^{L_s-2} & -\kappa^{L_s-3} & \cdots & m\kappa^{L_s}
  \end{array}\right)_{st},\\
  &&B(s,t) = A(t,s),
\end{eqnarray}
with $\kappa = (5-M_5)^{-1}$.

In a practical implementation of $W_5^{-1}$, it is convenient to use the LU decomposition.
Using the left and right representation of $W_5$,
\begin{equation}
  W_5(s,t) = P_R\big[ I - \kappa(\Delta + \Delta_m)\big]_{st}
           + P_L\big[ I - \kappa(\Delta^T + \Delta_m^T)\big]_{st}
\end{equation}
with 
\begin{equation}
  \Delta = \left(\begin{array}{ccccccc}
  0 &   &   &  & & & 0 \\
  1 & 0 & \\
    & 1 & 0 &  &   &   \\
    &   & \ddots& \ddots  \\
    &   &   &  & 1 & 0 &  \\
  0 &   &   &   &  & 1 & 0
  \end{array}\right),\quad
  \Delta_m = \left(\begin{array}{ccccccc}
   0& & & & -m \\
    & & & & \\
    & &\ddots & \\
    & & & & \\
   0& & & & 0
  \end{array}\right).
\end{equation}
We also know that the matrix without $P_R$ is represented as
\begin{equation}
  \big[ I - \kappa(\Delta + \Delta_m)\big]
   = \big(1-\kappa\Delta_m(I-\kappa\Delta)^{-1}\big) (I-\kappa\Delta),
\end{equation}
and 
\begin{equation}
  (I-\kappa\Delta)^{-1} = \left(\begin{array}{ccccccc}
  1 & & & & &  \\
  \kappa & 1 \\
  \vdots & & \ddots\\
  \kappa^{L_s-2} & \kappa^{L_s-3} & \cdots & \kappa   & 1 & \\ 
  \kappa^{L_s-1} & \kappa^{L_s-2} & \cdots & \kappa^2 & \kappa & 1 
  \end{array}\right).
\end{equation}
Thus we have
\begin{eqnarray}
&& \big[ I - \kappa(\Delta + \Delta_m)\big]\nonumber\\
&& = \left(\begin{array}{ccccccc}
  1+m\kappa^{L_s} & m\kappa^{L_s-1} & m\kappa^{L_s-2} & \cdots & m\kappa \\
  0 & 1 & 0 & \cdots & 0\\
  \vdots& & \ddots \\
  & &        & 1 & 0\\
  0 & & \cdots & 0 & 1 
  \end{array}\right)
  \left(\begin{array}{ccccccc}
  1 & 0  \\
  -\kappa & 1 & 0 \\
    & \ddots & \ddots&\ddots \\
  & &  -\kappa & 1 & 0\\
  & &  & -\kappa & 1 
  \end{array}\right).
\end{eqnarray}
Finally we obtain 
\begin{eqnarray}
&&A(s,t) = \nonumber\\
&&
  \left(\begin{array}{ccccccc}
  1 & & & & &  \\
  \kappa & 1 \\
  \vdots & & \ddots\\
  \kappa^{L_s-2} & \kappa^{L_s-3} & \cdots & \kappa   & 1 & \\ 
  \kappa^{L_s-1} & \kappa^{L_s-2} & \cdots & \kappa^2 & \kappa & 1 
  \end{array}\right)
  \left(\begin{array}{ccccccc}
  \frac{1}{1+m\kappa^{L_s}} & \frac{-m\kappa^{L_s-1}}{1+m\kappa^{L_s}} & \frac{-m\kappa^{L_s-2}}{1+m\kappa^{L_s}} & \cdots & \frac{-m\kappa}{1+m\kappa^{L_s}} \\
  0 & 1 & 0 & \cdots & 0\\
  \vdots& & \ddots \\
  & &        & 1 & 0\\
  0 & & \cdots & 0 & 1 
  \end{array}\right)
  .
\end{eqnarray}
Now the number of floating-point operations in the multiplication of 
$A(s, t)$ with a vector is reduced to $(L_s^2-1)/2$ from $L_s^2$, 
{\it i.e.} a gain of $(L_s^2+1)/2$.

$\gamma_5$-Hermiticity of the DWF operator is given by
\begin{equation}
  D_{DW}^\dag(s,t) = \sum_{s_1,t_1}\Gamma_5(s,s_1) D_{DW}(s_1,t_1)\Gamma_5(t_1,t),
  \label{eq:ddag}
\end{equation}
with $\Gamma_5(s,t) = \gamma_5\delta_{s,L_s-t+1}$, hence 
the Hermiticity of the even-odd preconditioned Domain-wall operator
\begin{equation}
  H_{ee} = \Gamma_5 D_{ee},
  \label{eq:Hee}
\end{equation}
follows from $D_{ee}$, $D_{ee}^\dag = \Gamma_5 D_{ee}\Gamma_5$,
since $\Gamma_5$ is a diagonal matrix at each 4D even-odd site.
The difference from DWF 5D even-odd preconditioning is 
that $H_{ee}$ can be represented as a single multiplication of $\Gamma_5$ without 
a flip of even-odd site. 
Eq.~(\ref{eq:Hee}) can be used in the Lanczos algorithm with 
$H=H_{ee}$ in Eq.~(\ref{eq:qH}) and (\ref{eq:qHmu}). 

\bibliography{ref.bib}

\begin{thebibliography}{45}
\expandafter\ifx\csname natexlab\endcsname\relax\def\natexlab#1{#1}\fi
\expandafter\ifx\csname bibnamefont\endcsname\relax
  \def\bibnamefont#1{#1}\fi
\expandafter\ifx\csname bibfnamefont\endcsname\relax
  \def\bibfnamefont#1{#1}\fi
\expandafter\ifx\csname citenamefont\endcsname\relax
  \def\citenamefont#1{#1}\fi
\expandafter\ifx\csname url\endcsname\relax
  \def\url#1{\texttt{#1}}\fi
\expandafter\ifx\csname urlprefix\endcsname\relax\def\urlprefix{URL }\fi
\providecommand{\bibinfo}[2]{#2}
\providecommand{\eprint}[2][]{\url{#2}}

\bibitem[{\citenamefont{Shintani et~al.}(2005)\citenamefont{Shintani, Aoki,
  Ishizuka, Kanaya, Kikukawa et~al.}}]{Shintani:2005xg}
\bibinfo{author}{\bibfnamefont{E.}~\bibnamefont{Shintani}},
  \bibinfo{author}{\bibfnamefont{S.}~\bibnamefont{Aoki}},
  \bibinfo{author}{\bibfnamefont{N.}~\bibnamefont{Ishizuka}},
  \bibinfo{author}{\bibfnamefont{K.}~\bibnamefont{Kanaya}},
  \bibinfo{author}{\bibfnamefont{Y.}~\bibnamefont{Kikukawa}},
  \bibnamefont{et~al.}, \bibinfo{journal}{Phys.Rev.}
  \textbf{\bibinfo{volume}{D72}}, \bibinfo{pages}{014504}
  (\bibinfo{year}{2005}), \eprint{hep-lat/0505022}.

\bibitem[{\citenamefont{Berruto et~al.}(2006)\citenamefont{Berruto, Blum,
  Orginos, and Soni}}]{Berruto:2005hg}
\bibinfo{author}{\bibfnamefont{F.}~\bibnamefont{Berruto}},
  \bibinfo{author}{\bibfnamefont{T.}~\bibnamefont{Blum}},
  \bibinfo{author}{\bibfnamefont{K.}~\bibnamefont{Orginos}}, \bibnamefont{and}
  \bibinfo{author}{\bibfnamefont{A.}~\bibnamefont{Soni}},
  \bibinfo{journal}{Phys. Rev.} \textbf{\bibinfo{volume}{D73}},
  \bibinfo{pages}{054509} (\bibinfo{year}{2006}), \eprint{hep-lat/0512004}.

\bibitem[{\citenamefont{Shintani et~al.}(2007)\citenamefont{Shintani, Aoki,
  Ishizuka, Kanaya, Kikukawa et~al.}}]{Shintani:2006xr}
\bibinfo{author}{\bibfnamefont{E.}~\bibnamefont{Shintani}},
  \bibinfo{author}{\bibfnamefont{S.}~\bibnamefont{Aoki}},
  \bibinfo{author}{\bibfnamefont{N.}~\bibnamefont{Ishizuka}},
  \bibinfo{author}{\bibfnamefont{K.}~\bibnamefont{Kanaya}},
  \bibinfo{author}{\bibfnamefont{Y.}~\bibnamefont{Kikukawa}},
  \bibnamefont{et~al.}, \bibinfo{journal}{Phys.Rev.}
  \textbf{\bibinfo{volume}{D75}}, \bibinfo{pages}{034507}
  (\bibinfo{year}{2007}), \eprint{hep-lat/0611032}.

\bibitem[{\citenamefont{Shintani et~al.}(2008)\citenamefont{Shintani, Aoki, and
  Kuramashi}}]{Shintani:2008nt}
\bibinfo{author}{\bibfnamefont{E.}~\bibnamefont{Shintani}},
  \bibinfo{author}{\bibfnamefont{S.}~\bibnamefont{Aoki}}, \bibnamefont{and}
  \bibinfo{author}{\bibfnamefont{Y.}~\bibnamefont{Kuramashi}},
  \bibinfo{journal}{Phys. Rev.} \textbf{\bibinfo{volume}{D78}},
  \bibinfo{pages}{014503} (\bibinfo{year}{2008}), \eprint{0803.0797}.

\bibitem[{\citenamefont{Blum et~al.}(2012{\natexlab{a}})\citenamefont{Blum,
  Hayakawa, and Izubuchi}}]{Blum:2013qu}
\bibinfo{author}{\bibfnamefont{T.}~\bibnamefont{Blum}},
  \bibinfo{author}{\bibfnamefont{M.}~\bibnamefont{Hayakawa}}, \bibnamefont{and}
  \bibinfo{author}{\bibfnamefont{T.}~\bibnamefont{Izubuchi}},
  \bibinfo{journal}{PoS} \textbf{\bibinfo{volume}{LATTICE2012}},
  \bibinfo{pages}{022} (\bibinfo{year}{2012}{\natexlab{a}}),
  \eprint{1301.2607}.

\bibitem[{\citenamefont{Christ et~al.}(2010)\citenamefont{Christ, Dawson,
  Izubuchi, Jung, Liu et~al.}}]{Christ:2010dd}
\bibinfo{author}{\bibfnamefont{N.}~\bibnamefont{Christ}},
  \bibinfo{author}{\bibfnamefont{C.}~\bibnamefont{Dawson}},
  \bibinfo{author}{\bibfnamefont{T.}~\bibnamefont{Izubuchi}},
  \bibinfo{author}{\bibfnamefont{C.}~\bibnamefont{Jung}},
  \bibinfo{author}{\bibfnamefont{Q.}~\bibnamefont{Liu}}, \bibnamefont{et~al.},
  \bibinfo{journal}{Phys.Rev.Lett.} \textbf{\bibinfo{volume}{105}},
  \bibinfo{pages}{241601} (\bibinfo{year}{2010}), \eprint{1002.2999}.

\bibitem[{\citenamefont{Blum et~al.}(2013)\citenamefont{Blum, Izubuchi, and
  Shintani}}]{Blum:2012uh}
\bibinfo{author}{\bibfnamefont{T.}~\bibnamefont{Blum}},
  \bibinfo{author}{\bibfnamefont{T.}~\bibnamefont{Izubuchi}}, \bibnamefont{and}
  \bibinfo{author}{\bibfnamefont{E.}~\bibnamefont{Shintani}},
  \bibinfo{journal}{Phys.Rev.} \textbf{\bibinfo{volume}{D88}},
  \bibinfo{pages}{094503} (\bibinfo{year}{2013}), \eprint{1208.4349}.

\bibitem[{\citenamefont{Giusti et~al.}(2003)\citenamefont{Giusti, Hoelbling,
  Luscher, and Wittig}}]{Giusti:2002sm}
\bibinfo{author}{\bibfnamefont{L.}~\bibnamefont{Giusti}},
  \bibinfo{author}{\bibfnamefont{C.}~\bibnamefont{Hoelbling}},
  \bibinfo{author}{\bibfnamefont{M.}~\bibnamefont{Luscher}}, \bibnamefont{and}
  \bibinfo{author}{\bibfnamefont{H.}~\bibnamefont{Wittig}},
  \bibinfo{journal}{Comput. Phys. Commun.} \textbf{\bibinfo{volume}{153}},
  \bibinfo{pages}{31} (\bibinfo{year}{2003}), \eprint{hep-lat/0212012}.

\bibitem[{\citenamefont{Giusti et~al.}(2004)\citenamefont{Giusti, Hernandez,
  Laine, Weisz, and Wittig}}]{Giusti:2004yp}
\bibinfo{author}{\bibfnamefont{L.}~\bibnamefont{Giusti}},
  \bibinfo{author}{\bibfnamefont{P.}~\bibnamefont{Hernandez}},
  \bibinfo{author}{\bibfnamefont{M.}~\bibnamefont{Laine}},
  \bibinfo{author}{\bibfnamefont{P.}~\bibnamefont{Weisz}}, \bibnamefont{and}
  \bibinfo{author}{\bibfnamefont{H.}~\bibnamefont{Wittig}},
  \bibinfo{journal}{JHEP} \textbf{\bibinfo{volume}{04}}, \bibinfo{pages}{013}
  (\bibinfo{year}{2004}), \eprint{hep-lat/0402002}.

\bibitem[{\citenamefont{DeGrand and Schaefer}(2004)}]{DeGrand:2004qw}
\bibinfo{author}{\bibfnamefont{T.~A.} \bibnamefont{DeGrand}} \bibnamefont{and}
  \bibinfo{author}{\bibfnamefont{S.}~\bibnamefont{Schaefer}},
  \bibinfo{journal}{Comput. Phys. Commun.} \textbf{\bibinfo{volume}{159}},
  \bibinfo{pages}{185} (\bibinfo{year}{2004}), \eprint{hep-lat/0401011}.

\bibitem[{\citenamefont{DeGrand and Schaefer}(2005)}]{DeGrand:2005vb}
\bibinfo{author}{\bibfnamefont{T.~A.} \bibnamefont{DeGrand}} \bibnamefont{and}
  \bibinfo{author}{\bibfnamefont{S.}~\bibnamefont{Schaefer}},
  \bibinfo{journal}{Phys. Rev.} \textbf{\bibinfo{volume}{D72}},
  \bibinfo{pages}{054503} (\bibinfo{year}{2005}), \eprint{hep-lat/0506021}.

\bibitem[{\citenamefont{Luscher}(2007)}]{Luscher:2007es}
\bibinfo{author}{\bibfnamefont{M.}~\bibnamefont{Luscher}},
  \bibinfo{journal}{JHEP} \textbf{\bibinfo{volume}{0712}}, \bibinfo{pages}{011}
  (\bibinfo{year}{2007}), \eprint{0710.5417}.

\bibitem[{\citenamefont{Fukaya et~al.}(2007)}]{Fukaya:2007fb}
\bibinfo{author}{\bibfnamefont{H.}~\bibnamefont{Fukaya}} \bibnamefont{et~al.}
  (\bibinfo{collaboration}{JLQCD}), \bibinfo{journal}{Phys. Rev. Lett.}
  \textbf{\bibinfo{volume}{98}}, \bibinfo{pages}{172001}
  (\bibinfo{year}{2007}), \eprint{hep-lat/0702003}.

\bibitem[{\citenamefont{Noaki et~al.}(2008)}]{Noaki:2008iy}
\bibinfo{author}{\bibfnamefont{J.}~\bibnamefont{Noaki}} \bibnamefont{et~al.}
  (\bibinfo{collaboration}{JLQCD and TWQCD}), \bibinfo{journal}{Phys. Rev.
  Lett.} \textbf{\bibinfo{volume}{101}}, \bibinfo{pages}{202004}
  (\bibinfo{year}{2008}), \eprint{0806.0894}.

\bibitem[{\citenamefont{Giusti and Necco}(2006)}]{Giusti:2005sx}
\bibinfo{author}{\bibfnamefont{L.}~\bibnamefont{Giusti}} \bibnamefont{and}
  \bibinfo{author}{\bibfnamefont{S.}~\bibnamefont{Necco}},
  \bibinfo{journal}{PoS} \textbf{\bibinfo{volume}{LAT2005}},
  \bibinfo{pages}{132} (\bibinfo{year}{2006}), \eprint{hep-lat/0510011}.

\bibitem[{\citenamefont{Li et~al.}(2010)}]{Li:2010pw}
\bibinfo{author}{\bibfnamefont{A.}~\bibnamefont{Li}} \bibnamefont{et~al.}
  (\bibinfo{collaboration}{xQCD}), \bibinfo{journal}{Phys. Rev.}
  \textbf{\bibinfo{volume}{D82}}, \bibinfo{pages}{114501}
  (\bibinfo{year}{2010}), \eprint{1005.5424}.

\bibitem[{\citenamefont{Bali et~al.}(2010{\natexlab{a}})\citenamefont{Bali,
  Castagnini, and Collins}}]{Bali:2010se}
\bibinfo{author}{\bibfnamefont{G.}~\bibnamefont{Bali}},
  \bibinfo{author}{\bibfnamefont{L.}~\bibnamefont{Castagnini}},
  \bibnamefont{and} \bibinfo{author}{\bibfnamefont{S.}~\bibnamefont{Collins}},
  \bibinfo{journal}{PoS} \textbf{\bibinfo{volume}{LATTICE2010}},
  \bibinfo{pages}{096} (\bibinfo{year}{2010}{\natexlab{a}}),
  \eprint{1011.1353}.

\bibitem[{\citenamefont{Gong et~al.}(2013)\citenamefont{Gong, Alexandru, Chen,
  Doi, Dong et~al.}}]{Gong:2013vja}
\bibinfo{author}{\bibfnamefont{M.}~\bibnamefont{Gong}},
  \bibinfo{author}{\bibfnamefont{A.}~\bibnamefont{Alexandru}},
  \bibinfo{author}{\bibfnamefont{Y.}~\bibnamefont{Chen}},
  \bibinfo{author}{\bibfnamefont{T.}~\bibnamefont{Doi}},
  \bibinfo{author}{\bibfnamefont{S.}~\bibnamefont{Dong}}, \bibnamefont{et~al.},
  \bibinfo{journal}{Phys.Rev.} \textbf{\bibinfo{volume}{D88}},
  \bibinfo{pages}{014503} (\bibinfo{year}{2013}), \eprint{1304.1194}.

\bibitem[{\citenamefont{Bali et~al.}(2010{\natexlab{b}})\citenamefont{Bali,
  Collins, and Schafer}}]{Bali:2009hu}
\bibinfo{author}{\bibfnamefont{G.~S.} \bibnamefont{Bali}},
  \bibinfo{author}{\bibfnamefont{S.}~\bibnamefont{Collins}}, \bibnamefont{and}
  \bibinfo{author}{\bibfnamefont{A.}~\bibnamefont{Schafer}},
  \bibinfo{journal}{Comput. Phys. Commun.} \textbf{\bibinfo{volume}{181}},
  \bibinfo{pages}{1570} (\bibinfo{year}{2010}{\natexlab{b}}),
  \eprint{0910.3970}.

\bibitem[{\citenamefont{Kolorenc and Mitas}(2011)}]{0034-4885-74-2-026502}
\bibinfo{author}{\bibfnamefont{J.}~\bibnamefont{Kolorenc}} \bibnamefont{and}
  \bibinfo{author}{\bibfnamefont{L.}~\bibnamefont{Mitas}},
  \bibinfo{journal}{Reports on Progress in Physics}
  \textbf{\bibinfo{volume}{74}}, \bibinfo{pages}{026502}
  (\bibinfo{year}{2011}),
  \urlprefix\url{http://stacks.iop.org/0034-4885/74/i=2/a=026502}.

\bibitem[{\citenamefont{Pollet}(2012)}]{0034-4885-75-9-094501}
\bibinfo{author}{\bibfnamefont{L.}~\bibnamefont{Pollet}},
  \bibinfo{journal}{Reports on Progress in Physics}
  \textbf{\bibinfo{volume}{75}}, \bibinfo{pages}{094501}
  (\bibinfo{year}{2012}),
  \urlprefix\url{http://stacks.iop.org/0034-4885/75/i=9/a=094501}.

\bibitem[{\citenamefont{Foulkes et~al.}(2001)\citenamefont{Foulkes, Mitas,
  Needs, and Rajagopal}}]{Foulkes:2001zz}
\bibinfo{author}{\bibfnamefont{W.}~\bibnamefont{Foulkes}},
  \bibinfo{author}{\bibfnamefont{L.}~\bibnamefont{Mitas}},
  \bibinfo{author}{\bibfnamefont{R.}~\bibnamefont{Needs}}, \bibnamefont{and}
  \bibinfo{author}{\bibfnamefont{G.}~\bibnamefont{Rajagopal}},
  \bibinfo{journal}{Rev.Mod.Phys.} \textbf{\bibinfo{volume}{73}},
  \bibinfo{pages}{33} (\bibinfo{year}{2001}).

\bibitem[{\citenamefont{Carlson et~al.}(2012)\citenamefont{Carlson, Gandolfi,
  and Gezerlis}}]{Carlson:2012mh}
\bibinfo{author}{\bibfnamefont{J.}~\bibnamefont{Carlson}},
  \bibinfo{author}{\bibfnamefont{S.}~\bibnamefont{Gandolfi}}, \bibnamefont{and}
  \bibinfo{author}{\bibfnamefont{A.}~\bibnamefont{Gezerlis}},
  \bibinfo{journal}{PTEP} \textbf{\bibinfo{volume}{2012}},
  \bibinfo{pages}{01A209} (\bibinfo{year}{2012}), \eprint{1210.6659}.

\bibitem[{\citenamefont{Leidemann and Orlandini}(2012)}]{Leidemann:2012hr}
\bibinfo{author}{\bibfnamefont{W.}~\bibnamefont{Leidemann}} \bibnamefont{and}
  \bibinfo{author}{\bibfnamefont{G.}~\bibnamefont{Orlandini}}
  (\bibinfo{year}{2012}), \eprint{1204.4617}.

\bibitem[{\citenamefont{Blum et~al.}(2012{\natexlab{b}})\citenamefont{Blum,
  Izubuchi, and Shintani}}]{Blum:2012my}
\bibinfo{author}{\bibfnamefont{T.}~\bibnamefont{Blum}},
  \bibinfo{author}{\bibfnamefont{T.}~\bibnamefont{Izubuchi}}, \bibnamefont{and}
  \bibinfo{author}{\bibfnamefont{E.}~\bibnamefont{Shintani}},
  \bibinfo{journal}{PoS} \textbf{\bibinfo{volume}{LATTICE2012}},
  \bibinfo{pages}{262} (\bibinfo{year}{2012}{\natexlab{b}}),
  \eprint{1212.5542}.

\bibitem[{\citenamefont{Aoki et~al.}(2011)}]{Aoki:2010dy}
\bibinfo{author}{\bibfnamefont{Y.}~\bibnamefont{Aoki}} \bibnamefont{et~al.}
  (\bibinfo{collaboration}{RBC Collaboration, UKQCD Collaboration}),
  \bibinfo{journal}{Phys.Rev.} \textbf{\bibinfo{volume}{D83}},
  \bibinfo{pages}{074508} (\bibinfo{year}{2011}), \eprint{1011.0892}.

\bibitem[{\citenamefont{Saad}(1984)}]{Saad:1984}
\bibinfo{author}{\bibfnamefont{Y.}~\bibnamefont{Saad}},
  \bibinfo{journal}{Math.Comp.} \textbf{\bibinfo{volume}{42}},
  \bibinfo{pages}{567} (\bibinfo{year}{1984}).

\bibitem[{\citenamefont{Sorensen}(1992)}]{Sorensen:1992:IAP:131879.131903}
\bibinfo{author}{\bibfnamefont{D.~C.} \bibnamefont{Sorensen}},
  \bibinfo{journal}{SIAM J. Matrix Anal. Appl.} \textbf{\bibinfo{volume}{13}},
  \bibinfo{pages}{357} (\bibinfo{year}{1992}), ISSN \bibinfo{issn}{0895-4798},
  \urlprefix\url{http://dx.doi.org/10.1137/0613025}.

\bibitem[{\citenamefont{Calvetti et~al.}(1994)\citenamefont{Calvetti, Reichel,
  and Sorensen}}]{calvetti1994implicitly}
\bibinfo{author}{\bibfnamefont{D.}~\bibnamefont{Calvetti}},
  \bibinfo{author}{\bibfnamefont{L.}~\bibnamefont{Reichel}}, \bibnamefont{and}
  \bibinfo{author}{\bibfnamefont{D.~C.} \bibnamefont{Sorensen}},
  \bibinfo{journal}{Electronic Transactions on Numerical Analysis}
  \textbf{\bibinfo{volume}{2}}, \bibinfo{pages}{21} (\bibinfo{year}{1994}).

\bibitem[{\citenamefont{Neff et~al.}(2001)\citenamefont{Neff, Eicker, Lippert,
  Negele, and Schilling}}]{Neff:2001zr}
\bibinfo{author}{\bibfnamefont{H.}~\bibnamefont{Neff}},
  \bibinfo{author}{\bibfnamefont{N.}~\bibnamefont{Eicker}},
  \bibinfo{author}{\bibfnamefont{T.}~\bibnamefont{Lippert}},
  \bibinfo{author}{\bibfnamefont{J.~W.} \bibnamefont{Negele}},
  \bibnamefont{and}
  \bibinfo{author}{\bibfnamefont{K.}~\bibnamefont{Schilling}},
  \bibinfo{journal}{Phys. Rev.} \textbf{\bibinfo{volume}{D64}},
  \bibinfo{pages}{114509} (\bibinfo{year}{2001}), \eprint{hep-lat/0106016}.

\bibitem[{\citenamefont{Yamazaki et~al.}(2009)\citenamefont{Yamazaki, Aoki,
  Blum, Lin, Ohta et~al.}}]{Yamazaki:2009zq}
\bibinfo{author}{\bibfnamefont{T.}~\bibnamefont{Yamazaki}},
  \bibinfo{author}{\bibfnamefont{Y.}~\bibnamefont{Aoki}},
  \bibinfo{author}{\bibfnamefont{T.}~\bibnamefont{Blum}},
  \bibinfo{author}{\bibfnamefont{H.-W.} \bibnamefont{Lin}},
  \bibinfo{author}{\bibfnamefont{S.}~\bibnamefont{Ohta}}, \bibnamefont{et~al.},
  \bibinfo{journal}{Phys.Rev.} \textbf{\bibinfo{volume}{D79}},
  \bibinfo{pages}{114505} (\bibinfo{year}{2009}), \eprint{0904.2039}.

\bibitem[{\citenamefont{Sasaki et~al.}(2002)\citenamefont{Sasaki, Blum, and
  Ohta}}]{Sasaki:2001nf}
\bibinfo{author}{\bibfnamefont{S.}~\bibnamefont{Sasaki}},
  \bibinfo{author}{\bibfnamefont{T.}~\bibnamefont{Blum}}, \bibnamefont{and}
  \bibinfo{author}{\bibfnamefont{S.}~\bibnamefont{Ohta}},
  \bibinfo{journal}{Phys.Rev.} \textbf{\bibinfo{volume}{D65}},
  \bibinfo{pages}{074503} (\bibinfo{year}{2002}), \eprint{hep-lat/0102010}.

\bibitem[{\citenamefont{Lin}(2011)}]{Lin:2011ti}
\bibinfo{author}{\bibfnamefont{H.-W.} \bibnamefont{Lin}},
  \bibinfo{journal}{Chin.J.Phys.} \textbf{\bibinfo{volume}{49}},
  \bibinfo{pages}{827} (\bibinfo{year}{2011}), \eprint{1106.1608}.

\bibitem[{\citenamefont{Brower et~al.}(2012)\citenamefont{Brower, Neff, and
  Orginos}}]{Brower:2012vk}
\bibinfo{author}{\bibfnamefont{R.~C.} \bibnamefont{Brower}},
  \bibinfo{author}{\bibfnamefont{H.}~\bibnamefont{Neff}}, \bibnamefont{and}
  \bibinfo{author}{\bibfnamefont{K.}~\bibnamefont{Orginos}}
  (\bibinfo{year}{2012}), \eprint{1206.5214}.

\bibitem[{\citenamefont{Borici}(2000)}]{Borici:1999zw}
\bibinfo{author}{\bibfnamefont{A.}~\bibnamefont{Borici}},
  \bibinfo{journal}{Nucl.Phys.Proc.Suppl.} \textbf{\bibinfo{volume}{83}},
  \bibinfo{pages}{771} (\bibinfo{year}{2000}), \eprint{hep-lat/9909057}.

\bibitem[{\citenamefont{Borici}(2004)}]{Borici:2004pn}
\bibinfo{author}{\bibfnamefont{A.}~\bibnamefont{Borici}}, pp.
  \bibinfo{pages}{25--39} (\bibinfo{year}{2004}), \eprint{hep-lat/0402035}.

\bibitem[{\citenamefont{Arthur et~al.}(2013)}]{Arthur:2012opa}
\bibinfo{author}{\bibfnamefont{R.}~\bibnamefont{Arthur}} \bibnamefont{et~al.}
  (\bibinfo{collaboration}{RBC Collaboration, UKQCD Collaboration}),
  \bibinfo{journal}{Phys.Rev.} \textbf{\bibinfo{volume}{D87}},
  \bibinfo{pages}{094514} (\bibinfo{year}{2013}), \eprint{1208.4412}.

\bibitem[{\citenamefont{Abdel-Rehim et~al.}(2013)\citenamefont{Abdel-Rehim,
  Stathopoulos, and Orginos}}]{Abdel-Rehim:2013xgv}
\bibinfo{author}{\bibfnamefont{A.}~\bibnamefont{Abdel-Rehim}},
  \bibinfo{author}{\bibfnamefont{A.}~\bibnamefont{Stathopoulos}},
  \bibnamefont{and} \bibinfo{author}{\bibfnamefont{K.}~\bibnamefont{Orginos}}
  (\bibinfo{year}{2013}), \eprint{1302.4077}.

\bibitem[{\citenamefont{Blum et~al.}(2014)\citenamefont{Blum, Boyle, Christ,
  Frison, Garron et~al.}}]{Blum:2014wsa}
\bibinfo{author}{\bibfnamefont{T.}~\bibnamefont{Blum}},
  \bibinfo{author}{\bibfnamefont{P.}~\bibnamefont{Boyle}},
  \bibinfo{author}{\bibfnamefont{N.}~\bibnamefont{Christ}},
  \bibinfo{author}{\bibfnamefont{J.}~\bibnamefont{Frison}},
  \bibinfo{author}{\bibfnamefont{N.}~\bibnamefont{Garron}},
  \bibnamefont{et~al.}, \bibinfo{journal}{PoS}
  \textbf{\bibinfo{volume}{LATTICE2013}}, \bibinfo{pages}{404}
  (\bibinfo{year}{2014}).

\bibitem[{\citenamefont{Yin and Mawhinney}(2011)}]{Yin:2011np}
\bibinfo{author}{\bibfnamefont{H.}~\bibnamefont{Yin}} \bibnamefont{and}
  \bibinfo{author}{\bibfnamefont{R.~D.} \bibnamefont{Mawhinney}},
  \bibinfo{journal}{PoS} \textbf{\bibinfo{volume}{LATTICE2011}},
  \bibinfo{pages}{051} (\bibinfo{year}{2011}), \eprint{1111.5059}.

\bibitem[{\citenamefont{Aoki et~al.}(2013)\citenamefont{Aoki, Shintani, and
  Soni}}]{Aoki:2013yxa}
\bibinfo{author}{\bibfnamefont{Y.}~\bibnamefont{Aoki}},
  \bibinfo{author}{\bibfnamefont{E.}~\bibnamefont{Shintani}}, \bibnamefont{and}
  \bibinfo{author}{\bibfnamefont{A.}~\bibnamefont{Soni}}
  (\bibinfo{year}{2013}), \eprint{1304.7424}.

\bibitem[{\citenamefont{Shintani et~al.}(2009)}]{Shintani:2008ga}
\bibinfo{author}{\bibfnamefont{E.}~\bibnamefont{Shintani}} \bibnamefont{et~al.}
  (\bibinfo{collaboration}{JLQCD Collaboration, TWQCD Collaboration}),
  \bibinfo{journal}{Phys.Rev.} \textbf{\bibinfo{volume}{D79}},
  \bibinfo{pages}{074510} (\bibinfo{year}{2009}), \eprint{0807.0556}.

\bibitem[{\citenamefont{Boyle et~al.}(2012)\citenamefont{Boyle, Del~Debbio,
  Kerrane, and Zanotti}}]{Boyle:2011hu}
\bibinfo{author}{\bibfnamefont{P.}~\bibnamefont{Boyle}},
  \bibinfo{author}{\bibfnamefont{L.}~\bibnamefont{Del~Debbio}},
  \bibinfo{author}{\bibfnamefont{E.}~\bibnamefont{Kerrane}}, \bibnamefont{and}
  \bibinfo{author}{\bibfnamefont{J.}~\bibnamefont{Zanotti}},
  \bibinfo{journal}{Phys.Rev.} \textbf{\bibinfo{volume}{D85}},
  \bibinfo{pages}{074504} (\bibinfo{year}{2012}), \eprint{1107.1497}.

\bibitem[{\citenamefont{Aubin and Blum}(2007)}]{Aubin:2006xv}
\bibinfo{author}{\bibfnamefont{C.}~\bibnamefont{Aubin}} \bibnamefont{and}
  \bibinfo{author}{\bibfnamefont{T.}~\bibnamefont{Blum}},
  \bibinfo{journal}{Phys.Rev.} \textbf{\bibinfo{volume}{D75}},
  \bibinfo{pages}{114502} (\bibinfo{year}{2007}), \eprint{hep-lat/0608011}.

\bibitem[{\citenamefont{Pochinsky}(2008)}]{Pochinsky}
\bibinfo{author}{\bibfnamefont{A.}~\bibnamefont{Pochinsky}}
  (\bibinfo{year}{2008}), \urlprefix\url{http://www.mit.edu/~avp/mdwf/}.

\end{thebibliography}
\end{document}